\documentclass[fleqn,usenatbib]{mnras}

\usepackage{newtxtext,newtxmath}
\usepackage{pgffor}
\usepackage{longtable}
\usepackage{booktabs}
\usepackage{ulem}

\usepackage[T1]{fontenc}

\DeclareRobustCommand{\VAN}[3]{#2}
\let\VANthebibliography\thebibliography
\def\thebibliography{\DeclareRobustCommand{\VAN}[3]{##3}\VANthebibliography}

\usepackage{graphicx}	
\usepackage{amsmath}	
\usepackage{tabularx}
\usepackage{multirow}
\usepackage{fix-cm}

\title[Diffuse radio emission in galaxy groups]{Bridging the mass gap: Diffuse radio emission in GAMA galaxy groups using EMU and DINGO survey data}

\author[Sai Wagh et al.]{
Sai Wagh,$^{1,2}$\thanks{E-mail: sai.wagh@research.uwa.edu.au}
Tessa Vernstrom,$^{2,1}$
Luke J. M. Davies,$^{1}$
Lister Staveley-Smith,$^{1}$
Stefan Duchesne,$^{2}$
\newauthor
Christopher J. Riseley,$^{3,4}$
Timothy J. Galvin,$^{2}$
Franco Vazza,$^{5}$
Konstantinos Kolokythas,$^{6,7,8}$ 
\newauthor
Andrew M. Hopkins,$^{9}$
Jonghwan Rhee,$^{1,2}$
Tobias Westmeier,$^{1}$
Pascal Jahan Elahi,$^{10}$
and Martin Meyer$^{1}$\\
$^{1}$International Centre for Radio Astronomy Research (ICRAR), University of Western Australia, 35 Stirling Hwy, Crawley, WA 6009, Australia\\
$^{2}$Australia Telescope National Facility, CSIRO Space \& Astronomy, PO Box 1130, Bentley, WA 6102, Australia\\
$^{3}$Astronomisches Institut der Ruhr-Universit\"{a}t Bochum (AIRUB), Universit\"{a}tsstra{\ss}e 150, 44801 Bochum, Germany\\
$^{4}$Ruhr Astroparticle and Plasma Physics Center (RAPP Center), 44780 Bochum, Germany\\
$^{5}$University of Bologna, Italy-40122\\
$^{6}$Centre for Radio Astronomy Techniques and Technologies, Department of Physics and Electronics, Rhodes University, Makhanda 6140, South Africa\\
$^{7}$South African Radio Astronomy Observatory, 2 Fir Street, Observatory
7925, South Africa\\
$^{8}$ INAF -- Istituto di Radioastronomia, via Gobetti 101, I–40129 Bologna, Italy \\
$^{9}$School of Mathematical and Physical Sciences, 12 Wally’s Walk, Macquarie University, NSW 2109, Australia\\
$^{10}$Pawsey Supercomputing Research Centre, 1 Bryce Avenue, 6151 WA
}

\date{Accepted XXX. Received YYY; in original form ZZZ}

\pubyear{\the\year{}}

\begin{document}
\label{firstpage}
\pagerange{\pageref{firstpage}--\pageref{lastpage}}
\maketitle

\begin{abstract}
Diffuse radio emission provides a powerful probe of non-thermal processes in the large-scale structure, yet its properties in galaxy groups remain poorly constrained. Using deep 943 MHz radio continuum data from the Evolutionary Map of the Universe (EMU) and 1.37 GHz data from the Deep Investigations of Neutral Gas Origins (DINGO) survey, we investigate diffuse radio emission in 400 galaxy groups selected from the GAMA survey at $z < 0.1$. We employ a multi-resolution filtering technique to suppress compact radio sources and enhance extended, low-surface-brightness emission associated with the intra group medium. Integrated flux densities are measured within group radii, and background fluctuations are quantified using random control regions. While most systems yield non-detections, we identify 46/400 galaxy groups with candidate diffuse emission, spanning radio powers of $10^{19}-10^{24}\,\mathrm{W\,Hz^{-1}}$. Stacked measurements reveal a weak positive trend between radio power and halo mass. The observed emission levels lie above simple extrapolations of cluster scaling relations, suggesting that different physical processes dominate in the group regime. Additionally, stellar mass ratios of the most massive galaxies in the group and Early Type Galaxy fractions suggest that these galaxy groups are relatively young, evolving systems where galaxy interactions and mergers may power the emission. Comparisons with Magneto Hydrodynamical simulations indicate shock acceleration alone cannot explain the observed emission, pointing to an important role for fossil plasma re-acceleration and group-scale dynamical activity. These results demonstrate diffuse radio emission is present in a non-negligible fraction of galaxy groups, providing new constraints on non-thermal processes in low-mass environments.

\end{abstract}




\section{Introduction}

Galaxy groups are the most common galaxy environments in the Universe, with halo masses ranging typically from 10$^{12}$ to 10$^{14}$ $M_\odot$. They occupy an intermediate position in the cosmic hierarchy, bridging the scale between isolated galaxies and massive clusters \citep[e.g.,][]{2000ARA&A..38..289M,Sun2012,2021Univ....7..139L,2021Galax...9...84N}.
They are embedded within the filaments of the cosmic web, and exhibit physical properties which are distinct from high mass galaxy clusters; specifically, galaxy groups possess lower characteristic thermal temperatures ($kT \lesssim 2$~keV), lower velocity dispersions ($\sigma_v \sim 100-500$~km~s$^{-1}$), and a central entropy excess indicative of significant non-gravitational feedback \citep{1999Natur.397..135P, 2021Univ....7..139L}. Galaxy groups are often found in non-virialised states with shallower gravitational potentials than galaxy clusters \citep{1993AJ....105.2035D}. They are dynamically active environments where processes such as mergers, supernova (SN) feedback, and black hole feedback play significant roles in shaping galaxy evolution \citep{1992ApJ...399..353H,2015A&A...573A.118L,2024Galax..12...24E}. They also host a higher fraction of Active Galactic Nuclei (AGN) compared to more massive systems \citep{2007MNRAS.380.1467G}. Thus, galaxy groups present a unique and compelling setting for investigating the interplay between galaxy evolution and the large-scale structure of the Universe \citep{2000ARA&A..38..289M}.

\par For many years, galaxy groups were not regarded as separate astrophysical systems but rather as the low-mass ends of the galaxy cluster population. Many efforts to distinguish groups from clusters have been made. For instance, \citet{2017MNRAS.471....2P} used cosmological hydrodynamic and N-body simulations to show that low-mass systems are unlikely to follow the self-similar scaling relations of clusters.
Moreover, AGN feedback has been proposed as a key factor differentiating clusters from groups, particularly in their baryonic properties \citep{2007MNRAS.376..193J, 2022arXiv220910554H,2024Galax..12...24E}. Despite being unique environments that differ substantially from high-mass galaxy clusters, the non-thermal properties of the Intra-Group Medium (IGrM) remain poorly understood \citep{2021Galax...9...84N}.
 
 \par In galaxy clusters, radio observations have revealed a population of non-thermal diffuse sources which are not directly linked to the radio emission from individual galaxies \citep{2004IJMPD..13.1549G,2019SSRv..215...16V}. These sources trace cosmic rays and magnetic fields in the Intra-Cluster Medium (ICM) and are typically categorized into three classes: radio halos, cluster radio shocks (relics), and revived AGN fossil plasma sources \citep{2009A&A...499..371G,2012ApJ...748....7M,2003AJ....125.1095S,2018MNRAS.477..957D,2019SSRv..215...16V}. The known major particle acceleration mechanisms are Diffusive Shock Acceleration \citep[DSA;][]{1983RPPh...46..973D} and Turbulent Re-acceleration (TRA) mechanism \citep{2007MNRAS.378..245B}. Recently, \cite{2025ApJ...979L..15G} have also shown the possibility of acceleration of particles in the ICM via magnetic reconnection.
As galaxy groups evolve through the hierarchical assembly of individual galaxies, the IGrM can be perturbed by merger induced events. However, such events may not necessarily create a system wide turbulent field. In the group regime, the diffuse emission ($<$500 kpc) is more likely to arise from individual galaxy-galaxy interactions or the stripping of tails, where the dissipated kinetic energy density is sufficient to sustain synchrotron emission over more modest physical scales than in cluster-scale counterparts \citep{2013MNRAS.431..781M,Kolokythasetal18,2024MNRAS.527.1062H}. In addition to this, such events are several orders of magnitude less energetic than the massive cluster-cluster mergers that generate Mpc-scale radio halos and relics \citep{2014IJMPD..2330007B, 2015A&A...580A..97C, 2019SSRv..215...16V}. The properties, prevalence, and underlying physics of this emission in the lower mass, dynamically younger environment of galaxy groups remain significant open questions. It is critical to note that while the mechanisms (TRA and DSA) are theoretically expected to scale across both regimes, the resulting emissions are physically distinct; cluster radio halos can span megaparsecs and are intrinsically more luminous than their fainter, group-scale counterparts. Therefore, a focused empirical study of diffuse emission in galaxy groups is necessary to test current theories under lower energy conditions and to accurately determine the nature of the processes in the IGrM.

\par Over the past four decades, studying diffuse radio emission in galaxy groups has been challenging due to several reasons. Unlike massive cluster mergers, which require only a fraction of their energy to power radio halos, the steep $P_{\mathrm{radio}} \propto {M_{500}}^{3-4}$ scaling relation predicts a precipitous drop in emission for lower-mass systems. Consequently, as mass decreases by an order of magnitude, radio power diminishes by a factor of 10$^{3}$-10$^{4}$, suggesting that the thermodynamic energy density in groups may be insufficient to sustain detectable diffuse sources \citep{2021A&A...647A..51C}. 
In addition to this, there are a few observational limitations too. First, nearby groups span angular scales larger than typical primary beams, requiring time-consuming mosaics. Second, distant groups are unresolved, blending member galaxies into single point sources. Conversely, groups in the intermediate regime suffer from a tension between resolution and sensitivity; the diffuse IGrM is often resolved out by interferometers, yet individual galaxies are confused by single-dish beams. Moreover, at low radio frequencies the large beam size reduces spatial resolution and at higher frequencies the synchrotron emission fades rapidly, making it more difficult to detect \citep{2021Galax...9...84N}. 

Nevertheless, there exists a few studies that established the groundwork for group-scale studies of non-thermal emission and magnetism, confirming the field's viability and providing initial observational constraints that precede current large-area surveys \citep{1985ApJ...296...60M,2011ApJ...732...95G}. Other searches for large-scale diffuse radio emission in X-ray luminous groups and low-mass clusters, such as the work by \citet[][]{2021A&A...646A.107M}, alongside systematic investigations by \citet[][]{2011ApJ...740L..28B}, \citet[][]{2015A&A...580A..97C}, and \citet[][]{2017ApJ...841...71G}, established important upper limits on the incidence and power of group-scale radio halos.
These foundational studies demonstrated the intrinsic link between the presence of diffuse radio emission and the host system’s dynamical state, a comparison later studied with the Complete Local Volume Groups Sample (CLoGS) series. 
In that context, the CLoGS sample studies have examined the X-ray \citep{OSullivanetal17}, radio \citep{Kolokythasetal18,2019MNRAS.489.2488K}, and multi-wavelength \citep{OSullivanetal18b,2022MNRAS.510.4191K} properties of 53 nearby brightest group-
dominant early-type galaxies (BGEs), analysing their energy output in an unbiased set of low-mass systems to understand the impact of AGN in the group environment. 
Group scaling relations are known to deviate from self-similarity, suggesting a strong influence of non-gravitational processes like AGN feedback in their evolution \citep{2002ApJ...578...74F,2003MNRAS.343..331P,2024Galax..12...24E}. 
A further study of radio emission in groups using the LOFAR Two-Metre Sky Survey (LoTSS) was conducted by \citet{2019A&A...622A..23N}. Their results yielded detection of radio emission in at least 73 out of 120 groups in their sample, with at least 17 groups showing extended radio emission. 
In addition to this, new generation interferometers have enabled successful detections of diffuse radio emission in several groups, including: HCG 15, which exhibits highly polarized emission likely associated with group members \citep{2025arXiv250308840R}; and HCG 92, which has bright, complex, polarised and extended radio structures, with significantly ordered intergalactic magnetic fields \citep{2013MNRAS.435..149N}.
\par The emergence of a new generation radio interferometers, such as the Low-Frequency Array (LOFAR), the upgraded Giant Metrewave Radio Telescope (uGMRT), the Murchison Widefield Array (MWA), and the Karoo Array Telescope (MeerKAT), has revolutionised the field by providing unprecedented sensitivity and resolution at both high and low frequencies \citep{2013A&A...556A...2V,2017JAI.....641011R,2013PASA...30....7T,2016mks..confE...1J}.
Now with the Australian Square Kilometre Array Pathfinder (ASKAP) telescope, detecting diffuse radio emission like fossil plasma and magnetized gas distribution in low-mass systems like galaxy groups in the southern sky is possible \citep{2024MNRAS.533.4068A}. Its design, combined with high-speed digital processing and high-performance computing (HPC), enables rapid and sensitive sky surveys \citep{2021PASA...38....9H,2021PASA...38...46N} with sufficient sensitivity to detect low-surface brightness structures in galaxy groups.

\par Here, we use ASKAP survey data at 0.8--1.4 GHz
to map the diffuse radio emission in galaxy groups. The aim of this study is to use an automated way to systematically identify and investigate diffuse radio emission across a large group sample using an automated approach. We investigate the average diffuse emission across the sample, and also study individual systems. The diffuse emission potentially originates from processes such as AGN activity, shock fronts, and turbulence within the IGrM. We characterise this emission and examine its relationship with group properties, with a particular focus on understanding the nature of diffuse radio emission and how the derived scaling relations compare to those observed in galaxy clusters. This paper is structured as follows: Section~\ref{section2} describes the optical and radio surveys, observation details, and data reduction and processing; Section~\ref{section3} describes the methodology; Section~\ref{section4} presents the results of our analysis; Section~\ref{section5} discusses possible physical mechanisms behind diffuse radio emission in groups and compares with simulations; and Section~\ref{section6} summarises our work and discusses future prospects.

Throughout this work, we consider a flat $\Lambda$ Cold Dark Matter (CDM) cosmology with $H_0$ = 70 km s$^{-1}$ Mpc$^{-1}$, $\Omega_\Lambda$ = 0.70, and $\Omega_\mathrm{M}$ = 0.3.
All physical parameters obtained from GAMA catalogues are also scaled to this cosmology.

\section{Data}\label{section2}
\subsection{GAMA: Group halo and galaxy properties}

\begin{figure*}
    \centering
    \includegraphics[width=1\linewidth]{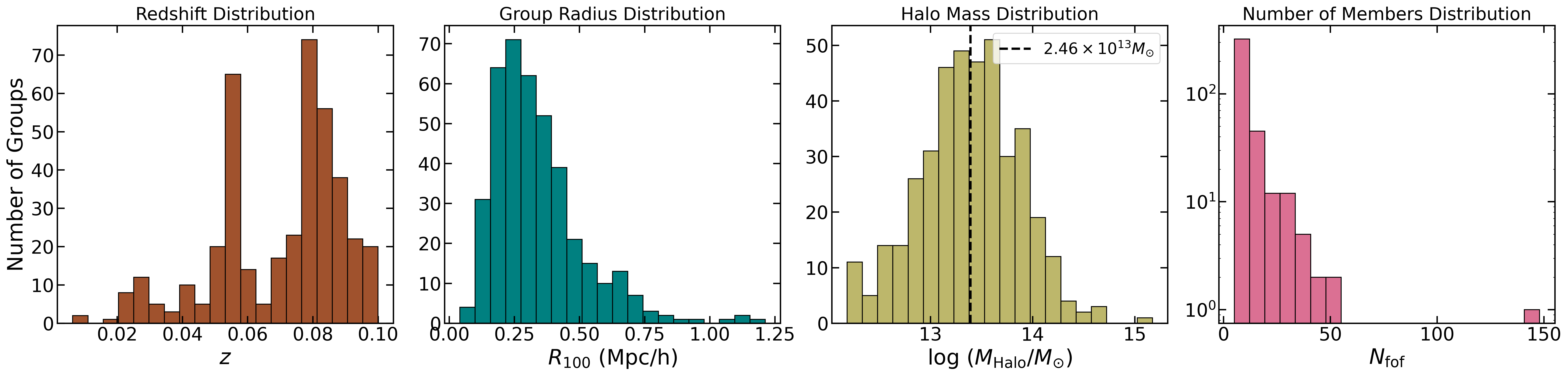}

    \caption{Redshift, radius, group halo mass and multiplicity distribution of all 400 GAMA groups in our sample. The black dashed line in halo mass distribution represents median of the whole sample, also referred as incompleteness limit. }
    \label{fig:SAMPLE}
\end{figure*}

Galaxy And Mass Assembly (GAMA) is multi-wavelength survey spanning across FUV - FIR using the 3.9 m Anglo-Australian Telescope and other telescope facilities for $\sim $ 300,000 galaxies \citep{2011MNRAS.413..971D,2013MNRAS.430.2047H,2015MNRAS.452.2087L,2022MNRAS.513..439D}. It is centered around RA of 02h (G02), 09h (G09), 12h (G12), 14h (G15),  23h (G23) regions. G09, G12 and G15 have 60 deg$^2$ area regions with  98\% completeness. While the G23 region is 50 deg$^2$ with 95\% completeness. Due to a mid-survey reconfiguration for high spectroscopic completeness G02 is restricted to the regions situated north of $-$6 deg declination \citep{2015MNRAS.452.2087L}. Consequently, our analysis is confined to G09, G12, G15 and G23 to achieve uniform sample depth. The main aim of the GAMA survey is to use the galaxy distribution and kinematics to test the CDM paradigm \citep{2011MNRAS.413..971D}.  
GAMA offers extensive optical coverage overlapping with major radio continuum surveys, enabling access to galaxy properties, spectroscopic redshifts, and environmental catalogues including groups, clusters, and filaments.
While the optical spectroscopic coverage from GAMA is highly uniform across all four GAMA regions, the available X-ray coverage exhibits significant heterogeneity. The G09 field benefits from the deep eROSITA Final Equatorial Depth Survey (eFEDS) \citep{2022A&A...661A...3S}. In contrast, the G12 and G15 fields are covered only by the initial shallow scans of the eROSITA All-Sky Survey \citep{2024A&A...682A..34M}, which is approximately 10x shallower than eFEDS. Furthermore, public X-ray data for the G23 field is currently unavailable due to proprietary restrictions on eROSITA data in the Eastern Galactic Hemisphere. To avoid introducing complex selection effects or biases arising from non-uniform X-ray depth, we restrict our analysis to optical properties only. Detailed investigation of the X-ray properties of this sample is reserved for a future study.

\subsection{Galaxy group sample}

For this study, we use galaxy group catalogue from the GAMA DR4 survey in the four (G09, G12, G15, G23)  regions. The GAMA galaxy group catalogue is generated using the Friends-of-Friends (FoF) algorithm \citep{2011MNRAS.416.2640R}. We select galaxy groups from this group catalogue, based on the following selection criteria:

1) Redshift limit: We select all galaxy groups residing at a redshift $z < 0.1$. This cutoff ensures that all galaxy groups are in the nearby universe and can be observed with higher resolution and better completeness. 
For $z \sim 0.1$, the GAMA group sample is complete to $M_{\rm halo} \sim$ 10$^{12.8}$ $M_\odot$ and $N_{\rm fof} \geq $  3 multiplicity systems \citep{2025MNRAS.541.3220D}.

2) Group membership: To ensure statistical significance, we include the groups having at least five members ($N_{\rm fof} $ > 4). This cut ensures that the halo masses are well estimated from the velocity dispersion ($\sigma_{\rm disp}$). It significantly reduces the impact of interloper galaxies on the mass estimation, ensuring that $\sigma_{\rm disp}$ serves as a robust proxy for the total gravitational potential of the halo.

3) Halo mass cutoff: We apply a minimum halo mass threshold of 10$^{12}$ $M_\odot$ to select sufficiently massive groups. Groups below this threshold may have halo mass estimates significantly dominated by a single group galaxy or field galaxies in the vicinity, which makes them less suitable to study group dynamics. We also have defined the median of our sample at 2.46 $\times$ 10$^{13}$ $M_\odot$ as the incompleteness limit -- see third panel in Figure~\ref{fig:SAMPLE}. Out of our total sample of 400 groups, 201 groups lie above this limit, while 199 groups populate the lower-mass regime down to our minimum cut of $10^{12}$~$M_\odot$. We retain the subsample below this incompleteness limit as they provide critical exploratory data and upper limits in a mass regime that is historically underrepresented in diffuse radio emission studies. To ensure transparency regarding completeness, this mass limit is explicitly demarcated (e.g., via dashed lines) on all relevant scaling relation plots. Below this limit our sample can be affected by incompleteness and above this limit our sample is robust.

4) Group edge: Groups detected near the survey boundaries could have missing group members, leading to biases in their derived group properties. This is evaluated using the Group Edge (GE) value in the GAMA catalogue which represents the fraction of the group lying within the survey volume. A value of 1 indicates that the group does not have any missing survey coverage and hence complete information on the group members. We exclude groups with a GE value $<$ 1 to avoid groups detected around survey boundaries.

\par Using these criteria, we selected 400 galaxy groups over all the GAMA fields. The distribution of this sample and its properties are given in Figure~\ref{fig:SAMPLE} and Table~\ref{tab:GGsample}.
Although galaxy groups and low mass clusters are said to be distinct systems, in practice there is no proper demarcation separating the two. Systems around 10$^{14}$ $M_\odot$ may be a transitional regime, where their properties can resemble either massive groups or emerging low mass galaxy clusters, and categorising such systems depends upon properties such as richness, dynamical state and physical extent. For example, systems having halo masses $\geq$ 10$^{14}$ $M_\odot$ may be low mass clusters if they have more than 100 galaxy members and radius greater than 500 kpc \citep{2021Univ....7..139L}. We have only one low mass cluster in our sample according to above criteria. This system has 148 galaxy members, a halo mass of 10$^{14.55}$ $M_\odot$ and radius of 1.6 Mpc. In addition, there are $\sim$ 50 systems with physical extent $>$500 kpc with a wide range of halo masses and number of galaxy members. 
Given the ambiguity in separating such systems, we may be including low-mass galaxy clusters in our sample. We retain such systems in our analysis, as they can provide valuable insights into the group-cluster transition regime.

\begin{table*}
    \centering
    \begin{tabular}{cccccccc}
    \hline
     GAMA regions & RA & Dec &No. of galaxy groups&Data &Survey code&Frequency  &RMS \\
          &  (deg)& (deg)&& && (MHz) & $\mu$Jy beam$^{-1}$\\

    \hline
    
    \vspace{0.1cm}
   G09  &129.0 to 141.0 & $-$2  to  +3 & 82 &EMU&AS201 &943  &33\\
    \vspace{0.1cm}
     G12  & 174.0  to  186.0 & $-$3  to  +2 & 139&EMU& AS201 &943&35 \\
    \vspace{0.1cm}

      G15 & 211.5  to  223.5 & $-$2  to  +3 & 98 &EMU& AS201&943  &36\\
    \vspace{0.1cm}

    G23 & 339.0  to  351.0 & $-$35  to $-$30 & 24&EMU &AS201 &943  &25  \\
    \vspace{0.1cm}

    & &  &57 & DINGO &
AS104 & 1367 & 13 \\
    \hline     

    \end{tabular}
    \caption{Description of the galaxy group sample and continuum data used. RMS values reported here are average RMS of the final mosaics of reprocessed EMU images.}
    \label{tab:GGsample}
\end{table*}

\subsection{EMU and DINGO: radio continuum data}
For radio continuum data, we use maps from two surveys of ASKAP. The Evolutionary Map of the Universe (EMU) is a radio continuum survey at 943 MHz with a bandwidth of 288 MHz. The main aim of the survey is to make deep continuum maps, achieving a sensitivity of 25-30 $\mu$Jy beam$^{-1}$ of the whole southern sky and also extending as far North as +7\degr declination \citep{2025PASA...42...71H}. EMU covers data in G09, G12, G15 GAMA fields and the western half of the G23 region, which we use in this paper. The eastern half of G23, not yet covered by EMU, is observed through the Deep Investigations of Neutral Gas Origins (DINGO) pilot survey. It is a deep HI survey spanning 0 $< z <$ 0.4, covering two ASKAP receiver bands: Band 2 at 1.367 GHz and Band 1 at 1.019 GHz \citep{2023MNRAS.518.4646R}. The pilot survey observed G23 region in one tile for a total of 100 hours, which is split into separate observations of 8 hours each \citep{2025arXiv251117307R}. We use the 100 hour image domain averaged DINGO radio continuum maps (Band 2) in our study.
\par Together, EMU and DINGO surveys constitute among the deepest large-area radio continuum surveys in the southern sky to date, with good sensitivity to low surface brightness emission. We apply diffuse filtering to both the EMU and DINGO maps, as detailed in following section. 

\subsection{Diffuse filtering code}\label{diffuse filtering}
Detection of faint diffuse emission in radio images is often hindered by the presence of compact sources that blend with the diffuse emission or bright compact emission that dominates the dynamic range of the image. Traditional methods of removing these sources range from subtraction in the $(u,v)$ plane via modeling, usually by imaging with short baselines removed to create a CLEAN component model of compact emission below a relevant angular scale \citep[e.g.][]{2021A&A...648A..11O}. This process has significant computational expense as multiple imaging rounds are required. Alternatively, point sources embedded in diffuse emission may have their contributions removed by subtracting their measured or modeled flux densities from an overall measurement of the diffuse source. This second method is prone to significant residual errors, depending on how well the point sources can be modeled and the relative brightness of the extended emission to the point sources \citep[e.g.][]{2021PASA...38...53D}, and does not help as much with the initial detection. We instead adopted an automated image-based angular scale filtering approach as an alternative.

\par  The diffuse filtering code is an implementation of the multi-resolution filtering scheme introduced by \citet{2002PASP..114..427R}.
It filters out emission on scales of $\leq 45\arcsec$ and $\geq 405\arcsec$. 
At our sample's median redshift of $z$ = 0.076, angular scales of 45\arcsec\ and 405\arcsec\ correspond to physical sizes of 66.09 kpc and 594.82 kpc, respectively. Although a minimum spatial filtering scale of 45\arcsec was applied, the final diffuse emission maps are retained at the native 15\arcsec resolution of the standard continuum images to facilitate direct morphological comparisons and to smooth out the sharp edges in the images caused by the choice of box filter used. Accordingly, the brightness units of the diffuse maps are expressed in Jy beam$^{-1}$, where the beam remains 15\arcsec.
This filtering code enhances low surface brightness structures in the presence of brighter/strong small-scale emissions -- see \citet[][]{2025PASA...42...71H} for more details. 
An example is shown in Figure \ref{fig:diffusefilteringexample}, where emission from bright point sources (from galaxies/AGN) is suppressed in the diffuse image, and only the low-surface brightness structures are retained. These images are helpful for detection of diffuse emission without having to model emission from specific galaxies or  subtract from the $(u,v)$ plane. There is some uncertainty in measurements of diffuse flux densities due to only approximate removal of compact emission close to the filtering scales.
When artefacts dominate, it becomes difficult to distinguish genuine diffuse emission from spurious features.

\subsection{EMU - Data reduction and imaging} 

EMU data is processed using the ASKAPsoft pipeline, which applies flagging, calibration, imaging and deconvolution routines. Phase self-calibration is performed using a sky model derived from RACS \citep{2020PASA...37...48M,2021PASA...38...58H}. For reliable flux scales and polarimetry, amplitude self-calibration is not performed.
Imaging and deconvolution is applied using a Multi-Scale Multi-Frequency Synthesis \citep[MS-MFS;][]{2011A&A...532A..71R} routine.
More details about ASKAPsoft can be found in  \citet[][]{2011PASA...28..215N}, \citet{2017ASPC..512..431W} and \citet{2025PASA...42...71H}. 

Data products are available on the CSIRO
ASKAP Science Data Archive (CASDA). EMU data is available under project AS201. EMU images have robust weighting of 0, with a common resolution of 15\arcsec\ $\times$ 15\arcsec. For the G09, G12 and G15 GAMA regions, EMU tiles are split into two separate observations of 5 hours each. This is because above Dec $-10$\degr, a full 10-hour observation is more difficult to schedule in a single run. The combination of equatorial location with separate imaging of shorter observations limits $(uv)$-coverage and cleaning depth, resulting in higher noise tiles.
Consequently, our EMU continuum maps have sensitivity of 35-39 $\mathrm{\mu}$Jy beam$^{-1}$ for equatorial fields and  25-30 $\mathrm{\mu}$Jy beam$^{-1}$ for G23.


\par Although the radio images generated using the ASKAPsoft pipeline are of high quality, they have certain limitations. One such limitation is the handling of \textit{w}-terms for sources located near the edge of, or beyond, the primary beam’s FWHM. This can result in artefacts around bright point sources. 
Another limitation is the absence of phase referencing, which is only partially mitigated by phase self-calibration. The success of this approach depends strongly on the quality of the deconvolution model; any errors in this model can introduce additional imaging artefacts. 
\begin{figure*}
    \centering
    \includegraphics[width = 0.79\linewidth]{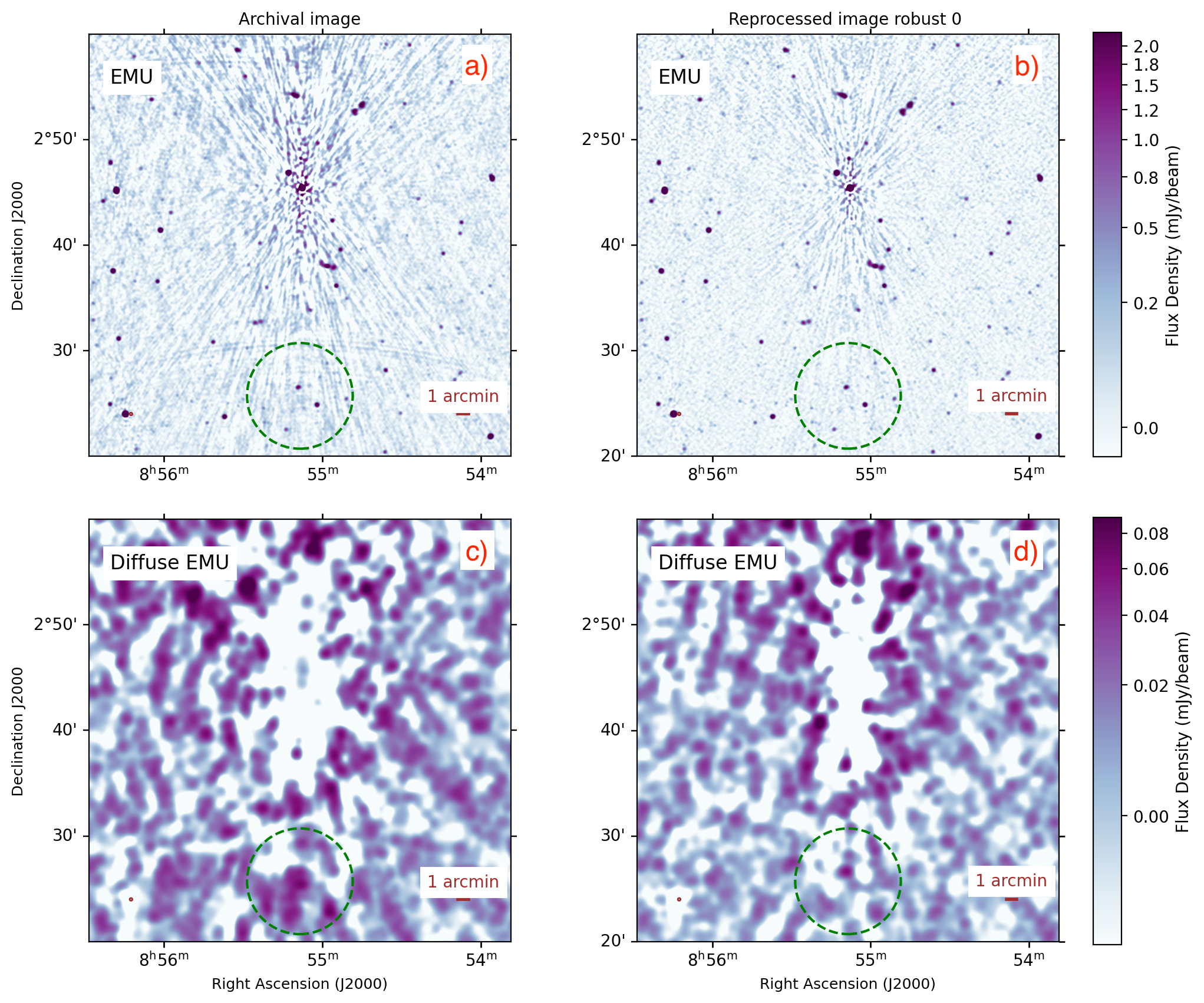}
     \includegraphics[width = 0.81\linewidth]{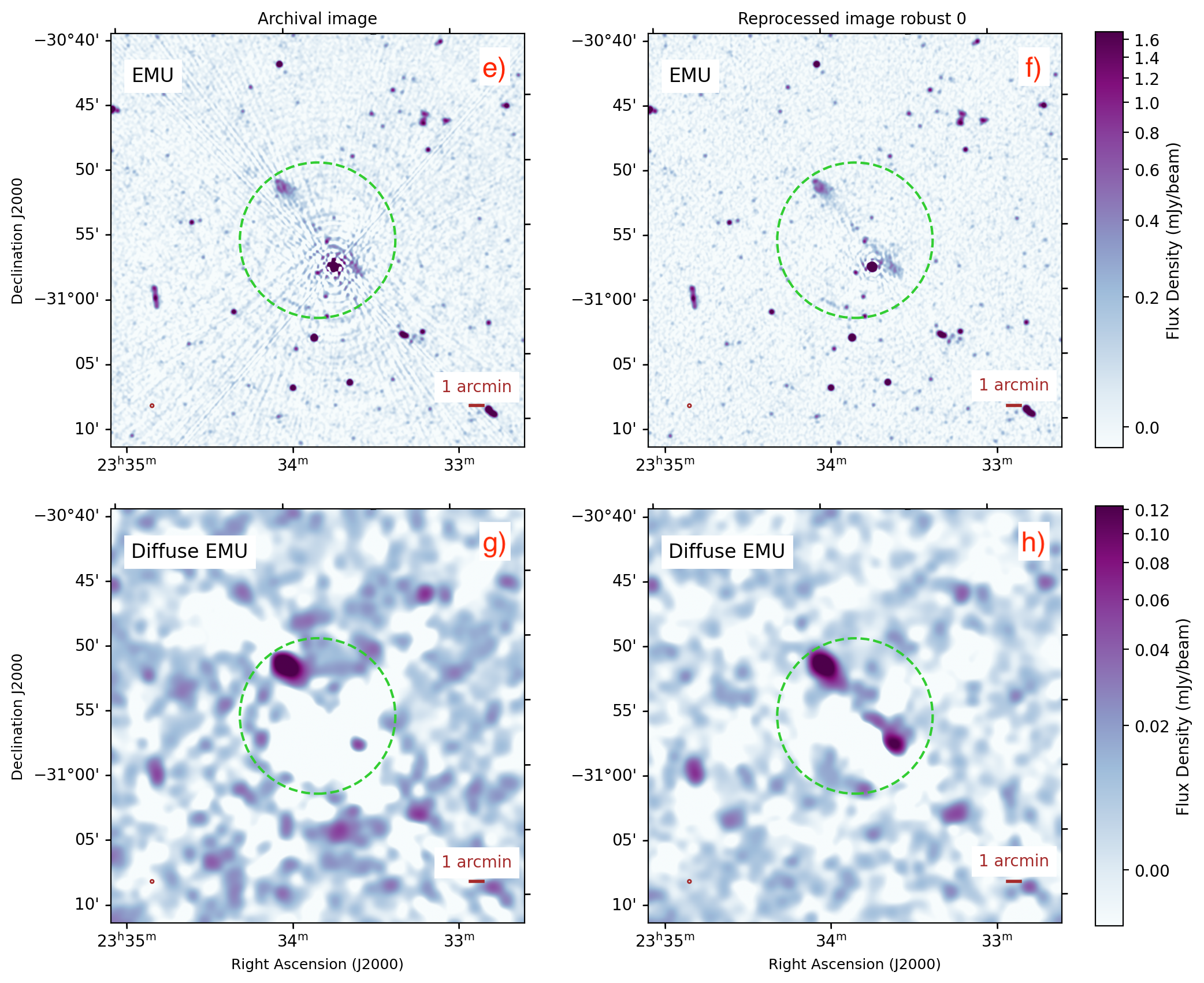}
    \caption{ G09 and G23 (upper and lower panel respectively)  regions in the EMU and diffuse EMU images with archival data (left column) and after reprocessing i.e., self calibration and imaging (right column).}
    \label{figure:reprocessing_images}
\end{figure*}
\par To address these issues, we reprocessed all EMU tiles using the Flint pipeline (Galvin et al., in prep) in order to reduce calibration artefacts from bright point sources in the field.\footnote{The \texttt{}{Flint} pipeline is available at \url{https://github.com/flint-crew/flint}.} Unlike the operational ASKAPsoft workflow, Flint employs clean masks between imaging rounds with heuristics to avoid incorporating potential calibration or imaging artefacts into the clean model. We use Flint to perform three rounds of self-calibration, i.e, two phase-only and one amplitude-phase self-calibration. For amplitude-phase self-calibration, we use a solution interval of 8 minutes, which provided ample signal-to-noise ratio to constrain the solver and fix slowly varying antenna errors. In addition to this, we also use a multiscale cleaning method for imaging with cleaning scales mentioned in Table~\ref{reprocessing_parameters}. We repeat this approach with different robust weighting parameters (-0.5 and 0); and for this study, we use robust 0 images, which provide a good compromise between sensitivity and resolution \citep{1995AAS...18711202B}. All the beams in one tile are convolved to a common resolution. For consistency, all the reprocessed EMU and DINGO images are convolved to a 15\arcsec\ $\times$ 15\arcsec\ beam size. After reprocessing, the RMS noise in the EMU equatorial maps is reduced from 37~$\mathrm{\mu}$Jy beam$^{-1}$ to 30~$\mathrm{\mu}$Jy beam$^{-1}$.
In addition to this, we conducted a quantitative comparison between the archival maps and the final reprocessed maps. We selected a control sample of 20 bright, isolated, and unresolved background point sources common to both datasets. Extracting the peak flux densities for these sources, we found a good agreement between the two iterations, yielding a median flux density ratio of $S_{\mathrm{reprocessed}} / S_{\mathrm{archival}} = 0.968 \pm 0.036$. The minor systemic offset ($\sim 3.2\%$ decrease) is an expected consequence of removing artificial background flux contributed by the artefacts in the original images. This test confirms that the reprocessing successfully suppresses imaging artefacts and improves local image fidelity while preserving the global flux density scale to within $\sim 4\%$. Consequently, no systematic flux correction factors are required for our downstream scaling relation analyses.
\par In Figure~\ref{figure:reprocessing_images}, Panel a shows the archival EMU image for a region (green circle) around a bright contaminating source in the north. Panel c shows the same region following the application of the diffuse filtering code. This could be mistaken as emission above 3$\sigma$ cutoff in the diffuse map (Panel c). However, after reprocessing (Panel b), the artefacts in the green circle are reduced, so the diffuse map has less emission above 3$\sigma$ (Panel d). In addition to this, the reprocessed maps are also able to recover diffuse emission which was earlier suppressed by artefacts in the field. 

In Panel e of Figure~\ref{figure:reprocessing_images}, a radio galaxy is in the vicinity of a bright point source. The artefacts suppress the second radio lobe and it is not significant in diffuse maps (Panel g). The lobe becomes visible when the image is self-calibrated and artefacts are reduced  (Panel h). 
Reprocessing also helps remove the w-terms errors which are taken care of by WSClean. An example of such errors can be seen in Panel a of Figure~\ref{figure:reprocessing_images}, where faint arc lines are seen over the green circle. After reprocessing, these errors are removed from images as seen in Panel b. Therefore, with reprocessing the EMU images, we minimize calibration and imaging artefacts and produce diffuse images which can capture more of the real extended radio emission. 

\par After reprocessing the EMU maps, we produced mosaic maps for equatorial GAMA regions to reduce the RMS noise at the edges of individual tiles. These regions have two separate 5 hour EMU observations with different Scheduling Block ID (SBIDs) for each tile. We average them together in the image domain to decrease the noise in the maps. We then generate RMS maps for these averaged image fits using Aegean - The Background and Noise Estimation (BANE) tool \citep{2012MNRAS.422.1812H,2018PASA...35...11H} and use these RMS maps to make weight maps for mosaicking. Image weights are calculated as 1/(RMS$^{2}$) over the whole image. Now with averaged EMU tiles and corresponding weight maps, we combine them using SWarp \citep{2002ASPC..281..228B}. The RMS improves by a factor of 5 at the edges of EMU tiles, from 130 $\mu$Jy beam$^{-1}$ to 24 $\mu$Jy beam$^{-1}$. These EMU mosaics are used as final image products for making the diffuse maps, flux density measurements, background variance calculations and making cutouts for individual groups for visual inspection. 
\subsection{DINGO - Data processing}
DINGO pilot survey data is available under project AS104 and it is processed using ASKAPsoft pipeline. However the processing follows standard HI data reduction procedures, which are slightly different from the ones used for EMU data processing. For example, here DINGO processing does not use the RACS sky model for self-calibration as used for EMU. DINGO uses robust weighting parameter of -0.5, scales are [0,6,15,30] for multiscale imaging, one phase self-calibration with solution interval of 200 seconds, etc \citep{2025arXiv251117307R}. We note that while internal RFI from on-dish calibration signals required flagging the 2-3 shortest baselines for roughly half of the integration time (50 out of 100 hours), this represents a tiny fraction of the 630 total baselines. The $uv$-plane remains densely sampled at the shortest spacings for the rest of the observation, meaning our sensitivity to large-scale diffuse emission is not systematically compromised.
Note that we have not reprocessed the DINGO continuum maps at all in this study because of good sensitivity data, less artefacts in the G23 tile and the large data volumes that make it computationally expensive. Therefore, we take the DINGO map which has a resolution of 12.2\arcsec\ $\times$ 8\arcsec\, at position angle of $-$54.45\degr and convolve it to a 15\arcsec\ $\times$ 15\arcsec\ beam size similar to EMU images. This DINGO map have sensitivity of 13 $\mathrm{\mu}$Jy beam$^{-1}$. We then run diffuse filtering code on this map to get final diffuse DINGO map used for measuring diffuse emission.
\begin{table}
\centering
\caption{Flint pipeline reprocessing details - self-calibration and imaging}
\begin{tabularx}{\linewidth}{ll}
\hline
\textbf{Number of tiles reprocessed} & 23 \\
\textbf{Self-calibration rounds} &  2-phase only and 1-amplitude-phase \\
\textbf{Solution intervals: phase cal} & 60s and 30s\\
\textbf{Solution intervals: amp-phase cal} & 480s\\
\textbf{Robust weighting} & 0\\
\textbf{UV range} &  50 $\lambda$\\
\textbf{Multiscale cleaning scales} & 0, 4, 8, 16, 24, 32, 48, 64, 92, 128, \\
\textbf{} &196, 512, 796, 1025 \\

\hline
\label{reprocessing_parameters}
\end{tabularx}
\end{table}

\begin{figure*}
    \centering
    \includegraphics[width=1\linewidth]{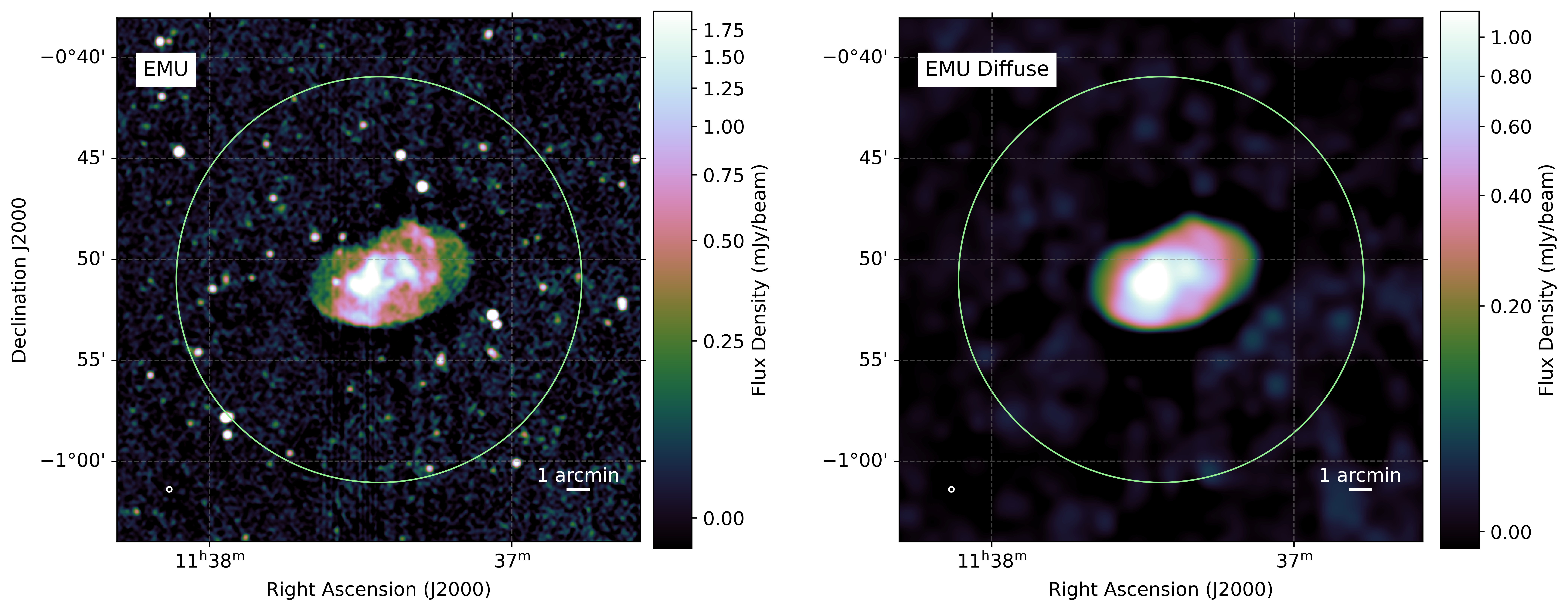}
    \caption{Example of how the diffuse filtering code suppresses emission from bright point sources and enhances low surface brightness emission. The region highlighted in the light green circle is GAMA group 200792 in this study, and the light green circle represents the group radius ($R_{100}$). The normal resolution EMU map has 15\arcsec\ resolution  and the diffuse filtered EMU map has 45\arcsec\ resolution but both of the images are in the units of Jy/15\arcsec\ beam shown by white circle in the lower left side of the image for direct comparison.}
    \label{fig:diffusefilteringexample}
\end{figure*}

\section{Methodology}\label{section3}

Here we use the diffuse EMU mosaics and diffuse DINGO maps to detect and measure diffuse emission in GAMA groups.
We also use optical images
(g, r, i band) from GAMA DR4 survey \citep{2020MNRAS.496.3235B} to make 3-color images using APLpy (rgb code).
First, we use a statistical approach to investigate a global diffuse emission signal from galaxy groups, in particular halo mass bins, using the diffuse maps, see Section~\ref{subsection4.1}. Using radio and 3-color images, we then visually inspected each group for significant diffuse emission, see Section~\ref{subsection4.2}.  
\label{sec:methodology} 

\subsection{Measuring diffuse emission in groups}
To measure the integrated diffuse radio emission, we adopt the $R_{100}$ radius defined as the projected distance to the most distant group member, as provided by the GAMA group catalog. While $R_{100}$ is naturally sensitive to spatial outliers, it provides the most conservative boundary to ensure we encapsulate any genuine peripheral diffuse emission (such as stripped tails or interaction-driven turbulence), as well as any physically bound group members that may lack formal spectroscopic confirmation. Local noise contributions are robustly quantified by evaluating the background RMS in random off-source regions of identical size to the $R_{100}$ aperture of each group.
We create a circular aperture on the diffuse maps using group position and radius ($R_{\mathrm{100}}$). For measuring integrated flux densities, first we measure the sum of flux density values in all pixels in group regions and then divide by a beam correction factor ($P_{\mathrm{beam}}$) of 50.36. The beam correction factor formula is:  
\begin{equation}\label{beam_correction_factor}
     P_\mathrm{beam} = \frac{1.1331 \times B_{\rm maj} \times B_{\rm min}}{\mathrm{pixel\ area}},
\end{equation}
where $B_{\rm maj} \times B_{\rm min}$ in all images is 15\arcsec\ $\times$ 15\arcsec\ with pixel area of 2.25\arcsec\ $\times$ 2.25\arcsec.
\par These measured integrated flux densities for galaxy groups are denoted by $F\mathrm{_{group}}$. We then use two methods to measure integrated flux densities in galaxy groups from the diffuse maps. First, we measure all emission inside the group region and second, we clip values below 3$\sigma$ in the group region and only measure emission above this threshold.

\subsection{Measuring background variance of the maps}

We estimate the background variation in the diffuse maps by measuring the variation across 100 random regions per galaxy group.
The radio maps consist of both true radio emission and instrumental noise, which varies across the image and is different for different ASKAP tiles. The background is not uniform, so using a single value for background subtraction might lead to biased flux estimates. By measuring background variations, we have a better estimate of true flux variations across each map and likely contributions from background sources which are not associated with a group.

For a galaxy group with radius ($R_{100}$), we create 100 random regions with same radius, ensured by following condition that they don't overlap with existing galaxy group regions.
To prevent overlap, we check the separation between the new random region and all galaxy groups. 
 A new region is accepted only if the separation satisfies:
\begin{equation}\label{overlapping_condition}
\centering
    { d \geq  R_{\text{new}} + R_{\text{group}}} ,
\end{equation}
where d is separation, $R\mathrm{_{new}}$ and $R\mathrm{_{group}}$ are the radii of the random region and the galaxy group, respectively.

\begin{figure*}
    \includegraphics[width=0.8\linewidth]{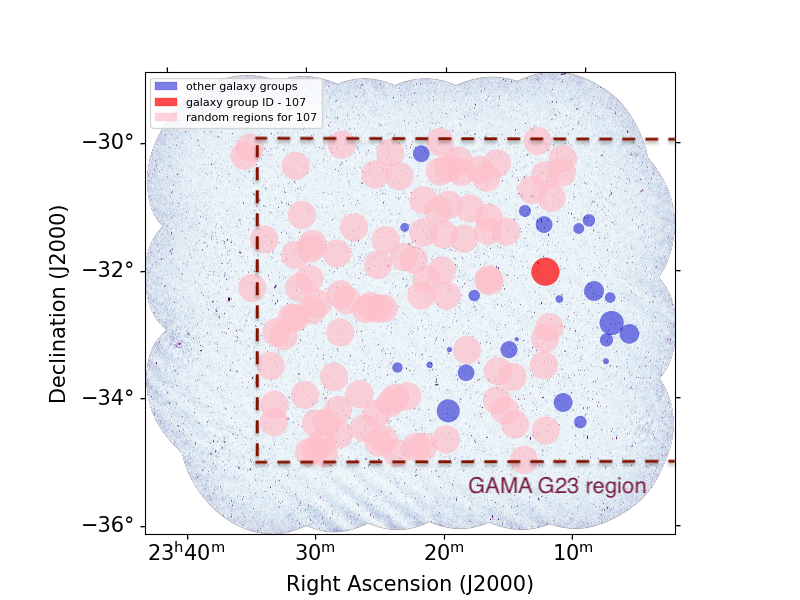}
    
    \caption{Example of random regions (pink circles) for a galaxy group (red circle) in EMU tile of G23 region generated using equation~\ref{overlapping_condition}. The blue circles are other galaxy groups lying in the G23 region. Brown dotted lines trace the GAMA G23 region extent. They are located on the western half side of the EMU tile because of the GAMA G23 region limit and the selection criteria we have used for our sample selection.} 
    \label{fig:random_regions}
\end{figure*}

If the generated random region does not overlap with any existing group, it is accepted. Otherwise, a new set of random regions are generated until the conditions are met. One example is shown in Figure~\ref{fig:random_regions}. While the background sampling algorithm strictly prevents overlaps with all group regions, partial overlaps between the random background regions themselves are permitted. However, the algorithm strictly prevents the duplication of exact central coordinates (RA/Dec), ensuring that all $n=$100 flux density measurements represent uniquely sampled spatial footprints within the local background. Allowing these partial overlaps is geometrically necessary to extract a statistically robust sample size from the immediate local vicinity of the group without expanding the search radius into disjointed noise regimes. We tested it on the G12 mosaic and it confirms that this partial inter-region overlap does not artificially suppress the integrated variance. Comparisons between uniquely overlapping ($n=$100) and strictly non-overlapping ($n=$14) local sampling routines demonstrate that the resulting $1\sigma$ background uncertainties remain similar ($\sim 2.30 $ mJy vs $\sim 3.14 $ mJy, respectively, derived via 68\% confidence intervals). This ensures our local variance estimates are robustly quantified by a large sample size and not severely underestimated by covariance. Using this method, we ensure that random regions are well-distributed within the radio continuum maps while avoiding contamination from known galaxy groups.
We also constrain the RA and Dec boundaries in the maps, given the potential presence of NaN values and higher noise floor near the edges, by restricting the selection to 70\% of the total RA and Dec range. We measure integrated flux densities of randoms in the selected RA and Dec region. Following this, we estimate background median flux density ($F\mathrm{_{bg}}$) and the background uncertainty is estimated from the standard deviation of the 100 measurements. Assuming a Gaussian noise distribution, this corresponds to a 1$\sigma$ (68\%) confidence interval (\(\sigma_{\text{bg}}\)).
The above steps are also repeated for measuring background variance in the 3$\sigma$ clipped method. 

\subsection{Background correction and power calculations}

For an integrated flux density $F_{\text{group}}$, the background-corrected flux density is defined as:
\begin{equation} 
    F_{\text{corr}} = F_{\text{group}} - F_{\text{bg}} \,,
\end{equation}
where $F_{\text{bg}}$ is the median background flux density measured in the local vicinity. The formal uncertainty on this background-corrected flux density, $\sigma_{F_{\text{corr}}}$, is calculated by adding the systematic and statistical errors in quadrature:

\begin{equation}
     \mathrm{\sigma_{F_{\text{corr}}} = \sqrt{\sigma_{\text{group}}^2 + \sigma_{\text{bg}}^2}}
\end{equation}

where,  \( \sigma_{\text{group}} \) is the measurement uncertainty in the group region.
 \( \sigma_{\text{bg}} \) accounts for the variations in the background.

\begin{figure*}
    \centering
    \includegraphics[width=1\linewidth]{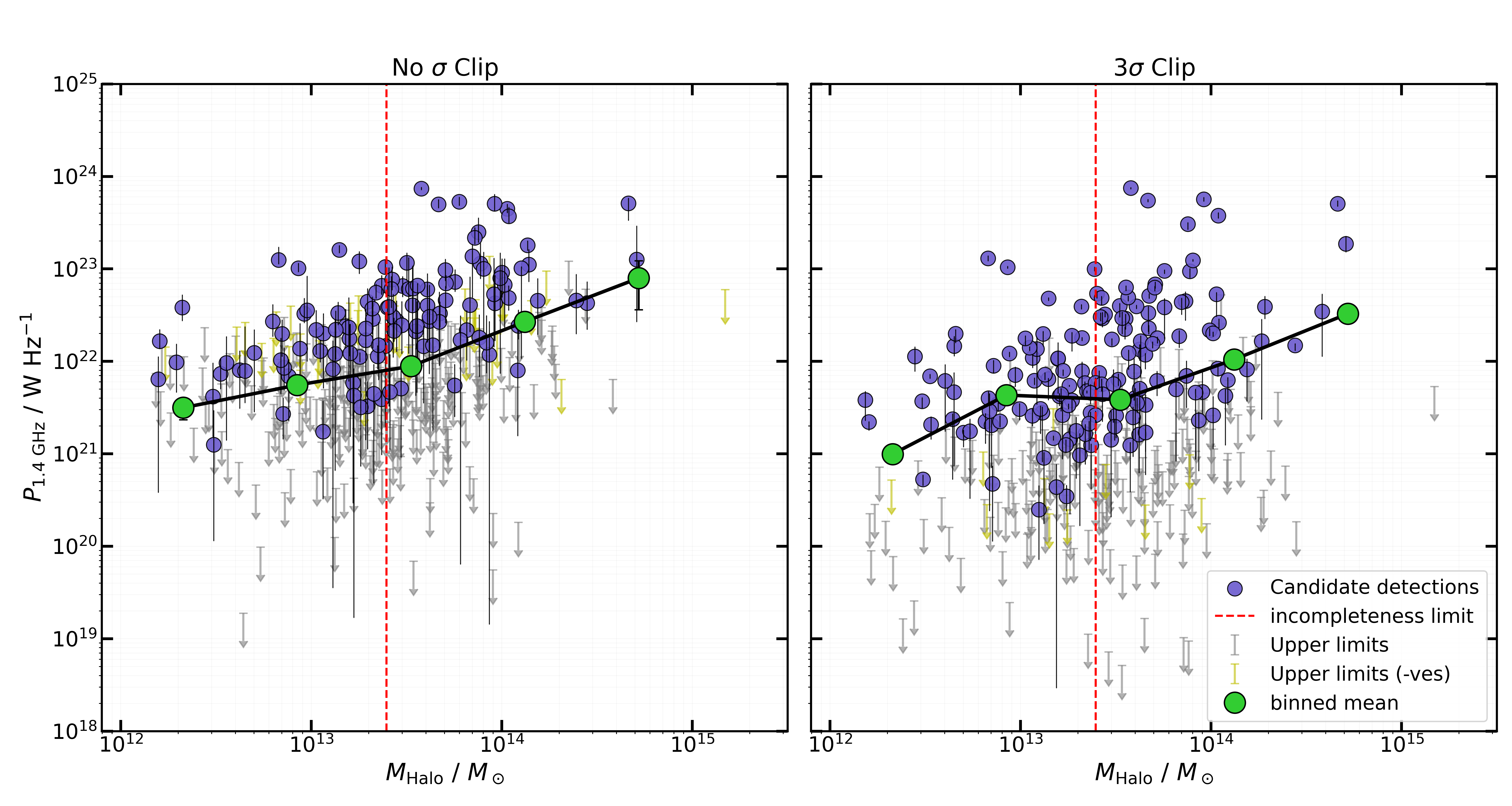}

    \caption{Radio power as a function of halo masses for all galaxy groups in our sample using two different methods to compute integrated flux densities from diffuse EMU maps. Purple data points represent groups with candidate detections, yellow and gray data points represent upper limits. Yellow data points are negative integrated flux densities, shown as their absolute values for visual representation only. Green data points in both the panels denote weighted average of all groups in our sample. The red dotted line represents the incompleteness limit of our sample.}

    \label{fig:diffuse emission from groups}
\end{figure*}

\par We extrapolate the background-corrected flux densities from EMU (943 MHz) and DINGO (1.367 GHz) to 1.4 GHz  using: 
\begin{equation}
    F_{\text{1.4}} =  F_{\nu } \times  \left( \frac{\rm 1.4~GHz}{\nu} \right)^{-1.3} ,
    \label{extrapolating}
\end{equation}
where $\nu=$  0.943 GHz for EMU data and 1.367 GHz for DINGO data. We use a spectral index of $-$1.3, typical for radio halos  detected in high-mass galaxy clusters \citep{2021PASA...38...53D,2024PASA...41...26D}. 
The radio power (W Hz$^{-1}$) at 1.4 GHz is calculated using:
\begin{equation}
   {P_{1.4}  = 4 \pi D_{\text{L}}^2  F_{1.4} } ,
\end{equation}\label{radiopower}
where $D_{\mathrm{L}}$ is the luminosity distance to the galaxy group.

\section{Results}\label{section4}


\setlength{\tabcolsep}{7pt}        
\begin{table*}
\centering
\caption{Properties of galaxy groups with candidate diffuse emission}
\begin{tabular}{ccccccccc}
\hline
\vspace{0.1cm}
Group ID & RA (J2000)  & Dec (J2000)   & $z$ & Halo mass & $N_{\rm fof}$ & Classification & Integrated flux density & Radio power \\
 & (degrees) & (degrees)&
 &(10$^{13}$$M_\odot$) & &  & (mJy) & (10$^{21}$ W Hz$^{-1}$)  \\[0.05cm]
\hline
21* & 343.116 & $-$34.077& 0.096 & 5.23 & 16 & cMH  & 0.54&12.27\\[0.05cm]

83* & 343.843 & $-$33.894 & 0.028 &4.53 & 19 & cMH/cMD  & 0.66&1.25 \\[0.05cm]
114* & 343.626 & $-$32.8187 & 0.098 &1.78& 8 & cMD&0.15&3.55\\[0.05cm]
234* & 342.839 & $-$31.928 & 0.078 &3.23& 6 & cMD &0.10 & 1.41\\[0.05cm]
316* & 343.045 & $-$31.197 & 0.092 &1.64 & 8 & cMD &0.07 & 1.49\\[0.05cm]
408* & 344.421 & $-$30.697 & 0.081 & 5.44 & 6 & cMD & 0.08& 1.26\\[0.05cm]

101106 & 129.267 & 2.754 & 0.083 & 2.80 & 5 & cMH  & 0.25
 & 2.54 \\[0.05cm]

& &  & &  & &cR/cAGN&0.76
 & 7.86 \\[0.05cm]
 &&&&&&  cR&1.37
 & 14.10 \\[0.05cm]
 &&&&&&  cMD& 0.49
 & 4.99 \\[0.05cm]
 & &&&&& cMD & 0.40
 & 4.15 \\[0.05cm]
100165 & 129.457 & $-$0.251 & 0.053 & 0.66 & 8 & cMH & 0.25 & 0.99 \\[0.05cm]
100146 & 130.143 & 2.611 & 0.051 & 12.21 & 9 & cMD & 0.27 & 0.96 \\[0.05cm]
&&&&& & cR & 1.03& 3.74 \\[0.05cm]
100072 & 131.127& 2.294 & 0.052 & 1.66 & 12 & cMH & 0.27 &1.02 \\[0.05cm]
100055 & 131.382 & 2.475 & 0.077 & 9.98 & 15 & cMH & 0.71 & 6.26 \\[0.05cm]
100084 & 132.387 & 1.728& 0.070 & 2.44 & 13 & cAGN/cMD & 0.49 & 3.46 \\[0.05cm]

100710 & 132.780 & $-$0.840 & 0.078 & 0.36 & 5 & cMD & 0.14 & 1.28\\[0.05cm]
100721 & 133.190 & $-$1.814 & 0.078 & 0.38 & 5 & cMH/cMD & 0.02 &0.15 \\[0.05cm]
100507 & 133.948 & 0.794 & 0.043 & 0.34 & 5 & cMH/cMD  & 1.10 & 2.80 \\[0.05cm]
100810 & 134.945 & $-$0.005 & 0.053 & 5.68 & 8 & cAGN/cMD & 23.19 & 95.50 \\[0.05cm]
100547 & 135.047 & 2.077& 0.057&4.38& 5 & cR/cAGN & 0.15 &0.69\\[0.05cm]
 & &&&&&cMD & 0.48 &2.23 \\[0.05cm]

100242 & 136.818 & 0.105 & 0.095 & 0.44 & 7 & cMD & 0.25 & 3.48\\[0.05cm]
100568 & 138.490 & 1.660 & 0.058 & 0.76 & 5 & cMD & 0.67 & 3.07 \\
100648 & 138.704& $-$0.592 & 0.055 & 2.88 & 5 & cMH &0.03
& 0.12\\[0.05cm]
 & &&&&&cMD & 0.03 &0.11 \\[0.05cm]
 101008& 139.67& 2.03 & 0.088 & 1.16 & 5 & cMH/cAGN&0.07& 0.82\\[0.05cm]

200084 & 181.008 & 1.444& 0.083 & 4.70 & 18 &cH \textsuperscript{\dag}/cMD & 2.53 & 27.01 \\[0.05cm]
200090 & 182.420 & 1.172& 0.079 & 5.23& 12 & cMD &0.66&6.08  \\[0.05cm]
200103 & 180.120 & $-$0.123 & 0.082 & 15.71& 17 & cAGN/cMD &0.24&2.41\\[0.05cm]

200140 & 177.281 & $-$1.366 & 0.082 & 5.41 & 20 & cMH/cMD & 2.03&20.40 \\[0.05cm]
200179 & 176.837 & $-$1.702 & 0.093 & 6.16 & 9 & cAGN/cMD &0.29 &3.86 \\[0.05cm]

200445 & 174.080 & $-$2.847 & 0.046 & 2.23 & 6 &cMD & 0.54&1.65 \\[0.05cm]
 & &&&&& cMD& 1.51 &4.59 \\[0.05cm]
200502 & 179.717& 1.728 & 0.079 &  1.77& 9 & cMH/cMD&0.62 &5.79\\[0.05cm]
200571 & 178.364 & $-$1.181 & 0.082 &3.44  & 5 & cMD & 0.95&9.25\\[0.05cm]
200669 & 185.726 & $-$2.669& 0.063 & 0.18 & 5 & cMH/cMD &0.94&5.47 \\[0.05cm]
200671 & 175.417 & $-$2.246 & 0.056 & 0.16 & 5 & cMD &1.97 & 8.65\\[0.05cm]
200792 & 174.364 & $-$0.846 & 0.047 & 3.78 & 8 & cAGN/cMD &231.80&739.65 \\[0.05cm]
201007 & 180.311 & 0.939 & 0.086 & 1.01 & 5 &cMD &0.38&4.24\\[0.05cm]
 & &&&&& cMD& 0.22 &2.44 \\[0.05cm]
 & &&&&& cMD& 0.12 &1.31 \\[0.05cm]

201057 & 174.352 & $-$2.097 & 0.080 & 0.17 & 5 & cMH &0.11&1.04\\[0.05cm]

201313 & 175.107& 1.984 & 0.077 &  1.80& 5 & cAGN/cMD &2.70 &23.32 \\[0.05cm]
300048 & 213.635 & 1.731 & 0.055 & 4.65 & 26 & cAGN/cMD &125.13 &518.04 \\[0.05cm]
300120 & 214.377 & 0.350 & 0.054 &38.28 & 37 & cMD &1.81 &7.63 \\[0.05cm]

\hline
\label{table-Group_results}
\end{tabular}
 \vspace{0.2cm}
\begin{minipage}{0.95\textwidth}
\small \textbf{Notes:} Group ID -- ID of each galaxy group as in GAMA's group catalog, RA -- Right Ascension of Brightest Central Galaxy (BCG) of the group, Dec -- Declination of Brightest Central Galaxy of the group, $z$ -- median redshift of the galaxy group, Halo mass -- unbiased halo mass estimate from GAMA's group catalog, $N_{\rm fof}$ -- number of group members, Classification -- type of diffuse emission from group; cH - candidate Halo, cMH - candidate Mini Halo, cR - candidate Relic, cMD - candidate Merger Driven, cAGN - candidate remnant AGN/irregular extended morphology, Integrated flux density -- 3$\sigma$ integrated flux densities as measured from diffuse maps for each candidate diffuse source in group at 943 MHz from EMU 
\newline * and  at 1.37GHz from DINGO , radio power -- diffuse radio power at 1.4 GHz extrapolated for both DINGO and EMU assuming a spectral index of -1.3.
\newline \dag The cH mentioned here is likely associated with a higher-redshift cluster along the same line of sight, rather than being related to the group itself. Furthermore, comparing the diffuse emission size to the overall group scales suggests that mini-halos may be a ubiquitous feature across the entire sample.

\end{minipage}
\end{table*}

\begin{table*}
\centering
\contcaption{Properties of galaxy groups with candidate diffuse emission}
\begin{tabular}{ccccccccc}
\hline
\vspace{0.1cm}
Group ID & RA (J2000)  & Dec (J2000)   & $z$ & Halo mass & $N_{\rm fof}$ & Classification & Integrated flux density & Radio power\\
 & (degrees) & (degrees)&  &(10$^{13}$$M_\odot$) & &  & (mJy) & (10$^{21}$ W Hz$^{-1}$)  \\[0.05cm]
\hline
300162 & 214.4571 & 0.510 & 0.054 & 12.21 & 12 & cAGN/cMD & 1.48& 5.89\\[0.05cm]
300206 & 211.715 & 0.850 & 0.048 &5.08 & 8 &cMD &0.10 &0.33 \\[0.05cm]
300439 & 220.109 & 1.258 & 0.078 &  0.53& 6 & cMD & 0.32&2.91\\[0.05cm]

300514 & 217.691 & 0.247 & 0.056 & 1.17& 6 & cMD &1.73&7.58\\[0.05cm]

300567 & 216.401 & 0.590 & 0.085 &1.60 & 5 & cMH & 0.02&0.24\\[0.05cm]
 & &&&&& cR& 0.51 &5.47 \\[0.05cm]

300594 & 221.036 & 1.195 & 0.028& 0.30& 5 & cMH &0.55 &0.62\\[0.05cm]
300667 & 218.189 & 1.877 & 0.031 &  0.23& 5 &  cMH &0.16 &0.20 \\[0.05cm]
 & &&&&& cR& 0.06 &0.07 \\[0.05cm]

300723 & 222.679 & $-$1.714 & 0.027 & 0.30 & 5 &  cMD &0.86 &0.87 \\[0.05cm]
 & &&&&& cMH/cMD& 3.46 &3.50 \\[0.05cm]

301021 & 221.864 & 1.978 & 0.035 & 4.05 & 10 & cAGN/cMD & 16.83&27.21\\[0.05cm]
 & &&&&& cAGN/cMD& 11.93 &19.20\\[0.05cm]
  & &&&&& cAGN/cMD& 6.50 &10.50 \\[0.05cm]
\hline
\end{tabular}

\vspace{0.2cm}
\begin{minipage}{0.95\textwidth}
\small \textbf{Notes:} Group ID -- ID of each galaxy group as in GAMA's group catalog, RA -- Right Ascension of Brightest Central Galaxy of the group, Dec -- Declination of Brightest Central Galaxy of the group, $z$ -- Median redshift of the galaxy group, Halo mass -- unbiased halo mass estimate from GAMA's group catalog, $N_{\rm fof}$ -- number of group members, Classification -- type of diffuse emission from group; cMH - candidate Mini Halo, cR - candidate Relic, cMD - candidate Merger Driven, cAGN - candidate remnant AGN/irregular extended morphology, Integrated flux density -- 3$\sigma$ integrated flux densities as measured from diffuse maps for each candidate diffuse source in group at 943 MHz from EMU 
\newline * and  at 1.37GHz from DINGO , radio power -- diffuse radio power at 1.4 GHz extrapolated for both DINGO and EMU assuming a spectral index of -1.3.

\end{minipage}
\end{table*}
\subsection{Global diffuse signal from galaxy groups}\label{subsection4.1}
Our main aim is to determine average diffuse emission as a function of halo mass in the galaxy group regime. Figure~\ref{fig:diffuse emission from groups} shows the relationship between radio power at 1.4 GHz and halo mass for our galaxy group sample, comparing measurements obtained without 3$\sigma$ clipping (left panel) and with a 3$\sigma$ clipping of pixel values (right panel). In both panels, detections are indicated by purple points with associated uncertainties, while non-detections are represented as upper limits (grey arrows), with negative integrated flux density measurements shown separately in yellow. We note that the negative integrated flux densities are non-physical and arise from residual imaging artefacts which imprint negative bowls in the images. In such specific groups, genuine diffuse emission may be present but is artificially suppressed by these surrounding negative artefacts.
 \par Without 3$\sigma$ clipping, the distribution of data points (left panel) appears systematically elevated, with a larger fraction of groups classified as detections. However, this method may overestimate the number of radio-luminous groups by including spurious contributions from noise and residual imaging artefacts.
Applying a 3$\sigma$ clip (right panel) yields a much more conservative set of detections. The majority of low-significance points are now treated as upper limits, substantially reducing the number of spurious positive measurements. Importantly, the detections that remain under this stricter criterion are generally higher-power systems, showing a trend with halo mass. This supports the interpretation that 3$\sigma$ clipping helps isolate real diffuse radio emission in group environments, although we are potentially excluding real but low-surface-brightness signals near the noise threshold. The $\sigma$-clipped analysis should be interpreted as a conservative lower limit on the incidence of diffuse emission.

\par Quantitatively, without $\sigma$-clipping, we identify 118 detections with a median radio power of $2.64 \times 10^{21}$ W Hz$^{-1}$, 282 upper limits (including sources whose error bars extend below zero) and
37 negative cases, re-interpreted as upper limits after taking the absolute value.
In the sample with $\sigma$-clipping, the relative fractions shift, with more detections i.e., 145 with a median radio power of $1.13 \times 10^{21}$ W Hz$^{-1}$, 244 upper limits and 11 negative cases. 
These statistics emphasize that our data is strongly affected by upper limits. It suggests that diffuse radio emission in galaxy groups is only detectable in the brightest systems with current sensitivity. 

\par We analyse the variation in diffuse radio power using inverse-variance weighted binning across five halo mass bins. These bins were initially defined to ensure a minimum of 30 groups per bin for statistical robustness. However, the final halo mass bin contains only five groups. Consequently, the average power in this high-mass regime is potentially biased by the presence of candidate detections. We also scale all negative flux density measurements to zero when performing weighted binning.
The weighted means are shown by the black data points in Figure~\ref{fig:diffuse emission from groups}. We observe a statistically significant underlying correlation between diffuse radio power and halo mass, albeit with substantial intrinsic scatter. To formally quantify this relationship, we performed Pearson, Spearman rank, and Kendall $\tau$ tests on the sample. Across both the unclipped and $3\sigma$-clipped data points, these tests yielded correlation coefficients in the range of $r \approx 0.2$ to $0.39$ with highly significant $p$-values ($p \sim 10^{-7}$). These quantitative metrics confirm that while the large inherent scatter in the group regime may drive down the correlation strength, the overarching physical scaling relationship between radio power and system mass remains robust. 
Despite the prevalence of upper limits within the dataset, this trend suggests a non-zero average diffuse radio power across the sampled galaxy groups.

\subsection{Visually inspected groups}\label{subsection4.2}
The results in the preceding section were found from the idea that most or all of the groups would not have diffuse emission above the detection limit. Thus a blind survey summing all the emission inside the group and averaging large numbers of groups together yielded a statistical average of the low level emission. However, it could be seen that within some groups there are larger regions of emission in the diffuse images. Additionally, in our flux density measurements, we find negative integrated fluxes in some groups. These non-physical values do not originate from real diffuse emission but from contamination by residual imaging artefacts, which can imprint negative flux regions in the maps. Yellow data points in the Figure~\ref{fig:diffuse emission from groups} are examples of such groups, where the real diffuse emission may be present, but is suppressed due to artefacts and the integrated flux  is negative. Furthermore, it is difficult to separate emission related to group environments and any background/foreground emission in an automated setup. Groups with unrelated emission may lie above our detection criterion, bias the scaling relations and thus our interpretation of IGrM. Consequently, even though we treat negative flux densities and measurements with larger errors as upper limits, we still can not be sure of detections in groups coming from purely group-related diffuse emission and such cases need to be studied with caution. In order to come up with candidates of real group diffuse emission, we carry out a visual inspection of all groups using GAMA's optical images overlaid onto radio images.
\begin{figure*}
    \centering        \includegraphics[width=0.9\linewidth]{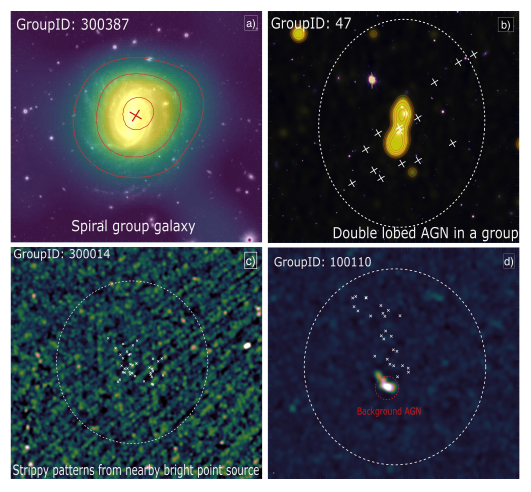}

    \caption{Example images of groups which are excluded from the final sample. Top left panel: Multicolor image of Group 300387, background image is the optical image from GAMA, blue color shows emission from the diffuse EMU map, red contours trace the normal EMU map radio emission and the red marker shows the central group galaxy. The radio emission covers the extent of spiral galaxy. Top right panel: Diffuse EMU image of Group  300014 with white markers showing group members and white dotted circle representing the group region, where group region is mainly dominated by artefacts.
    Bottom left panel: Multicolor image of Group 47, background image is the optical image from GAMA, yellow color shows emission from normal resolution EMU map, white markers show group galaxies and white dotted circle showing group region, showing a FR-II radio galaxy in the group. Bottom right panel: diffuse EMU image of Group  100110 where white markers are group members, white circle is the group region and red circle show background AGN which is not related to the group.}
    \label{fig:example_images}
\end{figure*}
 
\subsubsection{Selection criteria}
We identify and flag groups which are contaminated by obvious artefacts plus unrelated group emission to make final sample of candidate diffuse emission in groups. All groups were visual inspected by three co-authors to verify the presence of candidate diffuse radio emission.
 \par 
 The exclusion criteria are given below. 
 \begin{itemize}
    
    \item We exclude emission below 3$\sigma$ as it is potentially consistent with noise in the image.
    \item We exclude emission from nearby face-on spiral galaxies, see (Panel a) Figure~\ref{fig:example_images}. Such spirals may be group galaxies, although the radio emission is comparable in extent to the optical size of the galaxy, and hence likely not a diffuse emission.
    \item We exclude purely AGN and galaxy scale emission, for example, FR-I \& FR-II like radio galaxies. An example of such emission is seen in (Panel b) Figure~\ref{fig:example_images}, where galaxy Group - 47 hosts a double radio lobed AGN.
     \item We also exclude emission arising from image artefacts, see (Panel c) Figure~\ref{fig:example_images}. Here, the artefacts are seen as stripy patterns which get amplified in diffuse maps. Such artefacts suppress any real diffuse emission in these regions, if any present. 
     
     \item There are also background/unrelated group sources which could be mistaken as group related in the maps.  Such emissions may be blobs from background galaxies and AGN.  These blobs may be low-level artefacts amplified in diffuse maps or associated with background galaxies which is confirmed by looking at galaxy positions from GAMA. An example is shown in (Panel d) Figure~\ref{fig:example_images}, where emission from background AGN could be mistaken as relic emission, however is actually a background AGN at relatively higher redshift. 
     
 \end{itemize}
 
 The galaxy groups that we include in our final visually inspected sample are groups which show bright central blob in diffuse maps and this emission is well extended beyond the group galaxy sizes in the optical. In addition to this, we also include irregular shaped radio emission in diffuse maps, which do not seem to be associated with background galaxies. Lastly, we include diffuse emission which may look like remnant AGN activity, such emission also has irregular morphology but have 2-3 nearby galaxy group members. It is difficult to put constraints on the origin of emission like this without any X-ray or spectral indices information. It could be either the result of purely AGN activity, or be remnant AGN plasma that is subsequently re-accelerated by various interaction phenomena within the IGrM. Our final visually inspected sample includes all discrete diffuse sources in galaxy groups which come under the above criteria and are listed individually in Table~\ref{table-Group_results}.
 \subsection{Classification of candidate diffuse sources}
The candidate diffuse sources in groups are visually categorised based on their position, presence of group galaxies in the vicinity and shape, see Table~\ref{tab:diffuse_sources}. At this stage, all diffuse sources in our visually inspected sample are considered candidate detections. We categorize these into specific classes which are candidate mini-halos (cMH), candidate merger-driven emission (cMD), candidate relics (cR), and candidate AGN emission (cAGN); based on their morphology and their position in the group relative to optical emission. We assign this candidate status to sources that present marginal detections or ambiguous features that broadly align with the expected properties of these systems. This designation also includes sources that are difficult to classify robustly due to poor data quality, calibration artifacts, or severe contamination from superimposed compact sources. Confirming the true physical nature of these structures will require further diagnostics, such as spatially resolved spectral indices and deeper X-ray observations. 
  
\par Diffuse sources which are seen in the form a central blob are classified cMH. Their sizes are comparatively small (a few to tens of kpc) compared to group sizes and a central BCG is always present at the centre. Emission which is elongated and located at the periphery of the group is categorised as cR. cR are typically highly polarised sources and likely originated form merger-induced shocks. Examples of such emission can be seen in Groups  - 101106, 100146, etc. Some diffuse sources are likely the result of interacting group galaxies and re-acceleration of particles in IGrM due to merging activities, these are classified as cMD. Examples of cMD type of category can be seen in Groups  - 100710, 200445, 300514, etc. and examples of cMH category are Groups  - 100165, 100072, 201057, 300567, etc. cAGN emission may come from AGN-galaxy interaction component, remnant AGN plasma emission and a few cases of extended emission with irregular morphologies. An example of irregular extended emission is found in Group 200792 (see Figure~\ref{fig:diffusefilteringexample}), which is known to possess outer radio shells. It is likely originating from shock acceleration during group-scale merging activities (Koribalski et al., in prep). Additionally, Group 301021 demonstrates a different morphological class, where bright diffuse emission fills the group region.
 This will be a subject of future work.
\begin{table*}
    \centering
    \caption{Classification of candidate diffuse radio sources in our sample}
    \label{tab:diffuse_sources}
    \begin{tabularx}{\textwidth}{@{} l l >{\centering\arraybackslash}p{2.5cm} >{\centering\arraybackslash}X >{\centering\arraybackslash}X @{}}
        \toprule
        \textbf{Category} & \textbf{Label} & \textbf{Location} & \textbf{Features} & \textbf{Possible origin} \\ \midrule
        Mini Halo & cMH & Central & Central blobs  & Merger driven\\ \addlinespace
        Relic & cR & Periphery & Elongated shape at group edges & Shock acceleration\\ \addlinespace
        Merger Driven & cMD & Near group galaxies & Irregular morphologies & Linked to merging/galaxy interactions \\ \addlinespace
        AGN Emission & cAGN & Near AGN & Irregular morphologies & Remnant plasma/AGN-galaxy interactions \\ \bottomrule
    \end{tabularx}
    \begin{minipage}{\textwidth}
    \vspace{6pt}
    \footnotesize
    \textbf{Notes:} Labels are defined as follows: Candidate Mini Halo (cMH), Candidate Relic (cR), Candidate Merger Driven (cMD), and Candidate AGN-related emission (cAGN). The possible origin column describes the primary physical mechanism believed to be responsible for the observed radio emission in group-scale environments.
\end{minipage}
\end{table*}

\subsubsection{Statistics} 
\begin{figure*}
    \centering
    \includegraphics[width=1\linewidth]{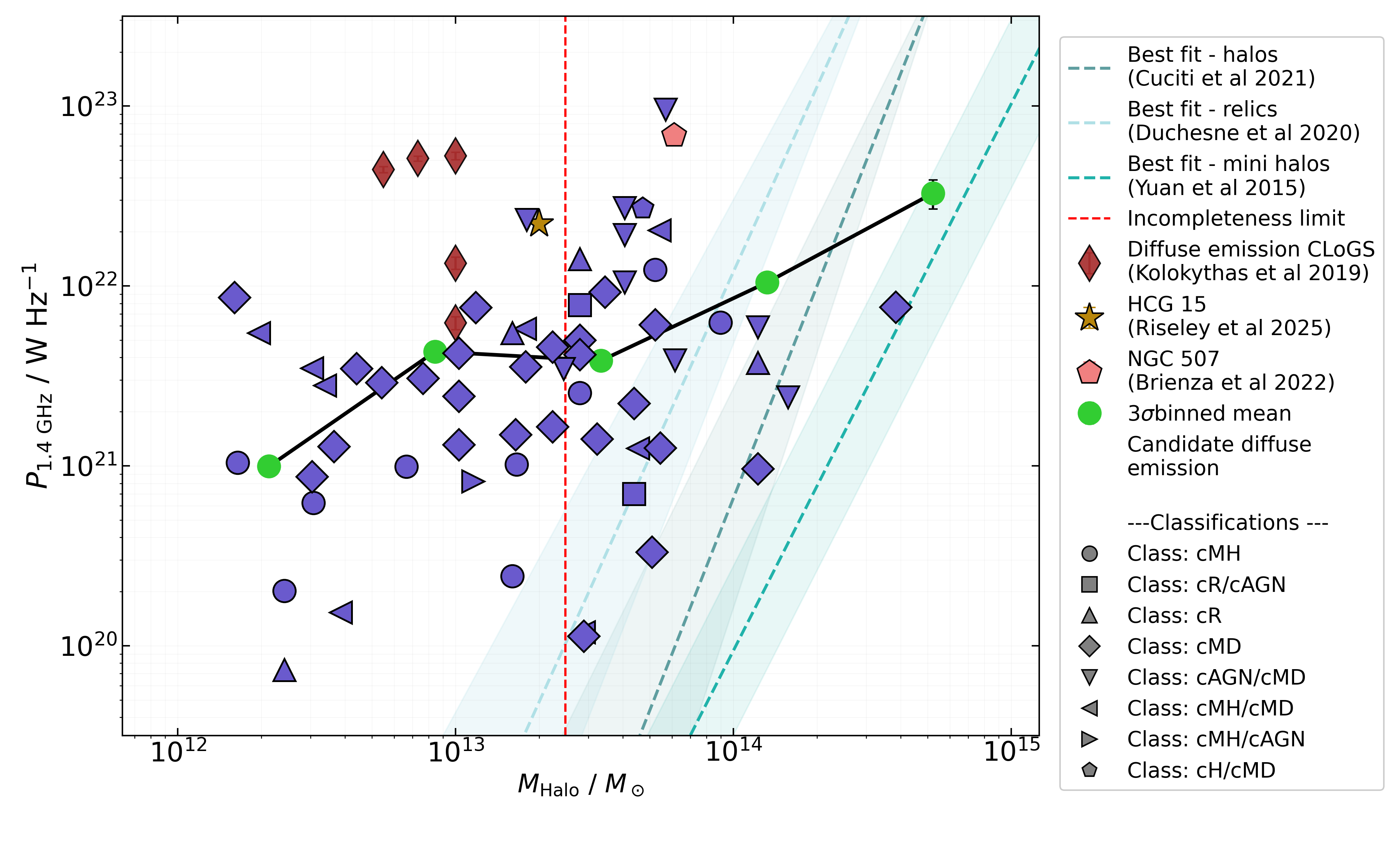}
    
    \caption{The radio power versus halo mass for all manually selected candidate diffuse emission in galaxy groups (purple markers). Red diamonds represent diffuse emission from CLoGS \citep{2019MNRAS.489.2488K}, the golden star indicates remnant plasma in HCG 15 \citep{2025arXiv250308840R} and orange marker shows NGC 507 \citep{2022A&A...661A..92B}. The red dotted line represents the incompleteness limit of our sample. Green data points are show 3$\sigma$ weighted average from all 400 galaxy groups.}
    \label{fig:MI_results}
\end{figure*}
\par After visually inspecting all of the galaxy groups in our sample, we have 46 groups which potentially host candidate diffuse emission and 61 diffuse sources (some groups have more than one diffuse radio emission source) in our final visually inspected sample. The integrated flux densities for diffuse sources which are above 3$\sigma$ in each galaxy group are measured (in contrast to the values in Figure~\ref{fig:diffuse emission from groups} which is sum of all pixels within the group radius). 
 The details of all candidate diffuse emissions and their group properties are listed in Table~\ref{table-Group_results} and the images of all can be seen in x~\ref{sec:appendix_plots}. 
 \par The candidate diffuse groups contribute to 11.50\% of the whole group sample and the diffuse radio powers of the visually inspected groups lie in the range of 10$^{19}$ - 10$^{24}$ W Hz$^{-1}$ with a median value of 3.48 $\times$ 10$^{21}$ W Hz$^{-1}$. Figure~\ref{fig:MI_results} shows the radio power of candidate diffuse emission sources as a function of group halo masses, compared to other group data points from the literature. We calculate the 1.4 GHz radio power assuming a fixed spectral index of $\alpha=-$1.3, assumed because group-scale phenomena often involve older relativistic electron populations; their true spectra may be steeper or curved. However, adopting a broader range of spectral indices between $-$1.0 and $-$1.7 yields negligible changes to the extrapolated power, indicating that our fiducial choice of $-$1.3 is reliable. For comparison, we plot the known scaling relations for giant radio halos \citep{2021A&A...647A..51C}, radio mini-halos \citep{2015ApJ...813...77Y} and radio relics \citep{2021PASA...38...53D} derived from massive galaxy clusters, extrapolated down to the group mass regime\footnote{A simple way to interpret this relation is to notice that the radio power produced by electrons accelerated by Diffusive Shock Acceleration should scale as $P_{\rm 1.4~GHz} \propto P_{\rm kin}[B^2/(B^2+B_{CMB}^2)] \propto \rho v_s^3 R^3 B^2$, i.e., the radio power scales with the kinetic power of shocks, and is additionally weighted by the magnetic field (as long as $B \ll B_{CMB}=3.27 \cdot (1+z)^2~\mu\mathrm{G}$, i.e. the magnetic field at shocks is smaller than the magnetic-field equivalent to the energy density of the CMB). From self-similarity, it follows that $v_s \propto M^{1/3}$, $R \propto M^{1/3}$ and $B \propto M^{1/3}$, hence putting everything together we get $P_{1.4 ~\rm GHz} \propto M^{8/3}$, which is only slightly flatter than the observed and simulated scaling relations. However, the most powerful emitters are expected to be merging clusters of galaxies, in which the departures from self-similarity can be significant and also the magnetic field amplification can be enhanced, which can overall produce a slightly steeper relation with mass \citep[][]{2011ApJ...735...96S,2014MNRAS.444.3130D,2024A&A...690A.146N}.}. We also include individual detections from recent studies of galaxy groups, including CLoGS \citep{2019MNRAS.489.2488K}, HCG 15 \citep{2025arXiv250308840R}, and NGC 507 \citep{2022A&A...661A..92B}.  
\par Our classification reveals a diverse landscape of diffuse radio emission in group-scale environments. We note that these classifications are not mutually exclusive; a single group can host multiple distinct emission features simultaneously. The sample is dominated by cMD and cMH features, which collectively appear in over 80\% of the sample ($n=51$). While cMD sources (present in 65.6\% of the groups) would typically trace the broader turbulent volume of a merging group, the cMH candidates (26.2\%) represent more compact, centrally located analogs to cluster radio mini-halos, often confined within the group's core potential. In contrast, cAGN accounts for 21.3\% of the sample, characterized by irregular, remnant plasma morphologies that suggest non-thermal energy injection from individual galaxy nuclei rather than global gravitational collapse. The least frequent category i.e, cR appears in only 9.8\% of the sources. Furthermore, overlapping emission origins are evident in sources such as Group - 200084, which exhibits cH/cMD emission. While this may arise from merging activity within the group, we note that there is a known galaxy cluster at a higher redshift residing at the same line of sight, and this complex emission may be partially related to it.

\par Collectively, candidate diffuse sources in groups largely fall above the extrapolated scaling relations for giant halos and relics in galaxy clusters. This suggests that diffuse emission in galaxy groups and low mass clusters is more luminous than what would be predicted by simply scaling down the relations found in massive galaxy clusters.
In addition to this, it may indicate that the efficiency of particle acceleration or the magnetic field strength in groups is higher than the predictions of  simple mass-scaling models. It challenges the idea that groups are just scaled-down galaxy clusters; they may have distinct dynamic processes (like highly efficient AGN feedback cycles) that boost their radio brightness.
Nevertheless, our candidate diffuse sources exhibit radio powers consistent with individually detected diffuse structures in galaxy groups, as well as those found in the CLoGS sample.

\begin{figure*}
    \centering
    \includegraphics[width=1\columnwidth]{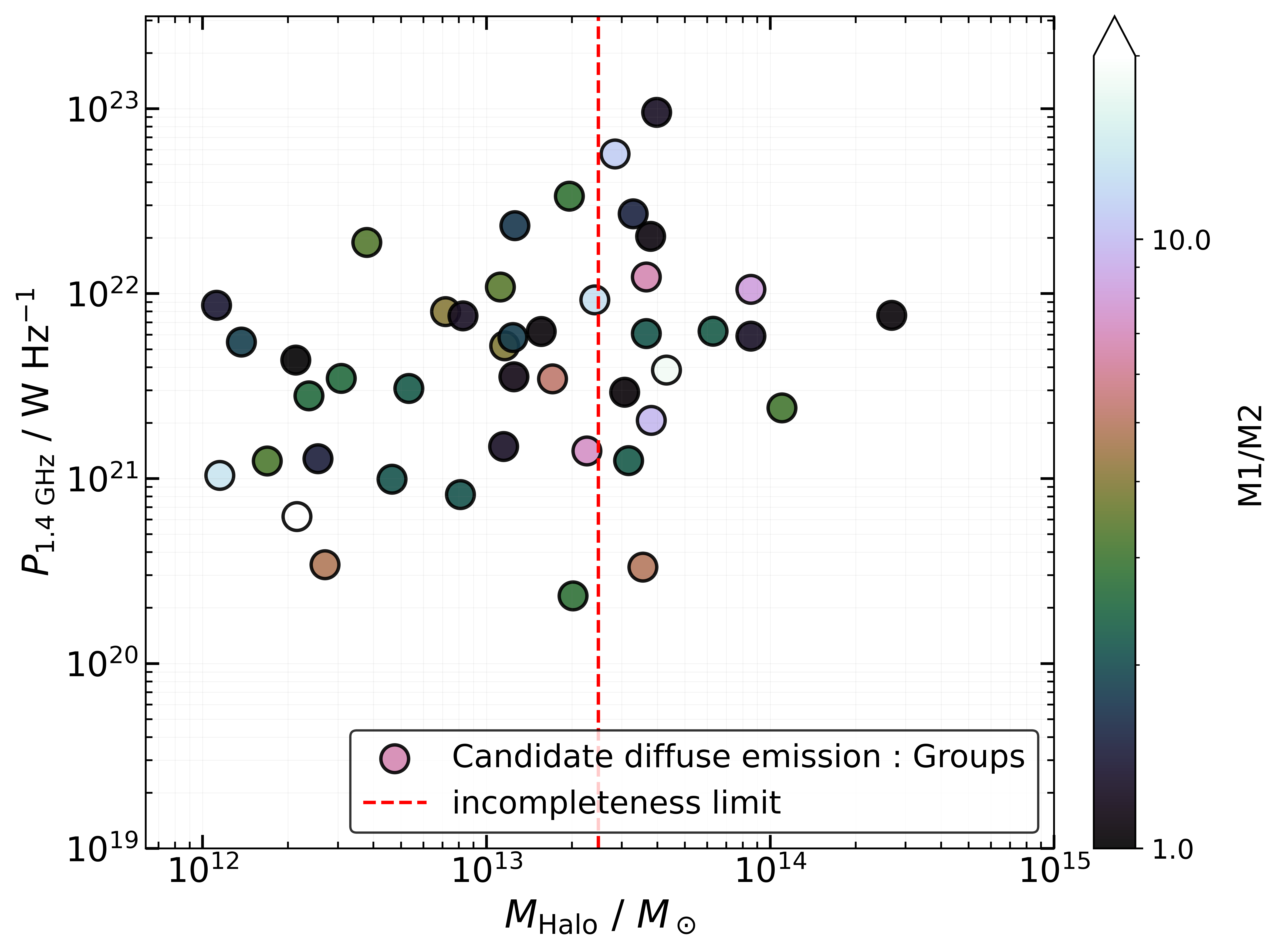}
    \includegraphics[width=1\columnwidth]{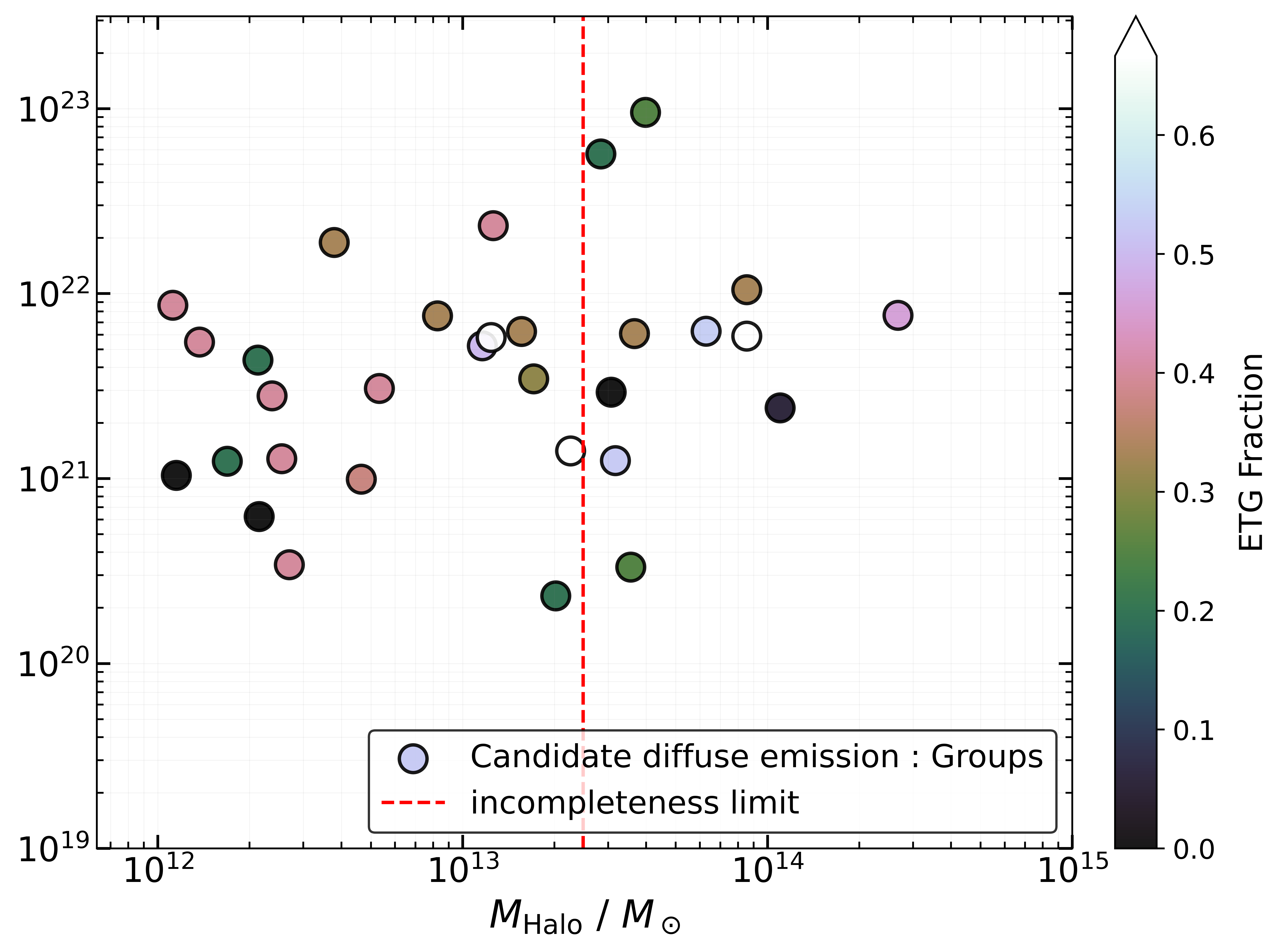}
    \includegraphics[width=1\columnwidth]{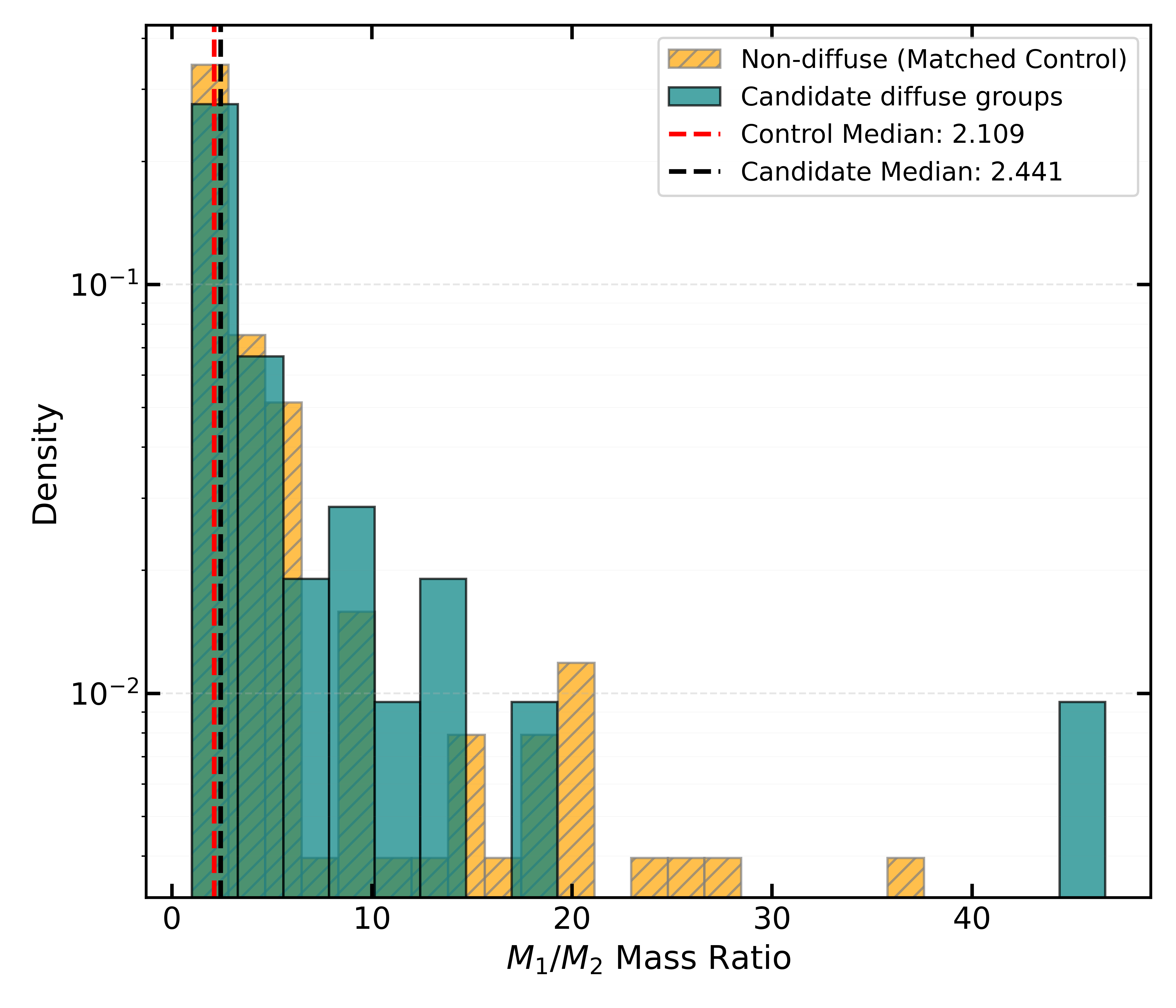}
    \includegraphics[width=1\columnwidth]{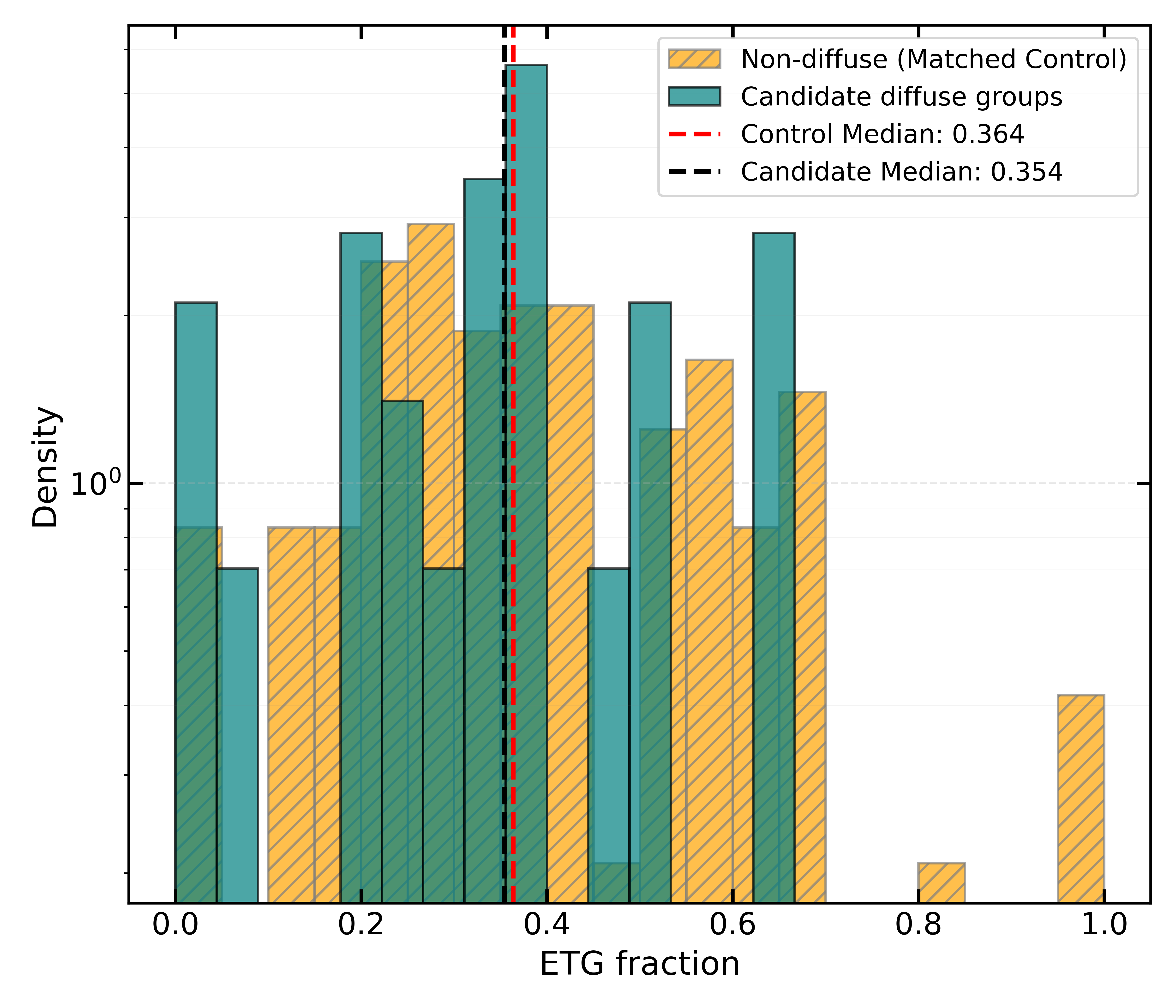}
    
    \caption{Top panels: The radio power versus halo mass for all manually selected candidate diffuse emission in galaxy groups color coded by group properties from GAMA survey. Top left panel : M1/M2 mass ratios in groups. Top right panel : ETG fraction in groups. In the top panels, the red dotted line represents the incompleteness limit of our sample.
    Bottom panels show histograms of M1/M2 mass ratios and ETG fraction for our candidate diffuse emission sample (teal color) as compared to control sample (yellow color), with black and red vertical lines showing the median values respectively. }
    \label{fig:MI_group_properties}
\end{figure*}

\subsubsection{Group properties}
 
The origin of diffuse emission in galaxy groups may depend upon the group environment, their evolutionary stages, merging activities and the presence of central AGN. To explore this, we use data from the GAMA survey which can serve as proxies for the properties of the group environment. Firstly, we calculate the M1/M2 ratio for our final sample, which is the ratio of stellar masses of the most massive galaxy to the second most massive galaxy in a group and serves as proxy for the age of the group. Higher M1/M2 would represent older groups, where the central galaxy is massive and satellite galaxies have previously merged into the central galaxy \citep{2019MNRAS.483.5444D}. The M1/M2 ratio for our sample lies between 1 - 45 with a median value of 2.40 , see Figure~\ref{fig:MI_group_properties}, suggesting the majority of the groups in our sample are relatively young. Secondly, we derive (Early Type Galaxy) ETG fraction for the galaxy groups, which is the ratio of number of ETGs to the total number of galaxies in a galaxy group. Generally, evolved groups are known to have ETGs as their central galaxies, which lack ongoing star formation activity \citep{2016ApJ...818..182V,2022MNRAS.510.4191K}. Therefore, ETG fraction can give us information about the evolutionary stage of galaxy groups. Morphology classifications for all GAMA galaxies only exist below redshift 0.08 \citep{2022MNRAS.513..439D} and therefore we could only calculate this fraction for 32 groups in our final sample. The ETG fractions lie between 0 to 0.7 with a median value of 0.3. We note that there is no statistically significant correlation observed between the radio power of diffuse emission candidates and either the mass ratios (Spearman $\rho=-$0.327, $p=0.07$) or the ETG fractions (Spearman $\rho=$0.261, $p=0.15$); Pearson tests similarly confirmed a lack of linear dependence ($p>$0.10 in both cases). 
\par We construct a control sample for candidate diffuse emission groups using a nearest-neighbour matching algorithm in the 2D log10($M_{\rm halo} - z$) plane. The resulting control population is statistically indistinguishable from the candidate diffuse emission group sample, as evidenced by KS-test p-values of 0.99 for both parameters. The small median offsets (0.038 dex in mass and 0.0017 in redshift) ensure that environmental effects tied to halo mass and cosmic time are effectively controlled, allowing for a direct comparison of morphological properties. The bottom panels of the Figure~\ref{fig:MI_group_properties} display the probability density of M1/M2 mass ratio and ETG fraction for the candidate diffuse emission sample compared to the control sample. 
\par An analysis of the overall characteristics of our candidate diffuse emission sample reveals a population dominated by unrelaxed, assembling systems. This conclusion is drawn from the median values of the evolutionary proxies of the sample, M1/M2 mass ratios show a low median value of $\sim 2.44$ and a low median ETG fraction of $\sim 0.35$. These values provide a preliminary indicator that the bulk of these systems have not yet formed a dominant central brightest group galaxy (BGG) and remain dynamically young, meaning galaxy interactions likely play a significant role. Notably, this aligns with the thresholds established by the CLoGs study, which classifies groups with a spiral fraction exceeding 0.75, equivalent to an ETG fraction below 0.25 as dynamically young and actively assembling \citep[]{2022MNRAS.510.4191K}. Our low median ETG fraction is consistent with this benchmark. It is important to note, however, that this demographic profile likely reflects the inherent sample bias of optically selected group catalogs, which are heavily populated by actively assembling systems at these mass scales.
\par Crucially, despite the overall unrelaxed nature of the sample, we observe no significant correlation between the diffuse radio power and either M1/M2 or the ETG fraction. This lack of correlation suggests the possibility that once diffuse emission is triggered, its luminosity might not scale simply with the dynamical maturity of the host group. This aligns with findings from CLoGs \citep[e.g.,][]{2018MNRAS.481.1550K,2019MNRAS.489.2488K}, which demonstrate that diffuse radio emission in groups can originate from a diverse mixture of physical mechanisms across various evolutionary stages. These origins range from active AGN jets triggered by merger-driven gas inflows, to unassociated diffuse remnants of past AGN outbursts, to minor-merger induced gas sloshing as observed in systems like NGC 1550 \citep{2022MNRAS.510.4191K}. If our sample encompasses a complex superposition of these diverse physical origins, it is possible that any underlying correlation between total diffuse radio power and a single evolutionary proxy may be washed out.
Given the current limited information on broader galaxy properties and morphological types, our analysis may fall short in providing exhaustive diagnostics for the nature of the group environment. Future studies will prioritize measuring missing group properties such as passive fractions, detailed morphologies of central dominant galaxies, radio spectral indices, and X-ray properties to provide a more complete picture of the true evolutionary state of these systems and its physical connection to diffuse radio emission.
\section{Discussion}\label{section5}
\subsection{Accuracy of the diffuse filtering code - simulated Gaussian}
A primary concern in deep continuum surveys like EMU and DINGO is the presence of residual calibration errors, such as negative bowls and stripey patterns around bright point sources, which can either mimic or suppress real diffuse structures. We test this with simple simulations of 2D-Gaussian profiles in the normal resolution EMU maps and study their behaviour in different artefact dominated regions in the diffuse EMU data. 

We simulated 2D-Gaussian profiles with a full-width at half-maximum (FWHM) of 60$\arcsec$, purposefully selected to be four times the synthesized beam of the EMU data (15$\arcsec$). Given that our diffuse filtering code utilizes a box filter size three times the beam size i.e, 45$\arcsec$, this ensures that the simulated Gaussian profiles are sufficiently extended to be retained by the pipeline.
The models were injected with an integrated flux density of 1.4 mJy, resulting in a peak flux density of approximately 88 $\mu$Jy beam$^{-1}$. This level corresponds to 2.5$\sigma_{RMS}$, representing the typical threshold for faint, candidate diffuse emission in our sample. These models were then placed into distinct artefact dominated regions in the EMU images.

Results from this test suggest that in artefact free regions filtering code successfully recovers the injected Gaussian models with high fidelity, confirming that the filtering code is calibrated for isolated diffuse structures. However in high artefact regions, which are dominated by negative troughs and high intensity stripey patterns from bright point sources, the simulated emission was frequently suppressed. The negative artefacts often cancel out the positive flux of the injected Gaussian, dropping it below the $-$4$\sigma$ threshold utilized by the filtering code for background subtraction. In such cases, the diffuse emission becomes undetectable, appearing as a negative region in the resulting diffuse maps. For more details, see Appendix~\ref{Appendix - numerical simulations}.

 \par The radio continuum images in the equatorial GAMA regions are dominated by artefacts from bright point sources which we have minimized with self-calibration and imaging. However due to poor uv sampling, we cannot completely remove them. Groups with real diffuse emission (if any) dominated with artefacts cannot be separated and we do not include such groups in our manual inspection analysis. Therefore, we may have missed a fraction of groups with real diffuse radio emission.
 
\subsection{Comparison with Simulations}

\begin{figure*}
    \centering
    \includegraphics[width=0.85\linewidth]{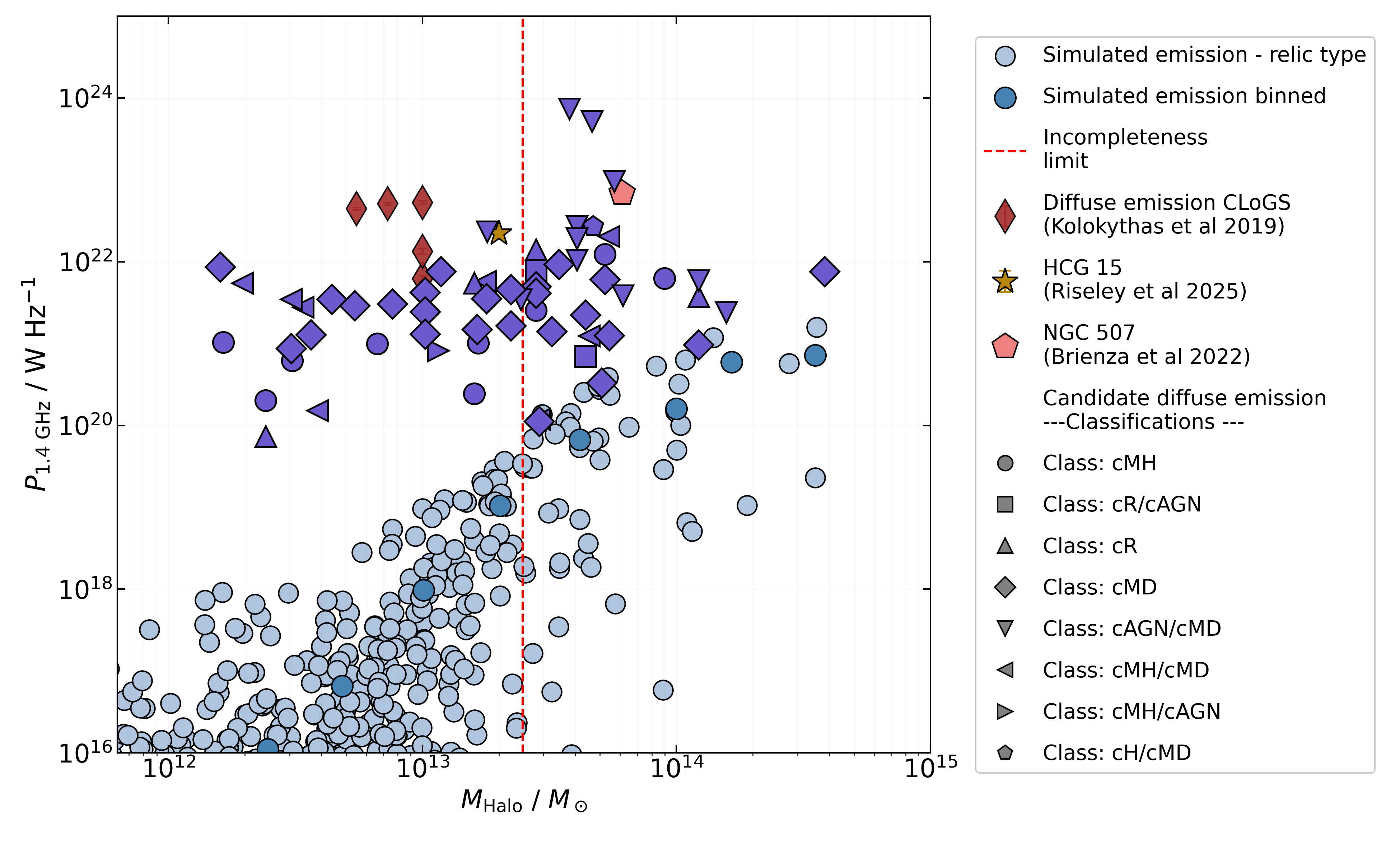}
 
    \caption{Radio power from simulated relic like emission for halo mass range 10$^{12}$ - 10$^{15}$ $M_\odot$ in comparison with our radio power in our sample. The red dotted line represents the incompleteness limit of our sample.}
    \label{fig:simulations_comparison}
\end{figure*}
While theory predicts that diffuse emission from turbulent re-acceleration should drop off at lower masses, Magneto Hydrodynamical (MHD) cosmological simulations do predict the existence of shock-induced diffuse emission extending into the low halo mass range. The simulated emission we use here is taken from a recent cosmological run. It is an ideal MHD cosmological simulation produced with a customised version of the ENZO code \footnote{\url{https://enzo-project.org}} as discussed in \citet[][]{2025A&A...696A..58V}. These new MHD simulations include a realistic view of cosmic magnetic fields, which are assumed to be seeded at run-time by galaxies (by magnetized winds and jets from active galactic nuclei) but are also introduced at the star of the simulation in the form of stochastic random magnetic fields drawn from a $P_B(k) dk \propto k^{-1} dk$ power spectrum, mimicking the generation by inflationary processes.

\par In the simulation, both processes of star formation and AGN feedback  release,  at run-time, additional magnetic fields,  corresponding to a fixed ($10\%$) fraction of their energy, mimicking the ubiquitous process of magnetisation by galaxies, which roughly follows the cosmic star formation history. The simulation tracks the propagation of a passive fluid of cosmic ray electrons (injected by shocks assuming DSA, AGN or star formation). The synchrotron emission is computed in post-processing, by parsing through pre-computed template spectra covering a large grid of possible density, temperature, magnetic field and redshift values, as detailed in \citet{2025A&A...696A..58V}. Here, we focus on the simulation of a comoving $(85\,\rm Mpc)^3$ volume simulated with $1024^3$ uniform cells, in which we extracted halos and their radio emission within $R_{100}$ at $z=0.02$ and at the observing frequency of $\nu=1.4$\,GHz.  


\par 
Figure~\ref{fig:simulations_comparison} illustrates the comparison between the candidate diffuse sources from our sample and simulated relic emission in galaxy groups. Our observational data points disagree with simulated predictions for diffuse relic-type emission. The simulated relic type diffuse emission (blue markers) for galaxy groups  lies predominantly below 10$^{21}$  W Hz$^{-1}$, whereas our observational data points (purple markers) generally sit above this threshold. This offset indicates that the diffuse emission predicted by current models using shock acceleration is largely below the detection threshold of these ASKAP observations, consistent with the high prevalence of non-detections in our sample. The mismatch is larger than the possible effect of a lack of magnetic field in simulated halos, which based on the comparison with \cite{2024MNRAS.533.4068A} would only require an upwards renormalisation by $\sim 2-3$ (yielding a $\sim 2^2-3^2$ renormalisation in synchrotron emission for the same objects). 
While this discrepancy may indicate residual contamination from compact AGN, it could alternatively point to a class of bright diffuse emission that has been previously uncharacterised in this mass regime. Furthermore, it raises the possibility that current simulation models lack specific physical aspects of re-acceleration processes which perhaps reproduce these higher luminosity events in group-scale environments.
\subsection{Underlying processes: group-scale diffuse emission}

We emphasize the fact that the physical origin of diffuse radio emission in the group and low-mass cluster regime does not appear to be a continuation of the scaling relations produced by massive galaxy clusters. In these systems, the thermal and non-thermal energy budgets are substantially reduced compared to high mass galaxy clusters. In addition, the IGrM is characterised by lower gas densities and magnetic field strengths. As a result, classical turbulent re-acceleration models predict that bright, Mpc scale halos should be rare below $M_{500}\sim$
few $\times 10^{14}\,M_\odot$, and that any diffuse synchrotron
emission present in groups will be fainter, more compact and typically steeper-spectrum than in rich clusters. 
Consequently, one should not expect a simple continuity of radio halo properties from clusters down to the group regime, nor the ubiquitous presence of halo analogues at lower masses. 
\par Instead, there may exist a transition mass scale below which the dominant emission mechanisms change, and where diffuse radio emission if present, is driven primarily by alternative processes such as re-acceleration of fossil plasma, weak shocks, or AGN-related activity. Our results suggest that galaxy groups occupy an intermediate regime in which diffuse radio emission is intermittent, morphologically diverse, and governed by different physical drivers than those operating in massive clusters. It is known that AGN feedback operates differently in groups , sometimes acting as a near-continuous bubbling mode with moderate thermal regulation \citep{Birzanetal12,Panagouliaetal14}, or in other cases manifesting as powerful AGN outbursts \citep{OSullivanetal11,Kolokythasetal15}. Such powerful outbursts, potentially extending beyond the virial radius, significantly influence the development of groups and shape their radio signatures, while also providing seed electrons for the observed diffuse radio emission.

Nevertheless, we suggest several plausible channels for generating observable diffuse radio emission in low-mass systems. The phenomenology of diffuse emission in these systems is governed by three primary channels: turbulent re-acceleration in the group, sloshing-induced activity in the core, and the re-acceleration of fossil AGN plasma.

\par \textbf{Turbulence and Merger Activity:} In the framework of turbulent re-acceleration models, the turbulent energy injection rate scales approximately as $M^{5/3}$ \citep{2003ApJ...584..190F,2007MNRAS.378..245B}. For low-mass clusters and groups the energy available to accelerate cosmic ray electrons (CRe) is substantially lower than in rich clusters. Therefore, group radio halos may look like patchy, irregular emissions filling a significant fraction of the virialized volume (up to $\sim$300 kpc), driven by group–group mergers or strong substructure infall. However, because the acceleration efficiency is low, these sources are predicted to exhibit ultra-steep spectra ($\alpha \leq -1.5$), particularly at GHz frequencies.
\par \textbf{Sloshing and Cool Cores:} In relaxed systems possessing cool cores, the primary driver of diffuse emission shifts from merger-driven turbulence to sloshing motions of the gas within the central potential well. Simulations of cluster cores suggest that even modest turbulence, when coupled with pre-existing CRe, can sustain mini-halos \citep{2011ApJ...743...16Z,2013ApJ...762...78Z}. Extrapolating to the group regime, we expect group mini-halos; diffuse synchrotron emitters on scales of 20–100 kpc. These sources are confined to the dense core where the magnetic field is relatively enhanced ( $B \sim$ a few $\mu$G) compared to the wider IGrM. Unlike their massive cluster counterparts, group mini-halos are physically smaller and likely fainter, driven by the interaction of sloshing fronts with seed electrons provided by repeated outbursts from the Brightest Group Galaxy (BGG).

\par \textbf{The Role of Fossil Plasma and Remnant AGN Activity:}
 Remnant AGN represent the fading lobes of past nuclear activity, whose emission persists after the central engine has switched off. While such sources are not intrinsically linked to the IGrM, their aged relativistic plasma can be subsequently re-energised by weak shocks, compression, or turbulence within the environment \citep{2001A&A...366...26E,2020A&A...634A...4M}. In this sense, fossil plasma can act as a seed population that blurs the distinction between classical AGN remnants and genuinely diffuse group-scale emission. This population of sources is driven by faded or reprocessed AGN plasma. In these systems, buoyant bubbles from past activity rise and mix with the IGrM, or fossil electrons are gently re-energized by weak sloshing turbulence and minor interactions. 
 
\par Observationally, this is seen as amorphous, fuzzy cocoons or ghost bubbles roughly centered on the BGG, or previously active AGN, but morphologically distinct from active jets. Because these structures rely on aged electron populations without strong shock re-acceleration, they typically exhibit steep, curved spectra with $\alpha \leq -1.3$ .
The diffuse emission in HCG 15 represents a striking example of magnetized fossil plasma. It is classified as a remnant source, likely originating from a past AGN outburst associated with the group member HCG 15-D roughly 80–86 Myr ago. The emission is believed to be re-energized or shaped by the draping of magnetic field lines in the IGrM, highlighting how group-scale weather can preserve old plasma \citep{2025arXiv250308840R}.

\par In the CLoGS sample, genuine diffuse radio emission is a rare phenomenon, detected in only about 10\% of the groups. These detections highlight a remarkable diversity in physical origins and morphologies. For instance, NGC 5903 serves as the primary example of this emission type in the low-richness subsample, hosting a prominent diffuse source approximately 65--75 kpc in size. With a high radio power of $1.50 \times 10^{23}$ W Hz$^{-1}$ at 235 MHz and a steep spectral index ($\alpha = -1.03$), its origin is attributed to a hybrid mechanism involving both an interaction-triggered AGN outburst and violent tidal collisions between galaxies. 

Other systems exhibit entirely different diffuse characteristics. ESO 507-25 presents asymmetric diffuse clumps separated from the central core, while in the X-ray faint system NGC 3078, the symmetric morphology and flat spectral index of the emission suggest it likely originates from a disrupted or decollimated jet, or potentially a galactic wind and outflow associated with a failed AGN outburst. Environmental interactions also play a critical role; in the case of NGC 1587, which resides in a dynamically active X-ray bright group, the emission is potentially powered by shock-driven re-acceleration triggered by its ongoing encounter with the companion galaxy NGC 1588, although the lack of visible X-ray shock fronts makes this interpretation tentative. Finally, for sources like NGC 677, the diffuse structures may represent aged fossil radio lobes from past activity that have been confined and distorted by the pressure of the intragroup medium (IGrM), gradually losing their distinct lobe-like shape over time.
\par 
Similarly, NGC 507 hosts a complex system of diffuse emission showing filamentary structures extending tens to hundreds of kpcs. This emission is interpreted as AGN remnant plasma transported and distorted by sloshing motions of the thermal gas, evidenced by its alignment with X-ray cold fronts. The source has a steep spectrum ($\alpha \leq$-1.2), typical of aging plasma that has lost its high energy electrons but is being gently re-accelerated or compressed by the group's dynamic environment \citep{2022A&A...661A..92B}.
\par The diverse physical mechanisms observed in systems like NGC 5903, NGC 507, and the broader CLoGS sample provide a vital framework for interpreting our 1.4 GHz group statistics. The rarity of diffuse emission in CLoGS is strongly reflected in our sample, though it is heavily dominated by upper limits. The nature of the mechanisms required to power these diffuse sources, such as interaction-triggered AGN outbursts, tidal collisions, and IGrM sloshing, suggests that the majority of our undetected groups are likely in dynamically relaxed states lacking recent injection or re-acceleration events. Furthermore, the steep spectral indices typical of aged fossil plasma ($\alpha \leq $ -1.0) mean that such structures fade rapidly at higher frequencies. 
\par At the sensitivities and angular resolution of EMU, we expect to detect only
the brightest examples of such emission on an object-by-object basis. Nearby($z \lesssim$ 0.1) cool-core groups with relatively strong core magnetic fields should host mini-halos with integrated flux densities of a few mJy at 1.3 GHz, corresponding to surface brightnesses of order tens of $\mu$Jy beam$^{-1}$ on 20–100 kpc scales. Similarly, low-mass clusters and rich groups undergoing particularly energetic mergers may host group-scale halos that are detectable in EMU as low-surface-brightness excesses extending beyond the central radio galaxy population. However, for the majority of groups the
expected halo powers are such that the corresponding surface brightness at 1.3 GHz falls near or below the nominal EMU detection threshold. In these systems, statistical techniques such as stacking as a function of halo mass, dynamical state and redshift will be required to constrain the average non-thermal emission.

\par EMU’s combination of wide sky coverage, uniform sensitivity and $\sim15\arcsec$ resolution is thus particularly well suited to defining the
bright end of the diffuse radio emission population in low-mass systems. Direct detections in individual groups will provide valuable laboratories for testing models of particle acceleration and magnetic field amplification in shallow potential wells, while stacked analyses of optically and X-ray selected group samples will probe the typical synchrotron emissivity in the IGrM an order of magnitude below the single-object detection threshold. Together, these measurements will link the non-thermal properties of groups to those of both massive clusters and the surrounding cosmic web, and will provide critical constraints for future low-frequency surveys with Square Kilometre Array-Low (SKA-Low) and its precursors.

In summary, the majority of diffuse emission observed in galaxy groups likely arises from a combination of weak turbulent re-acceleration, reprocessed AGN plasma, and intermittent group-scale interactions, rather than being a simple low-mass analogue of cluster radio halos.

\section{Conclusions}\label{section6}
We have conducted a systematic search for diffuse radio emission in 400 galaxy groups using ASKAP's EMU and DINGO data, providing one of the first statistical constraints on non-thermal emission in the group regime.
Diffuse emission is detected in a non-negligible fraction of systems ($\sim $11\%), with radio powers spanning 10$^{19}$-10$^{24}$ W Hz$^{-1}$. Both M1/M2 ratios and ETG fractions suggest that these galaxy groups are relatively young and evolving systems where the interaction between group galaxies and mergers may power the observed diffuse radio emission. While a weak positive trend between radio power and halo mass is observed, the majority of systems remain below the detection threshold, indicating that diffuse emission in groups is intrinsically faint and likely intermittent.
Crucially, the observed emission lies above simple extrapolations of cluster radio halo and relic scaling relations, demonstrating that galaxy groups are not merely scaled-down versions of clusters. Instead, the nature of diffuse emission in low-mass systems appears to be governed by different physical processes.
Comparison with MHD simulations shows that shock acceleration alone under-predicts the observed radio power, suggesting that additional mechanisms are required. The prevalence of irregular, merger-driven morphologies and centrally concentrated emission indicates that fossil plasma re-acceleration and AGN-related processes likely play a dominant role in powering diffuse emission in groups.
These results point to a transition in non-thermal physics between clusters and groups, where
\begin{itemize}
    \item turbulent re-acceleration becomes inefficient,
    \item shocks are weaker,
    \item and recycled relativistic plasma becomes increasingly important.
\end{itemize}

This work demonstrates the power of EMU for probing low-surface-brightness emission in large numbers of low-mass systems and highlights the need for deeper observations, spectral index measurements, and multi-wavelength constraints (e.g. X-ray, optical dynamics) to fully disentangle the origin of diffuse emission in the IGrM.
\par In future, high-sensitivity, low-frequency observations with next-generation instruments like the SKA-Low will be crucial for detecting the older, steep-spectrum fossil plasma currently hidden below our 1.4 GHz sensitivity limits. Additionally, deeper multi-wavelength follow-up with high-resolution X-ray imaging will be able to confirm the dynamical age of these groups and understand the physical mechanisms driving such emission. Finally, the offset between our observations and theoretical models highlights the need for refined MHD simulations that incorporate different ways of injecting AGN feedback in the IGrM which can be main driver for accelerating the plasma in the group medium.

\section*{Acknowledgements}
We thank the anonymous referee for a careful reading of the manuscript and highly constructive feedback that significantly improved the clarity and presentation of this work. This work was supported by resources provided by the Pawsey Supercomputing Centre under project code ja3 with funding from the Australian Government and the Government of Western Australia.
This scientific work uses data obtained from Inyarrimanha Ilgari Bundara, the CSIRO Murchison Radio-astronomy Observatory. We acknowledge the Wajarri Yamaji People as the Traditional Owners and native title holders of the Observatory site. CSIRO’s ASKAP radio telescope is part of the Australia Telescope National Facility (\url{https://ror.org/05qajvd42}). Operation of ASKAP is funded by the Australian Government with support from the National Collaborative Research Infrastructure Strategy. ASKAP uses the resources of the Pawsey Supercomputing Research Centre. Establishment of ASKAP, Inyarrimanha Ilgari Bundara, the CSIRO Murchison Radio-astronomy Observatory and the Pawsey Supercomputing Research Centre are initiatives of the Australian Government, with support from the Government of Western Australia and the Science and Industry Endowment Fund. This paper includes archived data obtained through the CSIRO ASKAP Science Data Archive, CASDA (\url{http://data.csiro.au}).
SW acknowledges AA, AD, DF, JP, JL, MM, SFR, TZ, TM for their useful suggestions and feedback during the preparation of this paper.
LJMD acknowledges support from the Australian Re-
search Council’s Future Fellowship scheme (FT200100055) and the Australian Research Council’s Dis-
covery Project scheme (DP250104611). 
CJR acknowledges financial support from the German Science Foundation DFG, via the Collaborative Research Center SFB1491 `Cosmic Interacting Matters – From Source to Signal'. KK acknowledges funding support from the South African Radio Astronomy Observatory (SARAO) and the National Research Foundation (NRF; grant UID: 97930).  FV acknowledges funding under the European Union’s  Horizon Europe program through the ERC Synergy Grant COSMOMAG (Project Id. 101224803). FV acknowledges the CINECA awards ``IscrB{\_}CREW" and ``IscrC{\_}UMAREL" under the ISCRA initiative, for the availability of high-performance computing resources and support, and the usage of online storage tools kindly provided by the INAF Astronomical Archive (IA2) initiative (http://www.ia2.inaf.it). 
SW also expresses deepest gratitude to TV, RW and HS for their constant encouragement and moral support over the course of this entire project.
\section*{Data Availability}

The radio continuum data from the EMU and DINGO surveys are available through the CSIRO ASKAP Science Data Archive (CASDA) at \url{http://data.csiro.au}. The optical group catalogs can be found on the GAMA survey database (\url{http://www.gama-survey.org/}). The reprocessed data generated in this research and other data such as final catalogs of candidate diffuse sources and calculated upper limits, can be shared upon request.



\bibliographystyle{mnras}
\bibliography{example} 

@ARTICLE{2019SSRv..215...16V,
       author = {{van Weeren}, R.~J. and {de Gasperin}, F. and {Akamatsu}, H. and {Br{\"u}ggen}, M. and {Feretti}, L. and {Kang}, H. and {Stroe}, A. and {Zandanel}, F.},
        title = "{Diffuse Radio Emission from Galaxy Clusters}",
      journal = {\ssr},
         year = 2019,
        month = feb,
       volume = {215},
       number = {1},
          eid = {16},
        pages = {16},
          doi = {10.1007/s11214-019-0584-z},
archivePrefix = {arXiv},
       eprint = {1901.04496},
 primaryClass = {astro-ph.HE},
       adsurl = {https://ui.adsabs.harvard.edu/abs/2019SSRv..215...16V}
}

@ARTICLE{1992ApJ...399..353H,
       author = {{Hickson}, Paul and {Mendes de Oliveira}, Claudia and {Huchra}, John P. and {Palumbo}, Giorgio G.},
        title = "{Dynamical Properties of Compact Groups of Galaxies}",
      journal = {\apj},
         year = 1992,
        month = nov,
       volume = {399},
        pages = {353},
          doi = {10.1086/171932},
       adsurl = {https://ui.adsabs.harvard.edu/abs/1992ApJ...399..353H}
}

@ARTICLE{2017MNRAS.471....2P,
       author = {{Paul}, S. and {John}, R.~S. and {Gupta}, P. and {Kumar}, H.},
        title = "{Understanding `galaxy groups' as a unique structure in the universe}",
      journal = {\mnras},
         year = 2017,
        month = oct,
       volume = {471},
       number = {1},
        pages = {2-11},
          doi = {10.1093/mnras/stx1488},
archivePrefix = {arXiv},
       eprint = {1706.01916},
 primaryClass = {astro-ph.CO},
       adsurl = {https://ui.adsabs.harvard.edu/abs/2017MNRAS.471....2P}
}

@ARTICLE{2000ARA&A..38..289M,
       author = {{Mulchaey}, John S.},
        title = "{X-ray Properties of Groups of Galaxies}",
      journal = {\araa},
         year = 2000,
        month = jan,
       volume = {38},
        pages = {289-335},
          doi = {10.1146/annurev.astro.38.1.289},
archivePrefix = {arXiv},
       eprint = {astro-ph/0009379},
 primaryClass = {astro-ph},
       adsurl = {https://ui.adsabs.harvard.edu/abs/2000ARA&A..38..289M}
}

@ARTICLE{2015A&A...573A.118L,
       author = {{Lovisari}, L. and {Reiprich}, T.~H. and {Schellenberger}, G.},
        title = "{Scaling properties of a complete X-ray selected galaxy group sample}",
      journal = {\aap},
         year = 2015,
        month = jan,
       volume = {573},
          eid = {A118},
        pages = {A118},
          doi = {10.1051/0004-6361/201423954},
archivePrefix = {arXiv},
       eprint = {1409.3845},
 primaryClass = {astro-ph.CO},
       adsurl = {https://ui.adsabs.harvard.edu/abs/2015A&A...573A.118L}
}

@ARTICLE{2007MNRAS.376..193J,
       author = {{Jetha}, N.~N. and {Ponman}, T.~J. and {Hardcastle}, M.~J. and {Croston}, J.~H.},
        title = "{Active galactic nuclei heating in the centres of galaxy groups: a statistical study}",
      journal = {\mnras},
         year = 2007,
        month = mar,
       volume = {376},
       number = {1},
        pages = {193-204},
          doi = {10.1111/j.1365-2966.2006.11407.x},
       adsurl = {https://ui.adsabs.harvard.edu/abs/2007MNRAS.376..193J}
}

@ARTICLE{2024MNRAS.533.4068A,
       author = {{Anderson}, Craig S. and {McClure-Griffiths}, N.~M. and {Rudnick}, L. and {Gaensler}, B.~M. and {O'Sullivan}, S.~P. and {Bradbury}, S. and {Akahori}, T. and {Baidoo}, L. and {Bruggen}, M. and {Carretti}, E. and {Duchesne}, S. and {Heald}, G. and {Jung}, S.~L. and {Kaczmarek}, J. and {Leahy}, D. and {Loi}, F. and {Ma}, Y.~K. and {Osinga}, E. and {Seta}, A. and {Stuardi}, C. and {Thomson}, A.~J.~M. and {Van Eck}, C. and {Vernstrom}, T. and {West}, J.},
        title = "{Probing the magnetized gas distribution in galaxy groups and the cosmic web with POSSUM Faraday rotation measures}",
      journal = {\mnras},
         year = 2024,
        month = oct,
       volume = {533},
       number = {4},
        pages = {4068-4080},
          doi = {10.1093/mnras/stae1954},
archivePrefix = {arXiv},
       eprint = {2407.20325},
 primaryClass = {astro-ph.GA},
       adsurl = {https://ui.adsabs.harvard.edu/abs/2024MNRAS.533.4068A}
}

@ARTICLE{2004IJMPD..13.1549G,
       author = {{Govoni}, Federica and {Feretti}, Luigina},
        title = "{Magnetic Fields in Clusters of Galaxies}",
      journal = {International Journal of Modern Physics D},
         year = 2004,
        month = jan,
       volume = {13},
       number = {8},
        pages = {1549-1594},
          doi = {10.1142/S0218271804005080},
archivePrefix = {arXiv},
       eprint = {astro-ph/0410182},
 primaryClass = {astro-ph},
       adsurl = {https://ui.adsabs.harvard.edu/abs/2004IJMPD..13.1549G}
}

@ARTICLE{2011MNRAS.416.2640R,
       author = {{Robotham}, A.~S.~G. and {Norberg}, P. and {Driver}, S.~P. and {Baldry}, I.~K. and {Bamford}, S.~P. and {Hopkins}, A.~M. and {Liske}, J. and {Loveday}, J. and {Merson}, A. and {Peacock}, J.~A. and {Brough}, S. and {Cameron}, E. and {Conselice}, C.~J. and {Croom}, S.~M. and {Frenk}, C.~S. and {Gunawardhana}, M. and {Hill}, D.~T. and {Jones}, D.~H. and {Kelvin}, L.~S. and {Kuijken}, K. and {Nichol}, R.~C. and {Parkinson}, H.~R. and {Pimbblet}, K.~A. and {Phillipps}, S. and {Popescu}, C.~C. and {Prescott}, M. and {Sharp}, R.~G. and {Sutherland}, W.~J. and {Taylor}, E.~N. and {Thomas}, D. and {Tuffs}, R.~J. and {van Kampen}, E. and {Wijesinghe}, D.},
        title = "{Galaxy and Mass Assembly (GAMA): the GAMA galaxy group catalogue (G$^{3}$Cv1)}",
      journal = {\mnras},
         year = 2011,
        month = oct,
       volume = {416},
       number = {4},
        pages = {2640-2668},
          doi = {10.1111/j.1365-2966.2011.19217.x},
archivePrefix = {arXiv},
       eprint = {1106.1994},
 primaryClass = {astro-ph.CO},
       adsurl = {https://ui.adsabs.harvard.edu/abs/2011MNRAS.416.2640R}
}

@ARTICLE{2021PASA...38...46N,
       author = {{Norris}, Ray P. and {Marvil}, Joshua and {Collier}, J.~D. and {Kapi{\'n}ska}, Anna D. and {O'Brien}, Andrew N. and {Rudnick}, L. and {Andernach}, Heinz and {Asorey}, Jacobo and {Brown}, Michael J.~I. and {Br{\"u}ggen}, Marcus and {Crawford}, Evan and {English}, Jayanne and {Rahman}, Syed Faisal ur and {Filipovi{\'c}}, Miroslav D. and {Gordon}, Yjan and {G{\"u}rkan}, G{\"u}lay and {Hale}, Catherine and {Hopkins}, Andrew M. and {Huynh}, Minh T. and {HyeongHan}, Kim and {James Jee}, M. and {Koribalski}, B{\"a}rbel S. and {Lenc}, Emil and {Luken}, Kieran and {Parkinson}, David and {Prandoni}, Isabella and {Raja}, Wasim and {Reiprich}, Thomas H. and {Riseley}, Christopher J. and {Shabala}, Stanislav S. and {Sheil}, Jaimie R. and {Vernstrom}, Tessa and {Whiting}, Matthew T. and {Allison}, James R. and {Anderson}, C.~S. and {Ball}, Lewis and {Bell}, Martin and {Bunton}, John and {Galvin}, T.~J. and {Gupta}, Neeraj and {Hotan}, Aidan and {Jacka}, Colin and {Macgregor}, Peter J. and {Mahony}, Elizabeth K. and {Maio}, Umberto and {Moss}, Vanessa and {Pandey-Pommier}, M. and {Voronkov}, Maxim A.},
        title = "{The Evolutionary Map of the Universe pilot survey}",
      journal = {\pasa},
         year = 2021,
        month = sep,
       volume = {38},
          eid = {e046},
        pages = {e046},
          doi = {10.1017/pasa.2021.42},
archivePrefix = {arXiv},
       eprint = {2108.00569},
 primaryClass = {astro-ph.CO},
       adsurl = {https://ui.adsabs.harvard.edu/abs/2021PASA...38...46N}
}

@ARTICLE{2011MNRAS.413..971D,
       author = {{Driver}, S.~P. and {Hill}, D.~T. and {Kelvin}, L.~S. and {Robotham}, A.~S.~G. and {Liske}, J. and {Norberg}, P. and {Baldry}, I.~K. and {Bamford}, S.~P. and {Hopkins}, A.~M. and {Loveday}, J. and {Peacock}, J.~A. and {Andrae}, E. and {Bland-Hawthorn}, J. and {Brough}, S. and {Brown}, M.~J.~I. and {Cameron}, E. and {Ching}, J.~H.~Y. and {Colless}, M. and {Conselice}, C.~J. and {Croom}, S.~M. and {Cross}, N.~J.~G. and {de Propris}, R. and {Dye}, S. and {Drinkwater}, M.~J. and {Ellis}, S. and {Graham}, Alister W. and {Grootes}, M.~W. and {Gunawardhana}, M. and {Jones}, D.~H. and {van Kampen}, E. and {Maraston}, C. and {Nichol}, R.~C. and {Parkinson}, H.~R. and {Phillipps}, S. and {Pimbblet}, K. and {Popescu}, C.~C. and {Prescott}, M. and {Roseboom}, I.~G. and {Sadler}, E.~M. and {Sansom}, A.~E. and {Sharp}, R.~G. and {Smith}, D.~J.~B. and {Taylor}, E. and {Thomas}, D. and {Tuffs}, R.~J. and {Wijesinghe}, D. and {Dunne}, L. and {Frenk}, C.~S. and {Jarvis}, M.~J. and {Madore}, B.~F. and {Meyer}, M.~J. and {Seibert}, M. and {Staveley-Smith}, L. and {Sutherland}, W.~J. and {Warren}, S.~J.},
        title = "{Galaxy and Mass Assembly (GAMA): survey diagnostics and core data release}",
      journal = {\mnras},
         year = 2011,
        month = may,
       volume = {413},
       number = {2},
        pages = {971-995},
          doi = {10.1111/j.1365-2966.2010.18188.x},
archivePrefix = {arXiv},
       eprint = {1009.0614},
 primaryClass = {astro-ph.CO},
       adsurl = {https://ui.adsabs.harvard.edu/abs/2011MNRAS.413..971D}
}

@ARTICLE{2012ApJ...748....7M,
       author = {{Menanteau}, Felipe and {Hughes}, John P. and {Sif{\'o}n}, Crist{\'o}bal and {Hilton}, Matt and {Gonz{\'a}lez}, Jorge and {Infante}, Leopoldo and {Barrientos}, L. Felipe and {Baker}, Andrew J. and {Bond}, John R. and {Das}, Sudeep and {Devlin}, Mark J. and {Dunkley}, Joanna and {Hajian}, Amir and {Hincks}, Adam D. and {Kosowsky}, Arthur and {Marsden}, Danica and {Marriage}, Tobias A. and {Moodley}, Kavilan and {Niemack}, Michael D. and {Nolta}, Michael R. and {Page}, Lyman A. and {Reese}, Erik D. and {Sehgal}, Neelima and {Sievers}, Jon and {Spergel}, David N. and {Staggs}, Suzanne T. and {Wollack}, Edward},
        title = "{The Atacama Cosmology Telescope: ACT-CL J0102-4915 ``El Gordo,'' a Massive Merging Cluster at Redshift 0.87}",
      journal = {\apj},
         year = 2012,
        month = mar,
       volume = {748},
       number = {1},
          eid = {7},
        pages = {7},
          doi = {10.1088/0004-637X/748/1/7},
archivePrefix = {arXiv},
       eprint = {1109.0953},
 primaryClass = {astro-ph.CO},
       adsurl = {https://ui.adsabs.harvard.edu/abs/2012ApJ...748....7M}
}

@ARTICLE{2024PASA...41...26D,
       author = {{Duchesne}, S.~W. and {Botteon}, A. and {Koribalski}, B.~S. and {Loi}, F. and {Rajpurohit}, K. and {Riseley}, C.~J. and {Rudnick}, L. and {Vernstrom}, T. and {Andernach}, H. and {Hopkins}, A.~M. and {Kapinska}, A.~D. and {Norris}, R.~P. and {Zafar}, T.},
        title = "{Evolutionary Map of the Universe (EMU): A pilot search for diffuse, non-thermal radio emission in galaxy clusters with the Australian SKA Pathfinder}",
      journal = {\pasa},
         year = 2024,
        month = jan,
       volume = {41},
          eid = {e026},
        pages = {e026},
          doi = {10.1017/pasa.2024.10},
archivePrefix = {arXiv},
       eprint = {2402.06192},
 primaryClass = {astro-ph.CO},
       adsurl = {https://ui.adsabs.harvard.edu/abs/2024PASA...41...26D}
}

@ARTICLE{2021Galax...9...84N,
       author = {{Nikiel-Wroczy{\'n}ski}, B{\l}a{\.z}ej},
        title = "{Somewhere in between: Tracing the Radio Emission from Galaxy Groups (or Why Does the Future of Observing Galaxy Groups with Radio Telescopes Look Promising?)}",
      journal = {Galaxies},
         year = 2021,
        month = oct,
       volume = {9},
       number = {4},
          eid = {84},
        pages = {84},
          doi = {10.3390/galaxies9040084},
archivePrefix = {arXiv},
       eprint = {2112.03635},
 primaryClass = {astro-ph.GA},
       adsurl = {https://ui.adsabs.harvard.edu/abs/2021Galax...9...84N}
}

@ARTICLE{2025arXiv250308840R,
       author = {{Riseley}, C.~J. and {Vernstrom}, T. and {Lovisari}, L. and {O'Sullivan}, E. and {Gastaldello}, F. and {Brienza}, M. and {Nayak}, Prasanta K. and {Bonafede}, A. and {Carretti}, E. and {Duchesne}, S.~W. and {Giacintucci}, S. and {Hopkins}, A.~M. and {Koribalski}, B.~S. and {Loi}, F. and {Pfrommer}, C. and {Raja}, W. and {Ross}, K. and {Rubinur}, K. and {Ruszkowski}, M. and {Shimwell}, T.~W. and {de Villiers}, M.~S. and {West}, J. and {Zovaro}, H.~R.~M. and {Akahori}, T. and {Anderson}, C.~S. and {Bomans}, D.~J. and {Drabent}, A. and {Rudnick}, L. and {Santra}, R.},
        title = "{Relighting the fire in Hickson Compact Group (HCG) 15: magnetised fossil plasma revealed by the SKA Pathfinders \& Precursors}",
      journal = {arXiv e-prints},
         year = 2025,
        month = mar,
          eid = {arXiv:2503.08840},
        pages = {arXiv:2503.08840},
          doi = {10.48550/arXiv.2503.08840},
archivePrefix = {arXiv},
       eprint = {2503.08840},
 primaryClass = {astro-ph.GA},
       adsurl = {https://ui.adsabs.harvard.edu/abs/2025arXiv250308840R}
}

@ARTICLE{2025PASA...42...71H,
       author = {{Hopkins}, Andrew and {Kapinska}, Anna and {Marvil}, Joshua and {Vernstrom}, Tessa and {Collier}, Jordan and {Norris}, Ray and {Gordon}, Yjan and {Duchesne}, Stefan and {Rudnick}, Lawrence and {Gupta}, Nikhel and {Carretti}, Ettore and {Anderson}, Craig and {Dai}, Shi and {G{\"u}rkan}, Gulay and {Parkinson}, David and {Prandoni}, Isabella and {Riggi}, Simone and {Shekhar Saraf}, Chandra and {Ma}, Yik Ki and {Filipovi{\'c}}, Miroslav D. and {Umana}, Grazia and {Bahr-Kalus}, Benedict and {Koribalski}, B{\"a}rbel Silvia and {Lenc}, Emil and {Ingallinera}, Adriano and {Afonso}, Jos{\'e} and {Ahmad}, Adeel and {Ahmed}, Ummee Tania and {Alexander}, Emma and {Andernach}, Heinz and {Asorey}, Jacobo and {Battisti}, Andrew J. and {Bilicki}, Maciej and {Botteon}, Andrea and {Brown}, Michael and {Br{\"u}ggen}, Marcus and {Cowley}, Michael and {Dage}, Kristen and {Hale}, Catherine Laura and {Hardcastle}, Martin and {Kothes}, Roland and {Lazarevi{\'c}}, Sanja and {Lin}, Yen-Ting and {Luken}, Kieran and {Moss}, Jeremy and {Prathap}, P.~K. Jahang and {ur Rahman}, Syed Faisal and {Reiprich}, Thomas and {Riseley}, Christopher and {Salvato}, Mara and {Seymour}, Nicholas and {Shabala}, Stanislav and {Smith}, Daniel and {Vaccari}, Mattia and {van Loon}, Jacco Th. and {Wong}, O. Ivy Ivy and {Zainal Alsaberi}, Rami and {Asher}, Albany and {Ball}, Brianna and {Barbosa}, Davi and {Biava}, Nadia and {Bradley}, Aaron and {Carvajal}, Rodrigo and {Crawford}, Evan J. and {Galvin}, Timothy James and {Huynh}, Minh and {Leahy}, Denis and {Matute}, Israel and {Moss}, Vanessa and {Pappalardo}, Ciro and {Smeaton}, Zachary and {Velovi{\'c}}, Velibor and {Zafar}, Tayyaba},
        title = "{The Evolutionary Map of the Universe: A new radio atlas for the southern hemisphere sky}",
      journal = {\pasa},
         year = 2025,
        month = may,
       volume = {42},
          eid = {e071},
        pages = {e071},
          doi = {10.1017/pasa.2025.10042},
archivePrefix = {arXiv},
       eprint = {2505.08271},
 primaryClass = {astro-ph.GA},
       adsurl = {https://ui.adsabs.harvard.edu/abs/2025PASA...42...71H}
}

@ARTICLE{2023MNRAS.518.4646R,
       author = {{Rhee}, Jonghwan and {Meyer}, Martin and {Popping}, Attila and {Bellstedt}, Sabine and {Driver}, Simon P. and {Robotham}, Aaron S.~G. and {Whiting}, Matthew and {Baldry}, Ivan K. and {Brough}, Sarah and {Brown}, Michael J.~I. and {Bunton}, John D. and {Dodson}, Richard and {Holwerda}, Benne W. and {Hopkins}, Andrew M. and {Koribalski}, B{\"a}rbel S. and {Lee-Waddell}, Karen and {L{\'o}pez-S{\'a}nchez}, {\'A}ngel R. and {Loveday}, Jon and {Mahony}, Elizabeth and {Roychowdhury}, Sambit and {Rozgonyi}, Krist{\'o}f and {Staveley-Smith}, Lister},
        title = "{Deep investigation of neutral gas origins (DINGO): H I stacking experiments with early science data}",
      journal = {\mnras},
         year = 2023,
        month = jan,
       volume = {518},
       number = {3},
        pages = {4646-4671},
          doi = {10.1093/mnras/stac3065},
archivePrefix = {arXiv},
       eprint = {2210.09697},
 primaryClass = {astro-ph.GA},
       adsurl = {https://ui.adsabs.harvard.edu/abs/2023MNRAS.518.4646R}
}

@ARTICLE{2011PASA...28..215N,
       author = {{Norris}, Ray P. and {Hopkins}, A.~M. and {Afonso}, J. and {Brown}, S. and {Condon}, J.~J. and {Dunne}, L. and {Feain}, I. and {Hollow}, R. and {Jarvis}, M. and {Johnston-Hollitt}, M. and {Lenc}, E. and {Middelberg}, E. and {Padovani}, P. and {Prandoni}, I. and {Rudnick}, L. and {Seymour}, N. and {Umana}, G. and {Andernach}, H. and {Alexander}, D.~M. and {Appleton}, P.~N. and {Bacon}, D. and {Banfield}, J. and {Becker}, W. and {Brown}, M.~J.~I. and {Ciliegi}, P. and {Jackson}, C. and {Eales}, S. and {Edge}, A.~C. and {Gaensler}, B.~M. and {Giovannini}, G. and {Hales}, C.~A. and {Hancock}, P. and {Huynh}, M.~T. and {Ibar}, E. and {Ivison}, R.~J. and {Kennicutt}, R. and {Kimball}, Amy E. and {Koekemoer}, A.~M. and {Koribalski}, B.~S. and {L{\'o}pez-S{\'a}nchez}, {\'A}. R. and {Mao}, M.~Y. and {Murphy}, T. and {Messias}, H. and {Pimbblet}, K.~A. and {Raccanelli}, A. and {Randall}, K.~E. and {Reiprich}, T.~H. and {Roseboom}, I.~G. and {R{\"o}ttgering}, H. and {Saikia}, D.~J. and {Sharp}, R.~G. and {Slee}, O.~B. and {Smail}, Ian and {Thompson}, M.~A. and {Urquhart}, J.~S. and {Wall}, J.~V. and {Zhao}, G. -B.},
        title = "{EMU: Evolutionary Map of the Universe}",
      journal = {\pasa},
         year = 2011,
        month = aug,
       volume = {28},
       number = {3},
        pages = {215-248},
          doi = {10.1071/AS11021},
archivePrefix = {arXiv},
       eprint = {1106.3219},
 primaryClass = {astro-ph.CO},
       adsurl = {https://ui.adsabs.harvard.edu/abs/2011PASA...28..215N}
}

@ARTICLE{2011A&A...532A..71R,
       author = {{Rau}, U. and {Cornwell}, T.~J.},
        title = "{A multi-scale multi-frequency deconvolution algorithm for synthesis imaging in radio interferometry}",
      journal = {\aap},
         year = 2011,
        month = aug,
       volume = {532},
          eid = {A71},
        pages = {A71},
          doi = {10.1051/0004-6361/201117104},
archivePrefix = {arXiv},
       eprint = {1106.2745},
 primaryClass = {astro-ph.IM},
       adsurl = {https://ui.adsabs.harvard.edu/abs/2011A&A...532A..71R}
}

@ARTICLE{2020PASA...37...48M,
       author = {{McConnell}, D. and {Hale}, C.~L. and {Lenc}, E. and {Banfield}, J.~K. and {Heald}, George and {Hotan}, A.~W. and {Leung}, James K. and {Moss}, Vanessa A. and {Murphy}, Tara and {O'Brien}, Andrew and {Pritchard}, Joshua and {Raja}, Wasim and {Sadler}, Elaine M. and {Stewart}, Adam and {Thomson}, Alec J.~M. and {Whiting}, M. and {Allison}, James R. and {Amy}, S.~W. and {Anderson}, C. and {Ball}, Lewis and {Bannister}, Keith W. and {Bell}, Martin and {Bock}, Douglas C. -J. and {Bolton}, Russ and {Bunton}, J.~D. and {Chippendale}, A.~P. and {Collier}, J.~D. and {Cooray}, F.~R. and {Cornwell}, T.~J. and {Diamond}, P.~J. and {Edwards}, P.~G. and {Gupta}, N. and {Hayman}, Douglas B. and {Heywood}, Ian and {Jackson}, C.~A. and {Koribalski}, B{\"a}rbel S. and {Lee-Waddell}, Karen and {McClure-Griffiths}, N.~M. and {Ng}, Alan and {Norris}, Ray P. and {Phillips}, Chris and {Reynolds}, John E. and {Roxby}, Daniel N. and {Schinckel}, Antony E.~T. and {Shields}, Matt and {Tremblay}, Chenoa and {Tzioumis}, A. and {Voronkov}, M.~A. and {Westmeier}, Tobias},
        title = "{The Rapid ASKAP Continuum Survey I: Design and first results}",
      journal = {\pasa},
         year = 2020,
        month = nov,
       volume = {37},
          eid = {e048},
        pages = {e048},
          doi = {10.1017/pasa.2020.41},
archivePrefix = {arXiv},
       eprint = {2012.00747},
 primaryClass = {astro-ph.IM},
       adsurl = {https://ui.adsabs.harvard.edu/abs/2020PASA...37...48M}
}

@ARTICLE{2021PASA...38...58H,
       author = {{Hale}, Catherine L. and {McConnell}, D. and {Thomson}, A.~J.~M. and {Lenc}, E. and {Heald}, G.~H. and {Hotan}, A.~W. and {Leung}, J.~K. and {Moss}, V.~A. and {Murphy}, T. and {Pritchard}, J. and {Sadler}, E.~M. and {Stewart}, A.~J. and {Whiting}, M.~T.},
        title = "{The Rapid ASKAP Continuum Survey Paper II: First Stokes I Source Catalogue Data Release}",
      journal = {\pasa},
         year = 2021,
        month = nov,
       volume = {38},
          eid = {e058},
        pages = {e058},
          doi = {10.1017/pasa.2021.47},
archivePrefix = {arXiv},
       eprint = {2109.00956},
 primaryClass = {astro-ph.GA},
       adsurl = {https://ui.adsabs.harvard.edu/abs/2021PASA...38...58H}
}

@INPROCEEDINGS{1995AAS...18711202B,
       author = {{Briggs}, D.~S.},
        title = "{High Fidelity Interferometric Imaging: Robust Weighting and NNLS Deconvolution}",
    booktitle = {American Astronomical Society Meeting Abstracts},
         year = 1995,
       series = {American Astronomical Society Meeting Abstracts},
       volume = {187},
        month = dec,
          eid = {112.02},
        pages = {112.02},
       adsurl = {https://ui.adsabs.harvard.edu/abs/1995AAS...18711202B}
}

@ARTICLE{2021PASA...38...53D,
       author = {{Duchesne}, S.~W. and {Johnston-Hollitt}, M. and {Bartalucci}, I.},
        title = "{Low-frequency integrated radio spectra of diffuse, steep-spectrum sources in galaxy clusters: palaeontology with the MWA and ASKAP}",
      journal = {\pasa},
         year = 2021,
        month = oct,
       volume = {38},
          eid = {e053},
        pages = {e053},
          doi = {10.1017/pasa.2021.45},
archivePrefix = {arXiv},
       eprint = {2106.12281},
 primaryClass = {astro-ph.CO},
       adsurl = {https://ui.adsabs.harvard.edu/abs/2021PASA...38...53D}
}

@ARTICLE{2012MNRAS.422.1812H,
       author = {{Hancock}, P.~J. and {Murphy}, T. and {Gaensler}, B.~M. and {Hopkins}, A. and {Curran}, J.~R.},
        title = "{Compact continuum source finding for next generation radio surveys}",
      journal = {\mnras},
         year = 2012,
        month = may,
       volume = {422},
       number = {2},
        pages = {1812-1824},
          doi = {10.1111/j.1365-2966.2012.20768.x},
archivePrefix = {arXiv},
       eprint = {1202.4500},
 primaryClass = {astro-ph.IM},
       adsurl = {https://ui.adsabs.harvard.edu/abs/2012MNRAS.422.1812H}
}

@ARTICLE{2018PASA...35...11H,
       author = {{Hancock}, Paul J. and {Trott}, Cathryn M. and {Hurley-Walker}, Natasha},
        title = "{Source Finding in the Era of the SKA (Precursors): Aegean 2.0}",
      journal = {\pasa},
         year = 2018,
        month = mar,
       volume = {35},
          eid = {e011},
        pages = {e011},
          doi = {10.1017/pasa.2018.3},
archivePrefix = {arXiv},
       eprint = {1801.05548},
 primaryClass = {astro-ph.IM},
       adsurl = {https://ui.adsabs.harvard.edu/abs/2018PASA...35...11H}
}

@INPROCEEDINGS{2002ASPC..281..228B,
       author = {{Bertin}, Emmanuel and {Mellier}, Yannick and {Radovich}, Mario and {Missonnier}, Gilles and {Didelon}, Pierre and {Morin}, Bertrand},
        title = "{The TERAPIX Pipeline}",
    booktitle = {Astronomical Data Analysis Software and Systems XI},
         year = 2002,
       editor = {{Bohlender}, David A. and {Durand}, Daniel and {Handley}, Thomas H.},
       series = {Astronomical Society of the Pacific Conference Series},
       volume = {281},
        month = jan,
        pages = {228},
       adsurl = {https://ui.adsabs.harvard.edu/abs/2002ASPC..281..228B}
}

@ARTICLE{1993AJ....105.2035D,
       author = {{Diaferio}, Antonaldo and {Ramella}, Massimo and {Geller}, Margaret J. and {Ferrari}, Attilio},
        title = "{Are Groups of Galaxies Virialized Systems?}",
      journal = {\aj},
         year = 1993,
        month = jun,
       volume = {105},
        pages = {2035},
          doi = {10.1086/116581},
       adsurl = {https://ui.adsabs.harvard.edu/abs/1993AJ....105.2035D}
}

@ARTICLE{2009A&A...499..371G,
       author = {{Govoni}, F. and {Murgia}, M. and {Markevitch}, M. and {Feretti}, L. and {Giovannini}, G. and {Taylor}, G.~B. and {Carretti}, E.},
        title = "{A search for diffuse radio emission in the relaxed, cool-core galaxy clusters A1068, A1413, A1650, A1835, A2029, and Ophiuchus}",
      journal = {\aap},
         year = 2009,
        month = may,
       volume = {499},
       number = {2},
        pages = {371-383},
          doi = {10.1051/0004-6361/200811180},
archivePrefix = {arXiv},
       eprint = {0901.1941},
 primaryClass = {astro-ph.CO},
       adsurl = {https://ui.adsabs.harvard.edu/abs/2009A&A...499..371G}
}

@ARTICLE{2003AJ....125.1095S,
       author = {{Subrahmanyan}, Ravi and {Beasley}, A.~J. and {Goss}, W.~M. and {Golap}, K. and {Hunstead}, R.~W.},
        title = "{PKS B1400-33: An Unusual Radio Relic in a Poor Cluster}",
      journal = {\aj},
         year = 2003,
        month = mar,
       volume = {125},
       number = {3},
        pages = {1095-1106},
          doi = {10.1086/367797},
archivePrefix = {arXiv},
       eprint = {astro-ph/0212154},
 primaryClass = {astro-ph},
       adsurl = {https://ui.adsabs.harvard.edu/abs/2003AJ....125.1095S}
}

@ARTICLE{2018MNRAS.477..957D,
       author = {{Dwarakanath}, K.~S. and {Parekh}, V. and {Kale}, R. and {George}, L.~T.},
        title = "{Twin radio relics in the nearby low-mass galaxy cluster Abell 168}",
      journal = {\mnras},
         year = 2018,
        month = jun,
       volume = {477},
       number = {1},
        pages = {957-963},
          doi = {10.1093/mnras/sty744},
archivePrefix = {arXiv},
       eprint = {1803.06129},
 primaryClass = {astro-ph.CO},
       adsurl = {https://ui.adsabs.harvard.edu/abs/2018MNRAS.477..957D}
}

@ARTICLE{2025ApJ...979L..15G,
       author = {{Ghosh}, Subham and {Bhat}, Pallavi},
        title = "{Magnetic Reconnection: An Alternative Explanation of Radio Emission in Galaxy Clusters}",
      journal = {\apjl},
         year = 2025,
        month = jan,
       volume = {979},
       number = {1},
          eid = {L15},
        pages = {L15},
          doi = {10.3847/2041-8213/ad9f2d},
archivePrefix = {arXiv},
       eprint = {2407.11156},
 primaryClass = {astro-ph.CO},
       adsurl = {https://ui.adsabs.harvard.edu/abs/2025ApJ...979L..15G}
}

@ARTICLE{1983RPPh...46..973D,
       author = {{Drury}, L. Oc.},
        title = "{REVIEW ARTICLE: An introduction to the theory of diffusive shock acceleration of energetic particles in tenuous plasmas}",
      journal = {Reports on Progress in Physics},
         year = 1983,
        month = aug,
       volume = {46},
       number = {8},
        pages = {973-1027},
          doi = {10.1088/0034-4885/46/8/002},
       adsurl = {https://ui.adsabs.harvard.edu/abs/1983RPPh...46..973D}
}

@ARTICLE{2007MNRAS.378..245B,
       author = {{Brunetti}, G. and {Lazarian}, A.},
        title = "{Compressible turbulence in galaxy clusters: physics and stochastic particle re-acceleration}",
      journal = {\mnras},
         year = 2007,
        month = jun,
       volume = {378},
       number = {1},
        pages = {245-275},
          doi = {10.1111/j.1365-2966.2007.11771.x},
archivePrefix = {arXiv},
       eprint = {astro-ph/0703591},
 primaryClass = {astro-ph},
       adsurl = {https://ui.adsabs.harvard.edu/abs/2007MNRAS.378..245B}
}

@ARTICLE{2013PASA...30....7T,
       author = {{Tingay}, S.~J. and {Goeke}, R. and {Bowman}, J.~D. and {Emrich}, D. and {Ord}, S.~M. and {Mitchell}, D.~A. and {Morales}, M.~F. and {Booler}, T. and {Crosse}, B. and {Wayth}, R.~B. and {Lonsdale}, C.~J. and {Tremblay}, S. and {Pallot}, D. and {Colegate}, T. and {Wicenec}, A. and {Kudryavtseva}, N. and {Arcus}, W. and {Barnes}, D. and {Bernardi}, G. and {Briggs}, F. and {Burns}, S. and {Bunton}, J.~D. and {Cappallo}, R.~J. and {Corey}, B.~E. and {Deshpande}, A. and {Desouza}, L. and {Gaensler}, B.~M. and {Greenhill}, L.~J. and {Hall}, P.~J. and {Hazelton}, B.~J. and {Herne}, D. and {Hewitt}, J.~N. and {Johnston-Hollitt}, M. and {Kaplan}, D.~L. and {Kasper}, J.~C. and {Kincaid}, B.~B. and {Koenig}, R. and {Kratzenberg}, E. and {Lynch}, M.~J. and {Mckinley}, B. and {Mcwhirter}, S.~R. and {Morgan}, E. and {Oberoi}, D. and {Pathikulangara}, J. and {Prabu}, T. and {Remillard}, R.~A. and {Rogers}, A.~E.~E. and {Roshi}, A. and {Salah}, J.~E. and {Sault}, R.~J. and {Udaya-Shankar}, N. and {Schlagenhaufer}, F. and {Srivani}, K.~S. and {Stevens}, J. and {Subrahmanyan}, R. and {Waterson}, M. and {Webster}, R.~L. and {Whitney}, A.~R. and {Williams}, A. and {Williams}, C.~L. and {Wyithe}, J.~S.~B.},
        title = "{The Murchison Widefield Array: The Square Kilometre Array Precursor at Low Radio Frequencies}",
      journal = {\pasa},
         year = 2013,
        month = jan,
       volume = {30},
          eid = {e007},
        pages = {e007},
          doi = {10.1017/pasa.2012.007},
archivePrefix = {arXiv},
       eprint = {1206.6945},
 primaryClass = {astro-ph.IM},
       adsurl = {https://ui.adsabs.harvard.edu/abs/2013PASA...30....7T}
}

@ARTICLE{2017JAI.....641011R,
       author = {{Reddy}, Suda Harshavardhan and {Kudale}, Sanjay and {Gokhale}, Upendra and {Halagalli}, Irappa and {Raskar}, Nilesh and {de}, Kishalay and {Gnanaraj}, Shelton and {Ajith Kumar}, B. and {Gupta}, Yashwant},
        title = "{A Wideband Digital Back-End for the Upgraded GMRT}",
      journal = {Journal of Astronomical Instrumentation},
         year = 2017,
        month = mar,
       volume = {6},
       number = {1},
          eid = {1641011-336},
        pages = {1641011-336},
          doi = {10.1142/S2251171716410117},
       adsurl = {https://ui.adsabs.harvard.edu/abs/2017JAI.....641011R}
}

@ARTICLE{2013A&A...556A...2V,
       author = {{van Haarlem}, M.~P. and {Wise}, M.~W. and {Gunst}, A.~W. and {Heald}, G. and {McKean}, J.~P. and {Hessels}, J.~W.~T. and {de Bruyn}, A.~G. and {Nijboer}, R. and {Swinbank}, J. and {Fallows}, R. and {Brentjens}, M. and {Nelles}, A. and {Beck}, R. and {Falcke}, H. and {Fender}, R. and {H{\"o}randel}, J. and {Koopmans}, L.~V.~E. and {Mann}, G. and {Miley}, G. and {R{\"o}ttgering}, H. and {Stappers}, B.~W. and {Wijers}, R.~A.~M.~J. and {Zaroubi}, S. and {van den Akker}, M. and {Alexov}, A. and {Anderson}, J. and {Anderson}, K. and {van Ardenne}, A. and {Arts}, M. and {Asgekar}, A. and {Avruch}, I.~M. and {Batejat}, F. and {B{\"a}hren}, L. and {Bell}, M.~E. and {Bell}, M.~R. and {van Bemmel}, I. and {Bennema}, P. and {Bentum}, M.~J. and {Bernardi}, G. and {Best}, P. and {B{\^\i}rzan}, L. and {Bonafede}, A. and {Boonstra}, A. -J. and {Braun}, R. and {Bregman}, J. and {Breitling}, F. and {van de Brink}, R.~H. and {Broderick}, J. and {Broekema}, P.~C. and {Brouw}, W.~N. and {Br{\"u}ggen}, M. and {Butcher}, H.~R. and {van Cappellen}, W. and {Ciardi}, B. and {Coenen}, T. and {Conway}, J. and {Coolen}, A. and {Corstanje}, A. and {Damstra}, S. and {Davies}, O. and {Deller}, A.~T. and {Dettmar}, R. -J. and {van Diepen}, G. and {Dijkstra}, K. and {Donker}, P. and {Doorduin}, A. and {Dromer}, J. and {Drost}, M. and {van Duin}, A. and {Eisl{\"o}ffel}, J. and {van Enst}, J. and {Ferrari}, C. and {Frieswijk}, W. and {Gankema}, H. and {Garrett}, M.~A. and {de Gasperin}, F. and {Gerbers}, M. and {de Geus}, E. and {Grie{\ss}meier}, J. -M. and {Grit}, T. and {Gruppen}, P. and {Hamaker}, J.~P. and {Hassall}, T. and {Hoeft}, M. and {Holties}, H.~A. and {Horneffer}, A. and {van der Horst}, A. and {van Houwelingen}, A. and {Huijgen}, A. and {Iacobelli}, M. and {Intema}, H. and {Jackson}, N. and {Jelic}, V. and {de Jong}, A. and {Juette}, E. and {Kant}, D. and {Karastergiou}, A. and {Koers}, A. and {Kollen}, H. and {Kondratiev}, V.~I. and {Kooistra}, E. and {Koopman}, Y. and {Koster}, A. and {Kuniyoshi}, M. and {Kramer}, M. and {Kuper}, G. and {Lambropoulos}, P. and {Law}, C. and {van Leeuwen}, J. and {Lemaitre}, J. and {Loose}, M. and {Maat}, P. and {Macario}, G. and {Markoff}, S. and {Masters}, J. and {McFadden}, R.~A. and {McKay-Bukowski}, D. and {Meijering}, H. and {Meulman}, H. and {Mevius}, M. and {Middelberg}, E. and {Millenaar}, R. and {Miller-Jones}, J.~C.~A. and {Mohan}, R.~N. and {Mol}, J.~D. and {Morawietz}, J. and {Morganti}, R. and {Mulcahy}, D.~D. and {Mulder}, E. and {Munk}, H. and {Nieuwenhuis}, L. and {van Nieuwpoort}, R. and {Noordam}, J.~E. and {Norden}, M. and {Noutsos}, A. and {Offringa}, A.~R. and {Olofsson}, H. and {Omar}, A. and {Orr{\'u}}, E. and {Overeem}, R. and {Paas}, H. and {Pandey-Pommier}, M. and {Pandey}, V.~N. and {Pizzo}, R. and {Polatidis}, A. and {Rafferty}, D. and {Rawlings}, S. and {Reich}, W. and {de Reijer}, J. -P. and {Reitsma}, J. and {Renting}, G.~A. and {Riemers}, P. and {Rol}, E. and {Romein}, J.~W. and {Roosjen}, J. and {Ruiter}, M. and {Scaife}, A. and {van der Schaaf}, K. and {Scheers}, B. and {Schellart}, P. and {Schoenmakers}, A. and {Schoonderbeek}, G. and {Serylak}, M. and {Shulevski}, A. and {Sluman}, J. and {Smirnov}, O. and {Sobey}, C. and {Spreeuw}, H. and {Steinmetz}, M. and {Sterks}, C.~G.~M. and {Stiepel}, H. -J. and {Stuurwold}, K. and {Tagger}, M. and {Tang}, Y. and {Tasse}, C. and {Thomas}, I. and {Thoudam}, S. and {Toribio}, M.~C. and {van der Tol}, B. and {Usov}, O. and {van Veelen}, M. and {van der Veen}, A. -J. and {ter Veen}, S. and {Verbiest}, J.~P.~W. and {Vermeulen}, R. and {Vermaas}, N. and {Vocks}, C. and {Vogt}, C. and {de Vos}, M. and {van der Wal}, E. and {van Weeren}, R. and {Weggemans}, H. and {Weltevrede}, P. and {White}, S. and {Wijnholds}, S.~J. and {Wilhelmsson}, T. and {Wucknitz}, O. and {Yatawatta}, S. and {Zarka}, P. and {Zensus}, A.},
        title = "{LOFAR: The LOw-Frequency ARray}",
      journal = {\aap},
         year = 2013,
        month = aug,
       volume = {556},
          eid = {A2},
        pages = {A2},
          doi = {10.1051/0004-6361/201220873},
archivePrefix = {arXiv},
       eprint = {1305.3550},
 primaryClass = {astro-ph.IM},
       adsurl = {https://ui.adsabs.harvard.edu/abs/2013A&A...556A...2V}
}

@INPROCEEDINGS{2016mks..confE...1J,
       author = {{Jonas}, J. and {MeerKAT Team}},
        title = "{The MeerKAT Radio Telescope}",
    booktitle = {MeerKAT Science: On the Pathway to the SKA},
         year = 2016,
        month = jan,
          eid = {1},
        pages = {1},
          doi = {10.22323/1.277.0001},
       adsurl = {https://ui.adsabs.harvard.edu/abs/2016mks..confE...1J}
}

@ARTICLE{2014MNRAS.444.3130D,
       author = {{de Gasperin}, F. and {van Weeren}, R.~J. and {Br{\"u}ggen}, M. and {Vazza}, F. and {Bonafede}, A. and {Intema}, H.~T.},
        title = "{A new double radio relic in PSZ1 G096.89+24.17 and a radio relic mass-luminosity relation}",
      journal = {\mnras},
         year = 2014,
        month = nov,
       volume = {444},
       number = {4},
        pages = {3130-3138},
          doi = {10.1093/mnras/stu1658},
archivePrefix = {arXiv},
       eprint = {1408.2677},
 primaryClass = {astro-ph.CO},
       adsurl = {https://ui.adsabs.harvard.edu/abs/2014MNRAS.444.3130D}
}

@ARTICLE{2011ApJ...735...96S,
       author = {{Skillman}, Samuel W. and {Hallman}, Eric J. and {O'Shea}, Brian W. and {Burns}, Jack O. and {Smith}, Britton D. and {Turk}, Matthew J.},
        title = "{Galaxy Cluster Radio Relics in Adaptive Mesh Refinement Cosmological Simulations: Relic Properties and Scaling Relationships}",
      journal = {\apj},
         year = 2011,
        month = jul,
       volume = {735},
       number = {2},
          eid = {96},
        pages = {96},
          doi = {10.1088/0004-637X/735/2/96},
archivePrefix = {arXiv},
       eprint = {1006.3559},
 primaryClass = {astro-ph.CO},
       adsurl = {https://ui.adsabs.harvard.edu/abs/2011ApJ...735...96S}
}

@ARTICLE{2024A&A...690A.146N,
       author = {{Nuza}, S.~E. and {Hoeft}, M. and {Contreras-Santos}, A. and {Knebe}, A. and {Yepes}, G.},
        title = "{The Three Hundred project: Radio luminosity evolution from merger-induced shock fronts in simulated galaxy clusters}",
      journal = {\aap},
         year = 2024,
        month = oct,
       volume = {690},
          eid = {A146},
        pages = {A146},
          doi = {10.1051/0004-6361/202450120},
archivePrefix = {arXiv},
       eprint = {2409.09422},
 primaryClass = {astro-ph.CO},
       adsurl = {https://ui.adsabs.harvard.edu/abs/2024A&A...690A.146N}
}

@ARTICLE{2013MNRAS.435..149N,
       author = {{Nikiel-Wroczy{\'n}ski}, B. and {Soida}, M. and {Urbanik}, M. and {Beck}, R. and {Bomans}, D.~J.},
        title = "{Intergalactic magnetic fields in Stephan's Quintet}",
      journal = {\mnras},
         year = 2013,
        month = oct,
       volume = {435},
       number = {1},
        pages = {149-157},
          doi = {10.1093/mnras/stt1263},
archivePrefix = {arXiv},
       eprint = {1307.3447},
 primaryClass = {astro-ph.CO},
       adsurl = {https://ui.adsabs.harvard.edu/abs/2013MNRAS.435..149N}
}

@ARTICLE{1985ApJ...296...60M,
       author = {{Menon}, T.~K. and {Hickson}, P.},
        title = "{Radio sources in dense groups.}",
      journal = {\apj},
         year = 1985,
        month = sep,
       volume = {296},
        pages = {60-64},
          doi = {10.1086/163419},
       adsurl = {https://ui.adsabs.harvard.edu/abs/1985ApJ...296...60M}
}

@ARTICLE{2011ApJ...732...95G,
       author = {{Giacintucci}, Simona and {O'Sullivan}, Ewan and {Vrtilek}, Jan and {David}, Laurence P. and {Raychaudhury}, Somak and {Venturi}, Tiziana and {Athreya}, Ramana M. and {Clarke}, Tracy E. and {Murgia}, Matteo and {Mazzotta}, Pasquale and {Gitti}, Myriam and {Ponman}, Trevor and {Ishwara-Chandra}, C.~H. and {Jones}, Christine and {Forman}, William R.},
        title = "{A Combined Low-radio Frequency/X-ray Study of Galaxy Groups. I. Giant Metrewave Radio Telescope Observations at 235 MHz AND 610 MHz}",
      journal = {\apj},
         year = 2011,
        month = may,
       volume = {732},
       number = {2},
          eid = {95},
        pages = {95},
          doi = {10.1088/0004-637X/732/2/95},
archivePrefix = {arXiv},
       eprint = {1103.1364},
 primaryClass = {astro-ph.CO},
       adsurl = {https://ui.adsabs.harvard.edu/abs/2011ApJ...732...95G}
}

@ARTICLE{2019A&A...622A..23N,
       author = {{Nikiel-Wroczy{\'n}ski}, B. and {Berger}, A. and {Herrera Ruiz}, N. and {Bomans}, D.~J. and {Blex}, S. and {Horellou}, C. and {Paladino}, R. and {Becker}, A. and {Miskolczi}, A. and {Beck}, R. and {Chy{\.z}y}, K. and {Dettmar}, R. -J. and {Heald}, G. and {Heesen}, V. and {Jamrozy}, M. and {Shimwell}, T.~W. and {Tasse}, C.},
        title = "{Exploring the properties of low-frequency radio emission and magnetic fields in a sample of compact galaxy groups using the LOFAR Two-Metre Sky Survey (LoTSS)}",
      journal = {\aap},
         year = 2019,
        month = feb,
       volume = {622},
          eid = {A23},
        pages = {A23},
          doi = {10.1051/0004-6361/201833934},
archivePrefix = {arXiv},
       eprint = {1810.08708},
 primaryClass = {astro-ph.GA},
       adsurl = {https://ui.adsabs.harvard.edu/abs/2019A&A...622A..23N}
}

@ARTICLE{2025A&A...696A..58V,
       author = {{Vazza}, F. and {Gheller}, C. and {Zanetti}, F. and {Tsizh}, M. and {Carretti}, E. and {Mtchedlidze}, S. and {Br{\"u}ggen}, M.},
        title = "{The evolution of cosmic ray electrons in the cosmic web: Seeding by active galactic nuclei, star formation, and shocks}",
      journal = {\aap},
         year = 2025,
        month = apr,
       volume = {696},
          eid = {A58},
        pages = {A58},
          doi = {10.1051/0004-6361/202451709},
archivePrefix = {arXiv},
       eprint = {2501.19041},
 primaryClass = {astro-ph.HE},
       adsurl = {https://ui.adsabs.harvard.edu/abs/2025A&A...696A..58V}
}

@ARTICLE{2022A&A...661A...3S,
       author = {{Salvato}, M. and {Wolf}, J. and {Dwelly}, T. and {Georgakakis}, A. and {Brusa}, M. and {Merloni}, A. and {Liu}, T. and {Toba}, Y. and {Nandra}, K. and {Lamer}, G. and {Buchner}, J. and {Schneider}, C. and {Freund}, S. and {Rau}, A. and {Schwope}, A. and {Nishizawa}, A. and {Klein}, M. and {Arcodia}, R. and {Comparat}, J. and {Musiimenta}, B. and {Nagao}, T. and {Brunner}, H. and {Malyali}, A. and {Finoguenov}, A. and {Anderson}, S. and {Shen}, Y. and {Ibarra-Medel}, H. and {Trump}, J. and {Brandt}, W.~N. and {Urry}, C.~M. and {Rivera}, C. and {Krumpe}, M. and {Urrutia}, T. and {Miyaji}, T. and {Ichikawa}, K. and {Schneider}, D.~P. and {Fresco}, A. and {Boller}, T. and {Haase}, J. and {Brownstein}, J. and {Lane}, R.~R. and {Bizyaev}, D. and {Nitschelm}, C.},
        title = "{The eROSITA Final Equatorial-Depth Survey (eFEDS). Identification and characterization of the counterparts to point-like sources}",
      journal = {\aap},
         year = 2022,
        month = may,
       volume = {661},
          eid = {A3},
        pages = {A3},
          doi = {10.1051/0004-6361/202141631},
archivePrefix = {arXiv},
       eprint = {2106.14520},
 primaryClass = {astro-ph.HE},
       adsurl = {https://ui.adsabs.harvard.edu/abs/2022A&A...661A...3S}
}

@ARTICLE{2024A&A...682A..34M,
       author = {{Merloni}, A. and {Lamer}, G. and {Liu}, T. and {Ramos-Ceja}, M.~E. and {Brunner}, H. and {Bulbul}, E. and {Dennerl}, K. and {Doroshenko}, V. and {Freyberg}, M.~J. and {Friedrich}, S. and {Gatuzz}, E. and {Georgakakis}, A. and {Haberl}, F. and {Igo}, Z. and {Kreykenbohm}, I. and {Liu}, A. and {Maitra}, C. and {Malyali}, A. and {Mayer}, M.~G.~F. and {Nandra}, K. and {Predehl}, P. and {Robrade}, J. and {Salvato}, M. and {Sanders}, J.~S. and {Stewart}, I. and {Tub{\'\i}n-Arenas}, D. and {Weber}, P. and {Wilms}, J. and {Arcodia}, R. and {Artis}, E. and {Aschersleben}, J. and {Avakyan}, A. and {Aydar}, C. and {Bahar}, Y.~E. and {Balzer}, F. and {Becker}, W. and {Berger}, K. and {Boller}, T. and {Bornemann}, W. and {Br{\"u}ggen}, M. and {Brusa}, M. and {Buchner}, J. and {Burwitz}, V. and {Camilloni}, F. and {Clerc}, N. and {Comparat}, J. and {Coutinho}, D. and {Czesla}, S. and {Dannhauer}, S.~M. and {Dauner}, L. and {Dauser}, T. and {Dietl}, J. and {Dolag}, K. and {Dwelly}, T. and {Egg}, K. and {Ehl}, E. and {Freund}, S. and {Friedrich}, P. and {Gaida}, R. and {Garrel}, C. and {Ghirardini}, V. and {Gokus}, A. and {Gr{\"u}nwald}, G. and {Grandis}, S. and {Grotova}, I. and {Gruen}, D. and {Gueguen}, A. and {H{\"a}mmerich}, S. and {Hamaus}, N. and {Hasinger}, G. and {Haubner}, K. and {Homan}, D. and {Ider Chitham}, J. and {Joseph}, W.~M. and {Joyce}, A. and {K{\"o}nig}, O. and {Kaltenbrunner}, D.~M. and {Khokhriakova}, A. and {Kink}, W. and {Kirsch}, C. and {Kluge}, M. and {Knies}, J. and {Krippendorf}, S. and {Krumpe}, M. and {Kurpas}, J. and {Li}, P. and {Liu}, Z. and {Locatelli}, N. and {Lorenz}, M. and {M{\"u}ller}, S. and {Magaudda}, E. and {Mannes}, C. and {McCall}, H. and {Meidinger}, N. and {Michailidis}, M. and {Migkas}, K. and {Mu{\~n}oz-Giraldo}, D. and {Musiimenta}, B. and {Nguyen-Dang}, N.~T. and {Ni}, Q. and {Olechowska}, A. and {Ota}, N. and {Pacaud}, F. and {Pasini}, T. and {Perinati}, E. and {Pires}, A.~M. and {Pommranz}, C. and {Ponti}, G. and {Poppenhaeger}, K. and {P{\"u}hlhofer}, G. and {Rau}, A. and {Reh}, M. and {Reiprich}, T.~H. and {Roster}, W. and {Saeedi}, S. and {Santangelo}, A. and {Sasaki}, M. and {Schmitt}, J. and {Schneider}, P.~C. and {Schrabback}, T. and {Schuster}, N. and {Schwope}, A. and {Seppi}, R. and {Serim}, M.~M. and {Shreeram}, S. and {Sokolova-Lapa}, E. and {Starck}, H. and {Stelzer}, B. and {Stierhof}, J. and {Suleimanov}, V. and {Tenzer}, C. and {Traulsen}, I. and {Tr{\"u}mper}, J. and {Tsuge}, K. and {Urrutia}, T. and {Veronica}, A. and {Waddell}, S.~G.~H. and {Willer}, R. and {Wolf}, J. and {Yeung}, M.~C.~H. and {Zainab}, A. and {Zangrandi}, F. and {Zhang}, X. and {Zhang}, Y. and {Zheng}, X.},
        title = "{The SRG/eROSITA all-sky survey. First X-ray catalogues and data release of the western Galactic hemisphere}",
      journal = {\aap},
         year = 2024,
        month = feb,
       volume = {682},
          eid = {A34},
        pages = {A34},
          doi = {10.1051/0004-6361/202347165},
archivePrefix = {arXiv},
       eprint = {2401.17274},
 primaryClass = {astro-ph.HE},
       adsurl = {https://ui.adsabs.harvard.edu/abs/2024A&A...682A..34M}
}

@ARTICLE{2019MNRAS.489.2488K,
       author = {{Kolokythas}, Konstantinos and {O'Sullivan}, Ewan and {Intema}, Huib and {Raychaudhury}, Somak and {Babul}, Arif and {Giacintucci}, Simona and {Gitti}, Myriam},
        title = "{The complete local volume groups sample - III. Characteristics of group central radio galaxies in the Local Universe}",
      journal = {\mnras},
         year = 2019,
        month = oct,
       volume = {489},
       number = {2},
        pages = {2488-2504},
          doi = {10.1093/mnras/stz2082},
archivePrefix = {arXiv},
       eprint = {1907.10768},
 primaryClass = {astro-ph.GA},
       adsurl = {https://ui.adsabs.harvard.edu/abs/2019MNRAS.489.2488K}
}

@ARTICLE{2021A&A...648A..11O,
       author = {{Osinga}, E. and {van Weeren}, R.~J. and {Boxelaar}, J.~M. and {Brunetti}, G. and {Botteon}, A. and {Br{\"u}ggen}, M. and {Shimwell}, T.~W. and {Bonafede}, A. and {Best}, P.~N. and {Bonato}, M. and {Cassano}, R. and {Gastaldello}, F. and {di Gennaro}, G. and {Hardcastle}, M.~J. and {Mandal}, S. and {Rossetti}, M. and {R{\"o}ttgering}, H.~J.~A. and {Sabater}, J. and {Tasse}, C.},
        title = "{Diffuse radio emission from galaxy clusters in the LOFAR Two-metre Sky Survey Deep Fields}",
      journal = {\aap},
         year = 2021,
        month = apr,
       volume = {648},
          eid = {A11},
        pages = {A11},
          doi = {10.1051/0004-6361/202039076},
archivePrefix = {arXiv},
       eprint = {2011.08249},
 primaryClass = {astro-ph.HE},
       adsurl = {https://ui.adsabs.harvard.edu/abs/2021A&A...648A..11O}
}

@ARTICLE{2022A&A...661A..92B,
       author = {{Brienza}, M. and {Lovisari}, L. and {Rajpurohit}, K. and {Bonafede}, A. and {Gastaldello}, F. and {Murgia}, M. and {Vazza}, F. and {Bonnassieux}, E. and {Botteon}, A. and {Brunetti}, G. and {Drabent}, A. and {Hardcastle}, M.~J. and {Pasini}, T. and {Riseley}, C.~J. and {R{\"o}ttgering}, H.~J.~A. and {Shimwell}, T. and {Simionescu}, A. and {van Weeren}, R.~J.},
        title = "{The galaxy group NGC 507: Newly detected AGN remnant plasma transported by sloshing}",
      journal = {\aap},
         year = 2022,
        month = may,
       volume = {661},
          eid = {A92},
        pages = {A92},
          doi = {10.1051/0004-6361/202142579},
archivePrefix = {arXiv},
       eprint = {2201.04591},
 primaryClass = {astro-ph.GA},
       adsurl = {https://ui.adsabs.harvard.edu/abs/2022A&A...661A..92B}
}

@ARTICLE{2015ApJ...813...77Y,
       author = {{Yuan}, Z.~S. and {Han}, J.~L. and {Wen}, Z.~L.},
        title = "{The Scaling Relations and the Fundamental Plane for Radio Halos and Relics of Galaxy Clusters}",
      journal = {\apj},
         year = 2015,
        month = nov,
       volume = {813},
       number = {1},
          eid = {77},
        pages = {77},
          doi = {10.1088/0004-637X/813/1/77},
archivePrefix = {arXiv},
       eprint = {1510.04980},
 primaryClass = {astro-ph.CO},
       adsurl = {https://ui.adsabs.harvard.edu/abs/2015ApJ...813...77Y}
}

@ARTICLE{2021A&A...647A..51C,
       author = {{Cuciti}, V. and {Cassano}, R. and {Brunetti}, G. and {Dallacasa}, D. and {de Gasperin}, F. and {Ettori}, S. and {Giacintucci}, S. and {Kale}, R. and {Pratt}, G.~W. and {van Weeren}, R.~J. and {Venturi}, T.},
        title = "{Radio halos in a mass-selected sample of 75 galaxy clusters. II. Statistical analysis}",
      journal = {\aap},
         year = 2021,
        month = mar,
       volume = {647},
          eid = {A51},
        pages = {A51},
          doi = {10.1051/0004-6361/202039208},
archivePrefix = {arXiv},
       eprint = {2101.01641},
 primaryClass = {astro-ph.CO},
       adsurl = {https://ui.adsabs.harvard.edu/abs/2021A&A...647A..51C}
}

@ARTICLE{2019MNRAS.483.5444D,
       author = {{Davies}, L.~J.~M. and {Robotham}, A.~S.~G. and {Lagos}, C. del P. and {Driver}, S.~P. and {Stevens}, A.~R.~H. and {Bah{\'e}}, Y.~M. and {Alpaslan}, M. and {Bremer}, M.~N. and {Brown}, M.~J.~I. and {Brough}, S. and {Bland-Hawthorn}, J. and {Cortese}, L. and {Elahi}, P. and {Grootes}, M.~W. and {Holwerda}, B.~W. and {Ludlow}, A.~D. and {McGee}, S. and {Owers}, M. and {Phillipps}, S.},
        title = "{Galaxy and Mass Assembly (GAMA): environmental quenching of centrals and satellites in groups}",
      journal = {\mnras},
         year = 2019,
        month = mar,
       volume = {483},
       number = {4},
        pages = {5444-5458},
          doi = {10.1093/mnras/sty3393},
archivePrefix = {arXiv},
       eprint = {1901.01640},
 primaryClass = {astro-ph.GA},
       adsurl = {https://ui.adsabs.harvard.edu/abs/2019MNRAS.483.5444D}
}

@ARTICLE{2022MNRAS.513..439D,
       author = {{Driver}, Simon P. and {Bellstedt}, Sabine and {Robotham}, Aaron S.~G. and {Baldry}, Ivan K. and {Davies}, Luke J. and {Liske}, Jochen and {Obreschkow}, Danail and {Taylor}, Edward N. and {Wright}, Angus H. and {Alpaslan}, Mehmet and {Bamford}, Steven P. and {Bauer}, Amanda E. and {Bland-Hawthorn}, Joss and {Bilicki}, Maciej and {Bravo}, Mat{\'\i}as and {Brough}, Sarah and {Casura}, Sarah and {Cluver}, Michelle E. and {Colless}, Matthew and {Conselice}, Christopher J. and {Croom}, Scott M. and {de Jong}, Jelte and {D'Eugenio}, Franceso and {De Propris}, Roberto and {Dogruel}, Burak and {Drinkwater}, Michael J. and {Dvornik}, Andrej and {Farrow}, Daniel J. and {Frenk}, Carlos S. and {Giblin}, Benjamin and {Graham}, Alister W. and {Grootes}, Meiert W. and {Gunawardhana}, Madusha L.~P. and {Hashemizadeh}, Abdolhosein and {H{\"a}u{\ss}ler}, Boris and {Heymans}, Catherine and {Hildebrandt}, Hendrik and {Holwerda}, Benne W. and {Hopkins}, Andrew M. and {Jarrett}, Tom H. and {Heath Jones}, D. and {Kelvin}, Lee S. and {Koushan}, Soheil and {Kuijken}, Konrad and {Lara-L{\'o}pez}, Maritza A. and {Lange}, Rebecca and {L{\'o}pez-S{\'a}nchez}, {\'A}ngel R. and {Loveday}, Jon and {Mahajan}, Smriti and {Meyer}, Martin and {Moffett}, Amanda J. and {Napolitano}, Nicola R. and {Norberg}, Peder and {Owers}, Matt S. and {Radovich}, Mario and {Raouf}, Mojtaba and {Peacock}, John A. and {Phillipps}, Steven and {Pimbblet}, Kevin A. and {Popescu}, Cristina and {Said}, Khaled and {Sansom}, Anne E. and {Seibert}, Mark and {Sutherland}, Will J. and {Thorne}, Jessica E. and {Tuffs}, Richard J. and {Turner}, Ryan and {van der Wel}, Arjen and {van Kampen}, Eelco and {Wilkins}, Steve M.},
        title = "{Galaxy And Mass Assembly (GAMA): Data Release 4 and the z < 0.1 total and z < 0.08 morphological galaxy stellar mass functions}",
      journal = {\mnras},
         year = 2022,
        month = jun,
       volume = {513},
       number = {1},
        pages = {439-467},
          doi = {10.1093/mnras/stac472},
archivePrefix = {arXiv},
       eprint = {2203.08539},
 primaryClass = {astro-ph.GA},
       adsurl = {https://ui.adsabs.harvard.edu/abs/2022MNRAS.513..439D}
}

@ARTICLE{2002PASP..114..427R,
       author = {{Rudnick}, L.},
        title = "{Simple Multiresolution Filtering and the Spectra of Radio Galaxies and Supernova Remnants}",
      journal = {\pasp},
         year = 2002,
        month = apr,
       volume = {114},
       number = {794},
        pages = {427-449},
          doi = {10.1086/342499},
       adsurl = {https://ui.adsabs.harvard.edu/abs/2002PASP..114..427R}
}

@ARTICLE{2020MNRAS.496.3235B,
       author = {{Bellstedt}, Sabine and {Driver}, Simon P. and {Robotham}, Aaron S.~G. and {Davies}, Luke J.~M. and {Bogue}, Kamran R.~J. and {Cook}, Robin H.~W. and {Hashemizadeh}, Abdolhosein and {Koushan}, Soheil and {Taylor}, Edward N. and {Thorne}, Jessica E. and {Turner}, Ryan J. and {Wright}, Angus H.},
        title = "{Galaxy And Mass Assembly (GAMA): assimilation of KiDS into the GAMA database}",
      journal = {\mnras},
         year = 2020,
        month = aug,
       volume = {496},
       number = {3},
        pages = {3235-3256},
          doi = {10.1093/mnras/staa1466},
archivePrefix = {arXiv},
       eprint = {2005.11215},
 primaryClass = {astro-ph.GA},
       adsurl = {https://ui.adsabs.harvard.edu/abs/2020MNRAS.496.3235B}
}

@ARTICLE{2025MNRAS.541.3220D,
       author = {{Davies}, L.~J.~M. and {Fuentealba-Fuentes}, M.~F. and {Wright}, R.~J. and {Bravo}, M. and {Wagh}, S. and {Siudek}, M.},
        title = "{Deep Extragalactic VIsible Legacy Survey (DEVILS): satellite quenching at intermediate redshift}",
      journal = {\mnras},
         year = 2025,
        month = aug,
       volume = {541},
       number = {4},
        pages = {3220-3235},
          doi = {10.1093/mnras/staf1205},
archivePrefix = {arXiv},
       eprint = {2507.20822},
 primaryClass = {astro-ph.GA},
       adsurl = {https://ui.adsabs.harvard.edu/abs/2025MNRAS.541.3220D}
}

@INPROCEEDINGS{2017ASPC..512..431W,
       author = {{Whiting}, M. and {Voronkov}, M. and {Mitchell}, D. and {Askap Team}},
        title = "{Early Science Pipelines for ASKAP}",
    booktitle = {Astronomical Data Analysis Software and Systems XXV},
         year = 2017,
       editor = {{Lorente}, N.~P.~F. and {Shortridge}, K. and {Wayth}, R.},
       series = {Astronomical Society of the Pacific Conference Series},
       volume = {512},
        month = dec,
        pages = {431},
       adsurl = {https://ui.adsabs.harvard.edu/abs/2017ASPC..512..431W}
}

@ARTICLE{2007MNRAS.380.1467G,
       author = {{Gilmour}, R. and {Gray}, M.~E. and {Almaini}, O. and {Best}, P. and {Wolf}, C. and {Meisenheimer}, K. and {Papovich}, C. and {Bell}, E.},
        title = "{Environmental dependence of active galactic nuclei activity in the supercluster A901/2}",
      journal = {\mnras},
         year = 2007,
        month = oct,
       volume = {380},
       number = {4},
        pages = {1467-1487},
          doi = {10.1111/j.1365-2966.2007.12127.x},
archivePrefix = {arXiv},
       eprint = {0707.1517},
 primaryClass = {astro-ph},
       adsurl = {https://ui.adsabs.harvard.edu/abs/2007MNRAS.380.1467G}
}

@ARTICLE{2016ApJ...818..182V,
       author = {{Vaddi}, Sravani and {O'Dea}, Christopher P. and {Baum}, Stefi A. and {Whitmore}, Samantha and {Ahmed}, Rabeea and {Pierce}, Katherine and {Leary}, Sara},
        title = "{Constraints on Feedback in the Local Universe: The Relation between Star Formation and AGN Activity in Early-type Galaxies}",
      journal = {\apj},
         year = 2016,
        month = feb,
       volume = {818},
       number = {2},
          eid = {182},
        pages = {182},
          doi = {10.3847/0004-637X/818/2/182},
archivePrefix = {arXiv},
       eprint = {1601.00344},
 primaryClass = {astro-ph.GA},
       adsurl = {https://ui.adsabs.harvard.edu/abs/2016ApJ...818..182V}
}

@ARTICLE{2021Univ....7..139L,
       author = {{Lovisari}, Lorenzo and {Ettori}, Stefano and {Gaspari}, Massimo and {Giles}, Paul A.},
        title = "{Scaling Properties of Galaxy Groups}",
      journal = {Universe},
         year = 2021,
        month = may,
       volume = {7},
       number = {5},
          eid = {139},
        pages = {139},
          doi = {10.3390/universe7050139},
archivePrefix = {arXiv},
       eprint = {2106.13256},
 primaryClass = {astro-ph.CO},
       adsurl = {https://ui.adsabs.harvard.edu/abs/2021Univ....7..139L}
}

@ARTICLE{2015MNRAS.452.2087L,
       author = {{Liske}, J. and {Baldry}, I.~K. and {Driver}, S.~P. and {Tuffs}, R.~J. and {Alpaslan}, M. and {Andrae}, E. and {Brough}, S. and {Cluver}, M.~E. and {Grootes}, M.~W. and {Gunawardhana}, M.~L.~P. and {Kelvin}, L.~S. and {Loveday}, J. and {Robotham}, A.~S.~G. and {Taylor}, E.~N. and {Bamford}, S.~P. and {Bland-Hawthorn}, J. and {Brown}, M.~J.~I. and {Drinkwater}, M.~J. and {Hopkins}, A.~M. and {Meyer}, M.~J. and {Norberg}, P. and {Peacock}, J.~A. and {Agius}, N.~K. and {Andrews}, S.~K. and {Bauer}, A.~E. and {Ching}, J.~H.~Y. and {Colless}, M. and {Conselice}, C.~J. and {Croom}, S.~M. and {Davies}, L.~J.~M. and {De Propris}, R. and {Dunne}, L. and {Eardley}, E.~M. and {Ellis}, S. and {Foster}, C. and {Frenk}, C.~S. and {H{\"a}u{\ss}ler}, B. and {Holwerda}, B.~W. and {Howlett}, C. and {Ibarra}, H. and {Jarvis}, M.~J. and {Jones}, D.~H. and {Kafle}, P.~R. and {Lacey}, C.~G. and {Lange}, R. and {Lara-L{\'o}pez}, M.~A. and {L{\'o}pez-S{\'a}nchez}, {\'A}. R. and {Maddox}, S. and {Madore}, B.~F. and {McNaught-Roberts}, T. and {Moffett}, A.~J. and {Nichol}, R.~C. and {Owers}, M.~S. and {Palamara}, D. and {Penny}, S.~J. and {Phillipps}, S. and {Pimbblet}, K.~A. and {Popescu}, C.~C. and {Prescott}, M. and {Proctor}, R. and {Sadler}, E.~M. and {Sansom}, A.~E. and {Seibert}, M. and {Sharp}, R. and {Sutherland}, W. and {V{\'a}zquez-Mata}, J.~A. and {van Kampen}, E. and {Wilkins}, S.~M. and {Williams}, R. and {Wright}, A.~H.},
        title = "{Galaxy And Mass Assembly (GAMA): end of survey report and data release 2}",
      journal = {\mnras},
         year = 2015,
        month = sep,
       volume = {452},
       number = {2},
        pages = {2087-2126},
          doi = {10.1093/mnras/stv1436},
archivePrefix = {arXiv},
       eprint = {1506.08222},
 primaryClass = {astro-ph.GA},
       adsurl = {https://ui.adsabs.harvard.edu/abs/2015MNRAS.452.2087L}
}

@ARTICLE{2025arXiv251117307R,
       author = {{Rhee}, Jonghwan and {Dodson}, Richard and {Williamson}, Alexander and {Meyer}, Martin and {Rozgony}, Krist{\'o}f and {Elahi}, Pascal J. and {Whiting}, Matthew and {Mitchell}, Daniel and {Westmeier}, Tobias},
        title = "{Deep Investigation of Neutral Gas Origins (DINGO): Options for robust Deep Spectral Line Imaging in the SKA-Era}",
      journal = {arXiv e-prints},
         year = 2025,
        month = nov,
          eid = {arXiv:2511.17307},
        pages = {arXiv:2511.17307},
          doi = {10.48550/arXiv.2511.17307},
archivePrefix = {arXiv},
       eprint = {2511.17307},
 primaryClass = {astro-ph.IM},
       adsurl = {https://ui.adsabs.harvard.edu/abs/2025arXiv251117307R}
}

@ARTICLE{2013MNRAS.430.2047H,
       author = {{Hopkins}, A.~M. and {Driver}, S.~P. and {Brough}, S. and {Owers}, M.~S. and {Bauer}, A.~E. and {Gunawardhana}, M.~L.~P. and {Cluver}, M.~E. and {Colless}, M. and {Foster}, C. and {Lara-L{\'o}pez}, M.~A. and {Roseboom}, I. and {Sharp}, R. and {Steele}, O. and {Thomas}, D. and {Baldry}, I.~K. and {Brown}, M.~J.~I. and {Liske}, J. and {Norberg}, P. and {Robotham}, A.~S.~G. and {Bamford}, S. and {Bland-Hawthorn}, J. and {Drinkwater}, M.~J. and {Loveday}, J. and {Meyer}, M. and {Peacock}, J.~A. and {Tuffs}, R. and {Agius}, N. and {Alpaslan}, M. and {Andrae}, E. and {Cameron}, E. and {Cole}, S. and {Ching}, J.~H.~Y. and {Christodoulou}, L. and {Conselice}, C. and {Croom}, S. and {Cross}, N.~J.~G. and {De Propris}, R. and {Delhaize}, J. and {Dunne}, L. and {Eales}, S. and {Ellis}, S. and {Frenk}, C.~S. and {Graham}, Alister W. and {Grootes}, M.~W. and {H{\"a}u{\ss}ler}, B. and {Heymans}, C. and {Hill}, D. and {Hoyle}, B. and {Hudson}, M. and {Jarvis}, M. and {Johansson}, J. and {Jones}, D.~H. and {van Kampen}, E. and {Kelvin}, L. and {Kuijken}, K. and {L{\'o}pez-S{\'a}nchez}, {\'A}. and {Maddox}, S. and {Madore}, B. and {Maraston}, C. and {McNaught-Roberts}, T. and {Nichol}, R.~C. and {Oliver}, S. and {Parkinson}, H. and {Penny}, S. and {Phillipps}, S. and {Pimbblet}, K.~A. and {Ponman}, T. and {Popescu}, C.~C. and {Prescott}, M. and {Proctor}, R. and {Sadler}, E.~M. and {Sansom}, A.~E. and {Seibert}, M. and {Staveley-Smith}, L. and {Sutherland}, W. and {Taylor}, E. and {Van Waerbeke}, L. and {V{\'a}zquez-Mata}, J.~A. and {Warren}, S. and {Wijesinghe}, D.~B. and {Wild}, V. and {Wilkins}, S.},
        title = "{Galaxy And Mass Assembly (GAMA): spectroscopic analysis}",
      journal = {\mnras},
         year = 2013,
        month = apr,
       volume = {430},
       number = {3},
        pages = {2047-2066},
          doi = {10.1093/mnras/stt030},
archivePrefix = {arXiv},
       eprint = {1301.7127},
 primaryClass = {astro-ph.CO},
       adsurl = {https://ui.adsabs.harvard.edu/abs/2013MNRAS.430.2047H}
}

@ARTICLE{OSullivanetal11,
   author = {{O'Sullivan}, E. and {Worrall}, D.~M. and {Birkinshaw}, M. and {Trinchieri}, G. and {Wolter}, A. and {Zezas}, A. and {Giacintucci}, S.},
    title = "{Interaction between the intergalactic medium and central radio source in the NGC 4261 group of galaxies}",
  journal = {mn},
     year = 2011,
    month = oct,
   volume = 416,
    pages = {2916},
      doi = {10.1111/j.1365-2966.2011.19239.x},
   adsurl = {http://adsabs.harvard.edu/abs/2011MNRAS.tmp.1180O},
 location = {my papers}}

@ARTICLE{OSullivanetal17,
   author = {{O'Sullivan}, E. and {Ponman}, T.~J. and {Kolokythas}, K. and {Raychaudhury}, S. and {Babul}, A. and {Vrtilek}, J.~M. and {David}, L.~P. and {Giacintucci}, S. and {Gitti}, M. and {Haines}, C.~P.},
    title = "{The Complete Local Volume Groups Sample -- I. Sample Selection and X-ray Properties of the High-Richness Subsample}",
  journal = {mn},
     year = 2017,
    month = dec,
   volume = 472,
    pages = {1482},
      doi = {10.1093/mnras/stx2078}}

@ARTICLE{OSullivanetal18b,
       author = {{O'Sullivan}, E. and {Combes}, F. and {Salom{\'e}}, P. and {David}, L.~P. and {Babul}, A. and {Vrtilek}, J.~M. and {Lim}, J. and {Olivares}, V. and {Raychaudhury}, S. and {Schellenberger}, G.},
        title = "{Cold gas in a complete sample of group-dominant early-type galaxies}",
      journal = {\aap},
         year = 2018,
        month = oct,
       volume = {618},
          eid = {A126},
        pages = {A126},
          doi = {10.1051/0004-6361/201833580},
archivePrefix = {arXiv},
       eprint = {1807.09110},
 primaryClass = {astro-ph.GA},
       adsurl = {https://ui.adsabs.harvard.edu/abs/2018A&A...618A.126O}
}

@article{Kolokythasetal18,
   author = {{Kolokythas}, K. and {O'Sullivan}, E. and {Raychaudhury}, S. and {Giacintucci}, S. and {Gitti}, M. and {Babul}, A.},
    title = "{The Complete Local Volume Groups Sample - II. A study of the central radio galaxies in the high-richness sample}",
  journal = {mn},
    month = dec,
   volume = 481,
    pages = {1550},
      doi = {10.1093/mnras/sty2030},
   adsurl = {http://adsabs.harvard.edu/abs/2018MNRAS.481.1550K},
     year = {2018}}

@ARTICLE{Kolokythasetal15,
   author = {{Kolokythas}, K. and {O'Sullivan}, E. and {Giacintucci}, S. and {Raychaudhury}, S. and {Ishwara-Chandra}, C.~H. and {Worrall}, D.~M. and {Birkinshaw}, M.},
    title = "{New insights into the evolution of the FR I radio galaxy 3C 270 (NGC 4261) from VLA and GMRT radio observations}",
  journal = {mn},
     year = 2015,
    month = jun,
   volume = 450,
    pages = {1732},
      doi = {10.1093/mnras/stv665},
   adsurl = {http://adsabs.harvard.edu/abs/2015MNRAS.450.1732K}}

@ARTICLE{2022MNRAS.510.4191K,
       author = {{Kolokythas}, Konstantinos and {Vaddi}, Sravani and {O'Sullivan}, Ewan and {Loubser}, Ilani and {Babul}, Arif and {Raychaudhury}, Somak and {Lagos}, Patricio and {Jarrett}, Thomas H.},
        title = "{The Complete Local-Volume Groups Sample - IV. Star formation and gas content in group-dominant galaxies}",
      journal = {\mnras},
         year = 2022,
        month = mar,
       volume = {510},
       number = {3},
        pages = {4191-4207},
          doi = {10.1093/mnras/stab3699},
archivePrefix = {arXiv},
       eprint = {2112.08498},
 primaryClass = {astro-ph.GA},
       adsurl = {https://ui.adsabs.harvard.edu/abs/2022MNRAS.510.4191K}
}

@ARTICLE{Birzanetal12,
   author = {{B{\^i}rzan}, L. and {Rafferty}, D.~A. and {Nulsen}, P.~E.~J. and {McNamara}, B.~R. and {R{\"o}ttgering}, H.~J.~A. and {Wise}, M.~W. and {Mittal}, R.},
    title = "{The duty cycle of radio-mode feedback in complete samples of clusters}",
  journal = {mn},
     year = 2012,
    month = dec,
   volume = 427,
    pages = {3468},
      doi = {10.1111/j.1365-2966.2012.22083.x}}

@ARTICLE{Panagouliaetal14,
   author = {{Panagoulia}, E.~K. and {Fabian}, A.~C. and {Sanders}, J.~S.},
    title = "{A volume-limited sample of X-ray galaxy groups and clusters - I. Radial entropy and cooling time profiles}",
  journal = {mn},
     year = 2014,
    month = mar,
   volume = 438,
    pages = {2341},
      doi = {10.1093/mnras/stt2349},
   adsurl = {http://adsabs.harvard.edu/abs/2014MNRAS.438.2341P}}

@ARTICLE{2024MNRAS.527.1062H,
       author = {{Hu}, Dan and {Zaja{\v{c}}ek}, Michal and {Werner}, Norbert and {Grossov{\'a}}, Romana and {J{\'a}chym}, Pavel and {Roberts}, Ian D. and {Ignesti}, Alessandro and {Kenney}, Jeffrey D.~P. and {Pl{\v{s}}ek}, Tom{\'a}{\v{s}} and {Breuer}, Jean-Paul and {Shimwell}, Timothy and {Tasse}, Cyril and {Zhu}, Zhenghao and {Wu}, Linhui},
        title = "{Ram-pressure stripped radio tail and two ULXs in the spiral galaxy HCG 97b}",
      journal = {\mnras},
         year = 2024,
        month = jan,
       volume = {527},
       number = {1},
        pages = {1062-1080},
          doi = {10.1093/mnras/stad3219},
archivePrefix = {arXiv},
       eprint = {2304.13066},
 primaryClass = {astro-ph.GA},
       adsurl = {https://ui.adsabs.harvard.edu/abs/2024MNRAS.527.1062H}
}

@ARTICLE{2013MNRAS.431..781M,
       author = {{Morsony}, Brian J. and {Miller}, Jacob J. and {Heinz}, Sebastian and {Freeland}, Emily and {Wilcots}, Eric and {Br{\"u}ggen}, Marcus and {Ruszkowski}, Mateusz},
        title = "{Simulations of bent-double radio sources in galaxy groups}",
      journal = {\mnras},
         year = 2013,
        month = may,
       volume = {431},
       number = {1},
        pages = {781-792},
          doi = {10.1093/mnras/stt210},
archivePrefix = {arXiv},
       eprint = {1210.1612},
 primaryClass = {astro-ph.CO},
       adsurl = {https://ui.adsabs.harvard.edu/abs/2013MNRAS.431..781M}
}

@ARTICLE{2024Galax..12...24E,
       author = {{Eckert}, Dominique and {Gastaldello}, Fabio and {O'Sullivan}, Ewan and {Finoguenov}, Alexis and {Brienza}, Marisa and {X-GAP Collaboration}},
        title = "{Galaxy Groups as the Ultimate Probe of AGN Feedback}",
      journal = {Galaxies},
         year = 2024,
        month = may,
       volume = {12},
       number = {3},
          eid = {24},
        pages = {24},
          doi = {10.3390/galaxies12030024},
archivePrefix = {arXiv},
       eprint = {2403.17145},
 primaryClass = {astro-ph.GA},
       adsurl = {https://ui.adsabs.harvard.edu/abs/2024Galax..12...24E}
}

@ARTICLE{2002ApJ...578...74F,
       author = {{Finoguenov}, A. and {Jones}, C. and {B{\"o}hringer}, H. and {Ponman}, T.~J.},
        title = "{ASCA Observations of Groups at Radii of Low Overdensity: Implications for the Cosmic Preheating}",
      journal = {\apj},
         year = 2002,
        month = oct,
       volume = {578},
       number = {1},
        pages = {74-89},
          doi = {10.1086/342472},
archivePrefix = {arXiv},
       eprint = {astro-ph/0206362},
 primaryClass = {astro-ph},
       adsurl = {https://ui.adsabs.harvard.edu/abs/2002ApJ...578...74F}
}

@ARTICLE{2003MNRAS.343..331P,
       author = {{Ponman}, T.~J. and {Sanderson}, A.~J.~R. and {Finoguenov}, A.},
        title = "{The Birmingham-CfA cluster scaling project - III. Entropy and similarity in galaxy systems}",
      journal = {\mnras},
         year = 2003,
        month = jul,
       volume = {343},
       number = {1},
        pages = {331-342},
          doi = {10.1046/j.1365-8711.2003.06677.x},
archivePrefix = {arXiv},
       eprint = {astro-ph/0304048},
 primaryClass = {astro-ph},
       adsurl = {https://ui.adsabs.harvard.edu/abs/2003MNRAS.343..331P}
}

@ARTICLE{2013ApJ...762...78Z,
       author = {{ZuHone}, J.~A. and {Markevitch}, M. and {Brunetti}, G. and {Giacintucci}, S.},
        title = "{Turbulence and Radio Mini-halos in the Sloshing Cores of Galaxy Clusters}",
      journal = {\apj},
         year = 2013,
        month = jan,
       volume = {762},
       number = {2},
          eid = {78},
        pages = {78},
          doi = {10.1088/0004-637X/762/2/78},
archivePrefix = {arXiv},
       eprint = {1203.2994},
 primaryClass = {astro-ph.CO},
       adsurl = {https://ui.adsabs.harvard.edu/abs/2013ApJ...762...78Z}
}

@ARTICLE{2011ApJ...743...16Z,
       author = {{ZuHone}, J.~A. and {Markevitch}, M. and {Lee}, D.},
        title = "{Sloshing of the Magnetized Cool Gas in the Cores of Galaxy Clusters}",
      journal = {\apj},
         year = 2011,
        month = dec,
       volume = {743},
       number = {1},
          eid = {16},
        pages = {16},
          doi = {10.1088/0004-637X/743/1/16},
archivePrefix = {arXiv},
       eprint = {1108.4427},
 primaryClass = {astro-ph.CO},
       adsurl = {https://ui.adsabs.harvard.edu/abs/2011ApJ...743...16Z}
}

@ARTICLE{2003ApJ...584..190F,
       author = {{Fujita}, Yutaka and {Takizawa}, Motokazu and {Sarazin}, Craig L.},
        title = "{Nonthermal Emissions from Particles Accelerated by Turbulence in Clusters of Galaxies}",
      journal = {\apj},
         year = 2003,
        month = feb,
       volume = {584},
       number = {1},
        pages = {190-202},
          doi = {10.1086/345599},
archivePrefix = {arXiv},
       eprint = {astro-ph/0210320},
 primaryClass = {astro-ph},
       adsurl = {https://ui.adsabs.harvard.edu/abs/2003ApJ...584..190F}
}

@ARTICLE{2020A&A...634A...4M,
       author = {{Mandal}, S. and {Intema}, H.~T. and {van Weeren}, R.~J. and {Shimwell}, T.~W. and {Botteon}, A. and {Brunetti}, G. and {de Gasperin}, F. and {Br{\"u}ggen}, M. and {Di Gennaro}, G. and {Kraft}, R. and {R{\"o}ttgering}, H.~J.~A. and {Hardcastle}, M. and {Tasse}, C.},
        title = "{Revived fossil plasma sources in galaxy clusters}",
      journal = {\aap},
         year = 2020,
        month = feb,
       volume = {634},
          eid = {A4},
        pages = {A4},
          doi = {10.1051/0004-6361/201936560},
archivePrefix = {arXiv},
       eprint = {1911.02034},
 primaryClass = {astro-ph.CO},
       adsurl = {https://ui.adsabs.harvard.edu/abs/2020A&A...634A...4M}
}

@ARTICLE{2001A&A...366...26E,
       author = {{En{\ss}lin}, T.~A. and {Gopal-Krishna}},
        title = "{Reviving fossil radio plasma in clusters of galaxies by adiabatic compression in environmental shock waves}",
      journal = {\aap},
         year = 2001,
        month = jan,
       volume = {366},
        pages = {26-34},
          doi = {10.1051/0004-6361:20000198},
archivePrefix = {arXiv},
       eprint = {astro-ph/0011123},
 primaryClass = {astro-ph},
       adsurl = {https://ui.adsabs.harvard.edu/abs/2001A&A...366...26E}
}

@article{Sun2012,
  author  = {Sun, Ming},
  title   = {Hot Gas in Galaxy Groups: Recent Observations},
  journal = {New Journal of Physics},
  year    = {2012},
  volume  = {14},
  number  = {4},
  pages   = {045004},
  doi     = {10.1088/1367-2630/14/4/045004}
}

@article{2022arXiv220910554H,
  author = {{Hlavacek-Larrondo}, Julie and {Li}, Yuan and {Churazov}, Eugene},
  title = "{AGN Feedback in Groups and Clusters of Galaxies}",
  journal = {arXiv e-prints},
  year = 2022,
  eid = {arXiv:2209.10554},
  pages = {arXiv:2209.10554},
  doi = {10.48550/arXiv.2209.10554}
}

@article{1999Natur.397..135P,
       author = {{Ponman}, T.~J. and {Cannon}, D.~B. and {Navarro}, J.~F.},
        title = "{The thermal imprint of galaxy formation on X-ray clusters}",
      journal = {nature},
         year = 1999,
        month = jan,
       volume = {397},
       number = {6715},
        pages = {135-137},
          doi = {10.1038/16418}
}

@article{2014IJMPD..2330007B,
  author  = {{Brunetti}, G. and {Jones}, T.~W.},
  title   = "{Cosmic Rays in Galaxy Clusters and Their Nonthermal Emission}",
  journal = {International Journal of Modern Physics D},
  year    = {2014},
  volume  = {23},
  number  = {4},
  pages   = {1430007},
  doi     = {10.1142/S021827181430007X}
}

@article{2015A&A...580A..97C,
  author  = {{Cuciti}, V. and {Cassano}, R. and {Brunetti}, G. and {Dallas}, R. and {Kale}, R. and {Venturi}, T.},
  title   = "{The occurrence of radio halos in galaxy clusters: The low-mass and high-redshift limits}",
  journal = {Astronomy \& Astrophysics},
  year    = {2015},
  volume  = {580},
  pages   = {A97},
  doi     = {10.1051/0004-6361/201526149}
}

@article{2011ApJ...740L..28B,
       author = {{Brown}, Shea and {Emerick}, Andrew and {Rudnick}, Lawrence and {Brunetti}, Gianfranco},
        title = "{On the Evolution of Cluster-scale Radio Emission}",
      journal = {ApJ},
         year = 2011,
       volume = {740},
        pages = {L28},
          doi = {10.1088/2041-8205/740/1/L28}
}

@article{2021A&A...646A.107M,
       author = {{Mandolesi}, S. and {Brunetti}, G. and {Cassano}, R. and {Cuciti}, V. and {Botteon}, A. and {Br{\"u}ggen}, M. and {Eckert}, D. and {Gastaldello}, F. and {Rojas-Bolivar}, R. and {van Weeren}, R.~J.},
        title = "{First upper limits on the diffuse radio emission in galaxy groups with LOFAR}",
      journal = {A\&A},
         year = 2021,
        month = feb,
       volume = {646},
          eid = {A107},
        pages = {A107},
          doi = {10.1051/0004-6361/202039234},
}

@article{2017ApJ...841...71G,
       author = {{Giacintucci}, Simona and {Markevitch}, Maxim and {Cassano}, Rossella and {Venturi}, Tiziana and {Clarke}, Tracy E. and {Brunetti}, Gianfranco},
        title = "{Occurrence of Radio Minihalos in a Mass-limited Sample of Galaxy Clusters}",
      journal = {ApJ},
         year = 2017,
       volume = {841},
        pages = {71},
          doi = {10.3847/1538-4357/aa6d60}
}

@article{2018MNRAS.481.1550K,
       author = {{Kolokythas}, K. and {O'Sullivan}, E. and {Raychaudhury}, S. and {Ishwara-Chandra}, C.~H. and {Kantharia}, N.~G. and {Giacintucci}, S.},
        title = "{Complete Local-volume Groups Sample - II. A study of the central radio galaxies in the high-richness sub-sample}",
      journal = {\mnras},
         year = 2018,
       volume = {481},
       number = {2},
        pages = {1550-1577},
          doi = {10.1093/mnras/sty2030}
}

@ARTICLE{2021PASA...38....9H,
       author = {{Hotan}, A.~W. and {Bunton}, J.~D. and {Chippendale}, A.~P. and {Whiting}, M. and {Tuthill}, J. and {Moss}, V.~A. and {McConnell}, D. and {Amy}, S.~W. and {Huynh}, M.~T. and {Allison}, J.~R. and {Anderson}, C.~S. and {Bannister}, K.~W. and {Bastholm}, E. and {Beresford}, R. and {Bock}, D.~C.-J. and {Bolton}, R. and {Chapman}, J.~M. and {Chow}, K. and {Collier}, J.~D. and {Cooray}, F.~R. and {Cornwell}, T.~J. and {Diamond}, P.~J. and {Edwards}, P.~G. and {Feain}, I.~J. and {Franzen}, T.~M.~O. and {George}, D. and {Gupta}, N. and {Hampson}, G.~A. and {Harvey-Smith}, L. and {Hayman}, D.~B. and {Heywood}, I. and {Jacka}, C. and {Jackson}, C.~A. and {Jackson}, S. and {Jeganathan}, K. and {Johnston}, S. and {Kesteven}, M. and {Kleiner}, D. and {Koribalski}, B.~S. and {Lee-Waddell}, K. and {Lenc}, E. and {Lensson}, E.~S. and {Mackay}, S. and {Mahony}, E.~K. and {McClure-Griffiths}, N.~M. and {McConigley}, R. and {Mirtschin}, P. and {Ng}, A.~K. and {Norris}, R.~P. and {Pearce}, S.~E. and {Phillips}, C. and {Pilawa}, M.~A. and {Raja}, W. and {Reynolds}, J.~E. and {Roberts}, P. and {Roxby}, D.~N. and {Sadler}, E.~M. and {Shields}, M. and {Schinckel}, A.~E.~T. and {Serra}, P. and {Shaw}, R.~D. and {Sweetnam}, T. and {Troup}, E.~R. and {Tzioumis}, A. and {Voronkov}, M.~A. and {Westmeier}, T.},
        title = "{Australian square kilometre array pathfinder: I. system description}",
      journal = {\pasa},
         year = 2021,
        month = mar,
       volume = {38},
          eid = {e009},
        pages = {e009},
          doi = {10.1017/pasa.2021.1},
archivePrefix = {arXiv},
       eprint = {2102.01870},
 primaryClass = {astro-ph.IM},
       adsurl = {https://ui.adsabs.harvard.edu/abs/2021PASA...38....9H}
}



\appendix

\section{Multicolor images of diffuse emission in galaxy groups}
\label{sec:appendix_plots}

\begin{figure*}
    \centering
    \includegraphics[width=0.32\textwidth]{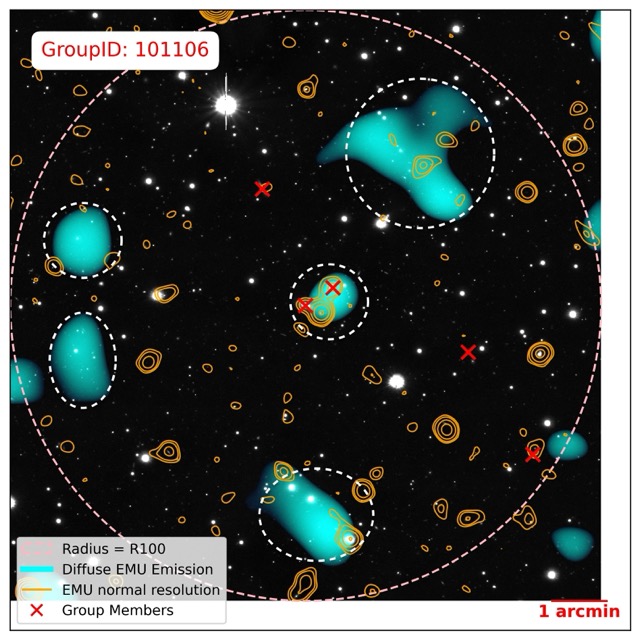}
    \includegraphics[width=0.32\textwidth]{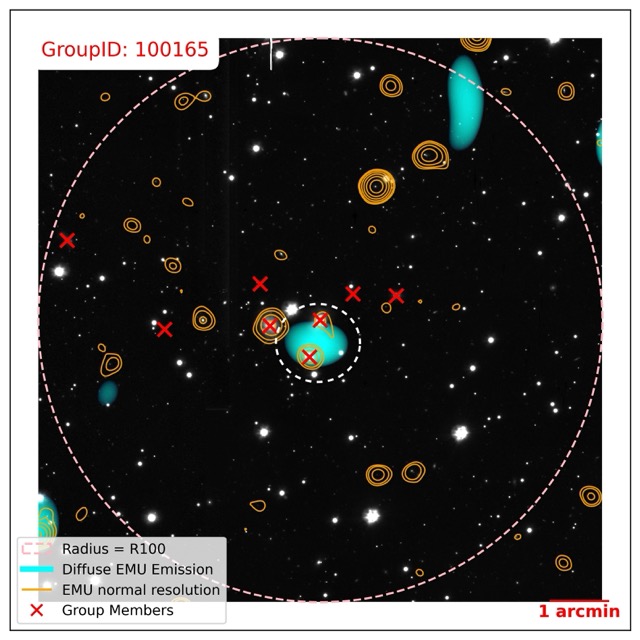}
    \includegraphics[width=0.32\textwidth]{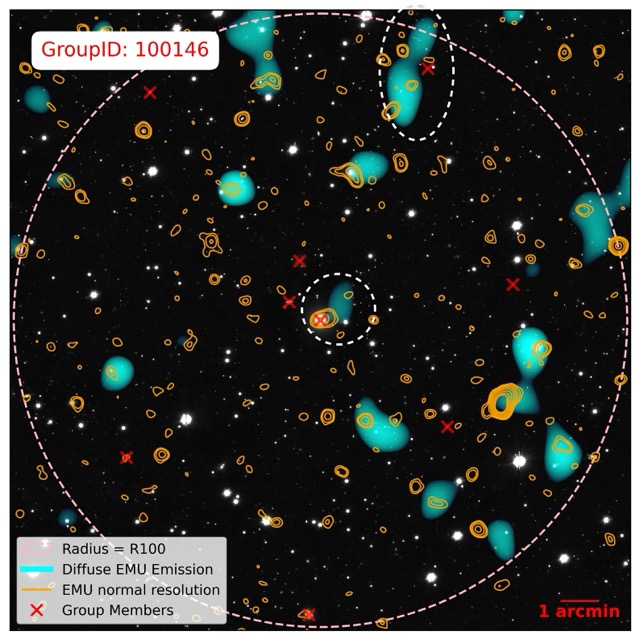}
    \\[1mm] 
    \includegraphics[width=0.32\textwidth]{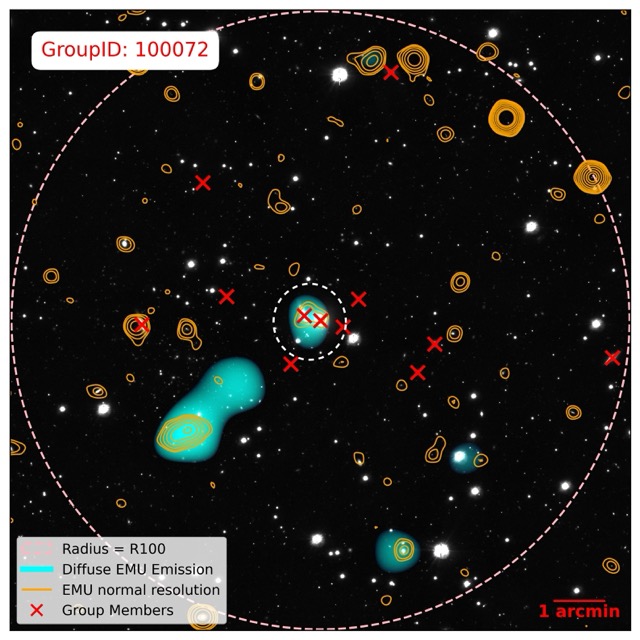}
    \includegraphics[width=0.32\textwidth]{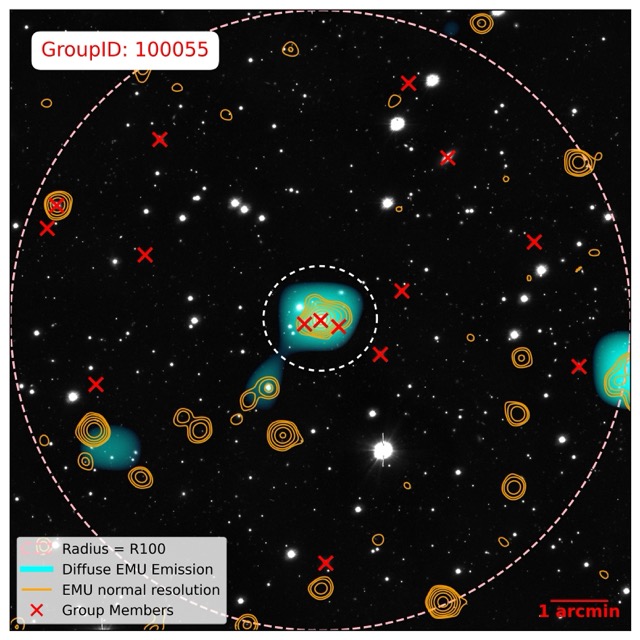}
    \includegraphics[width=0.32\textwidth]{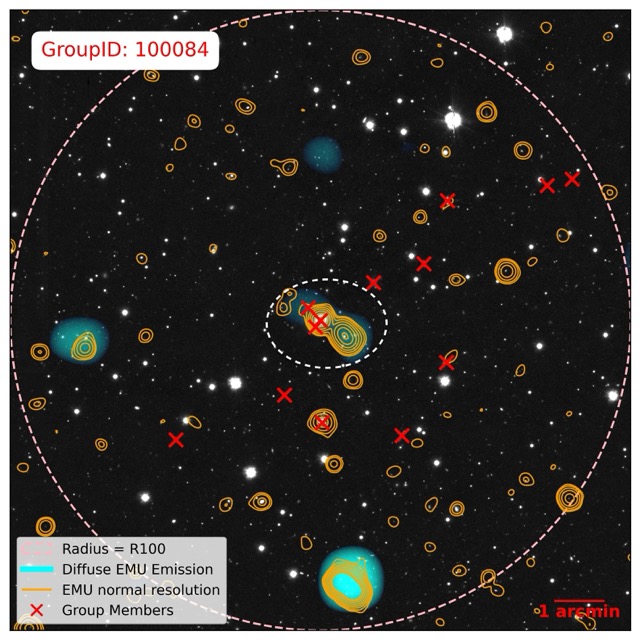}
    \\[1mm] 
    \includegraphics[width=0.32\textwidth]{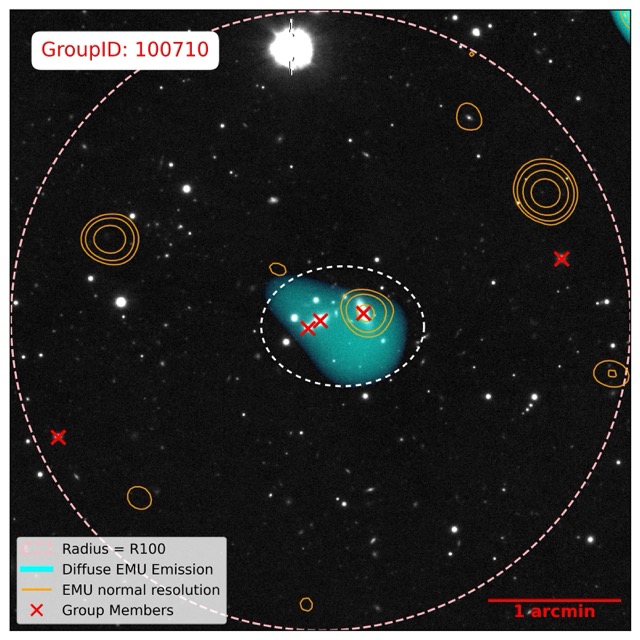}
    \includegraphics[width=0.32\textwidth]{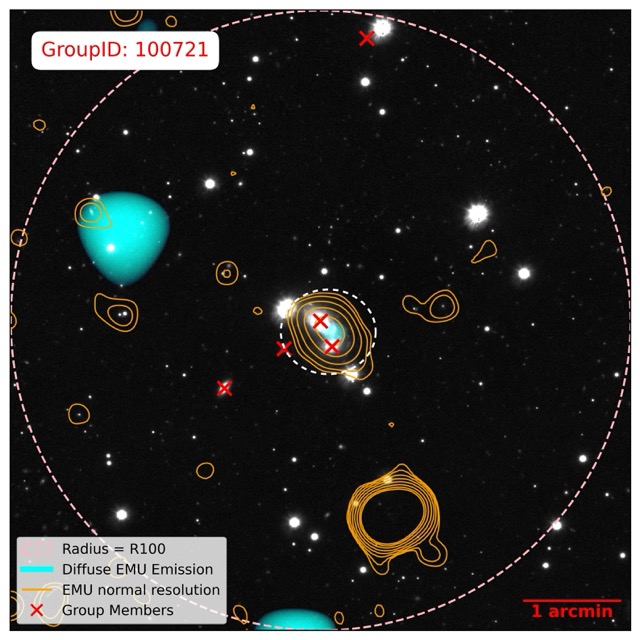}
    \includegraphics[width=0.32\textwidth]{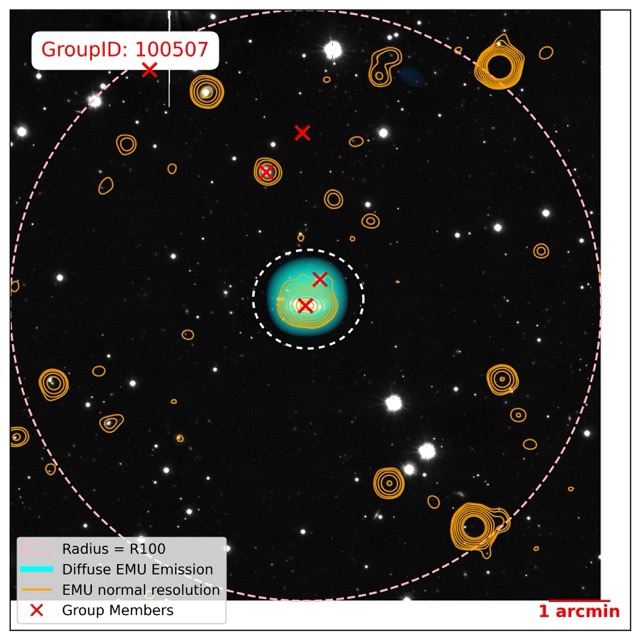}
    \\[1mm] 
    \includegraphics[width=0.32\textwidth]{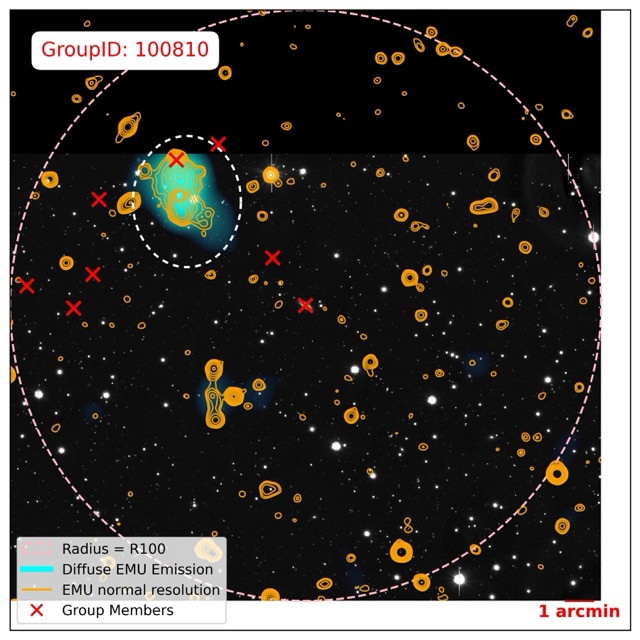}
    \includegraphics[width=0.32\textwidth]{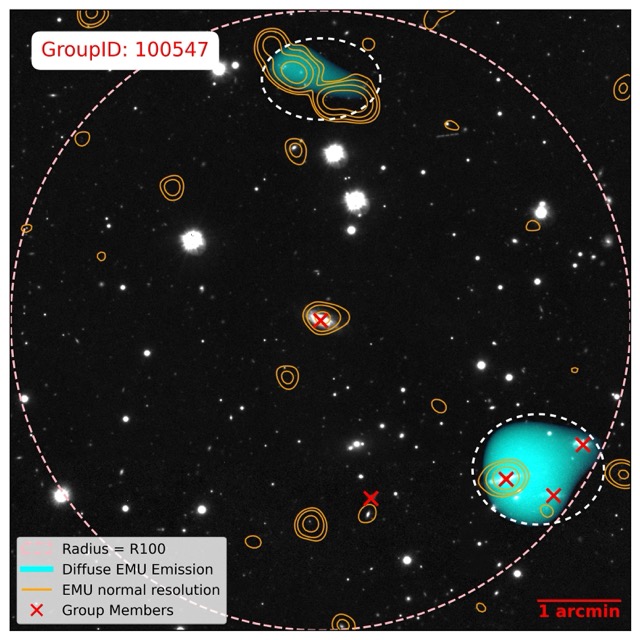}
    \includegraphics[width=0.32\textwidth]{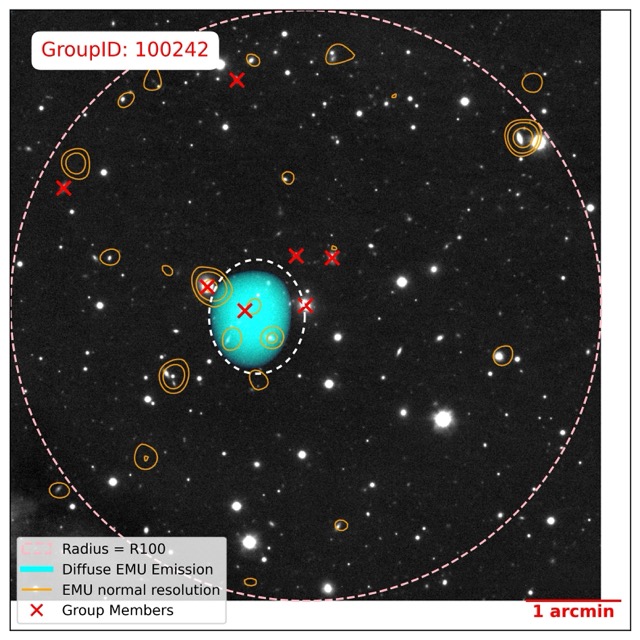}
    \\[1mm] 

    \caption{Multicolor images of candidate diffuse emission groups. Background image is grayscale GAMA map, blue color shows diffuse emission and orange contours show emission from 15$\arcsec$ EMU maps. Red markers show the positions of group galaxies and white circle highlights the diffuse emission region.}
    \label{fig:appendix_plots_1}
\end{figure*}

\clearpage 

\begin{figure*}
    \centering
    \includegraphics[width=0.32\textwidth]{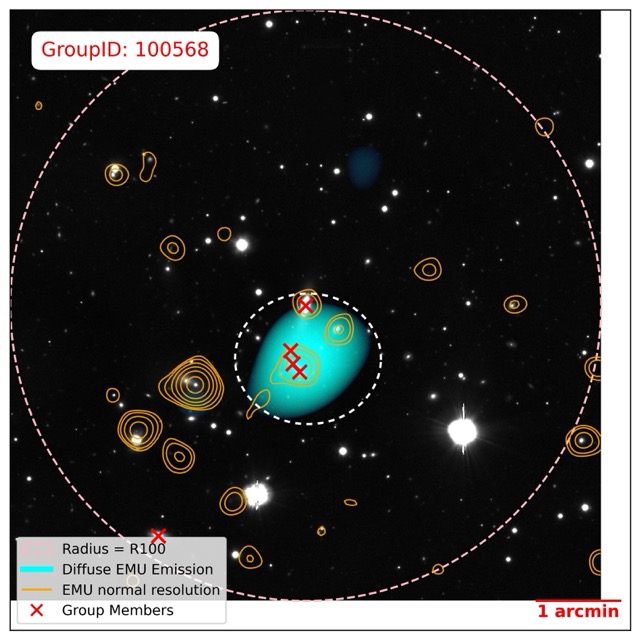}
    \includegraphics[width=0.32\textwidth]{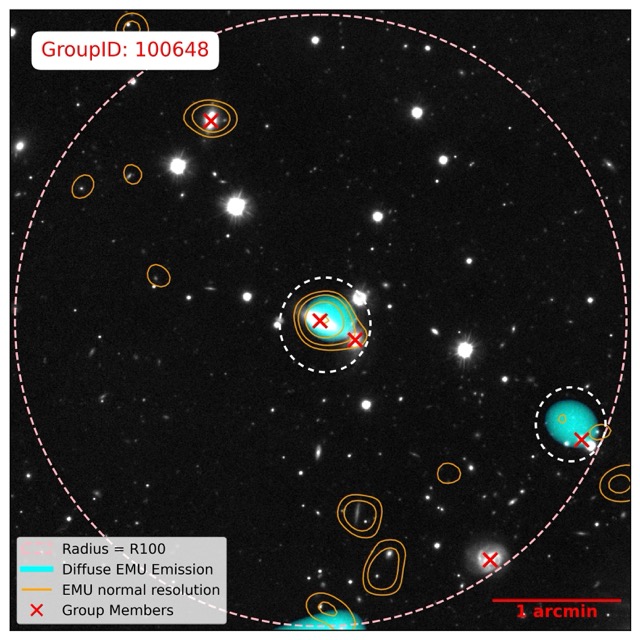}
    \includegraphics[width=0.32\textwidth]{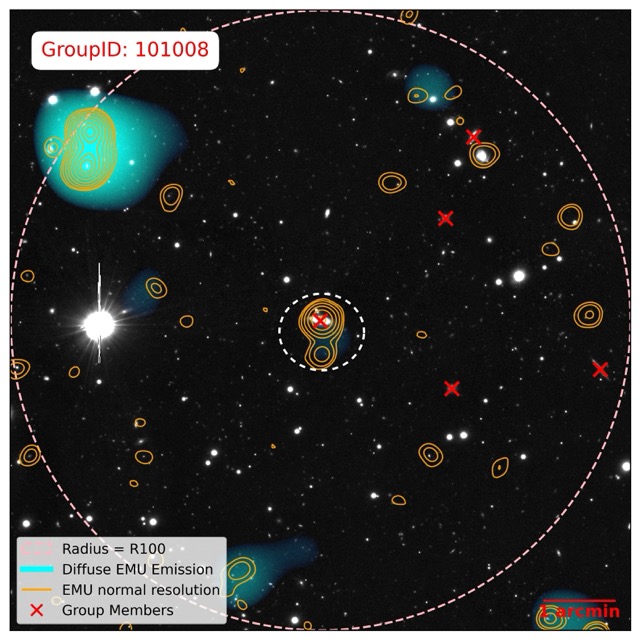}
    \\[1mm] 

    \includegraphics[width=0.32\textwidth]{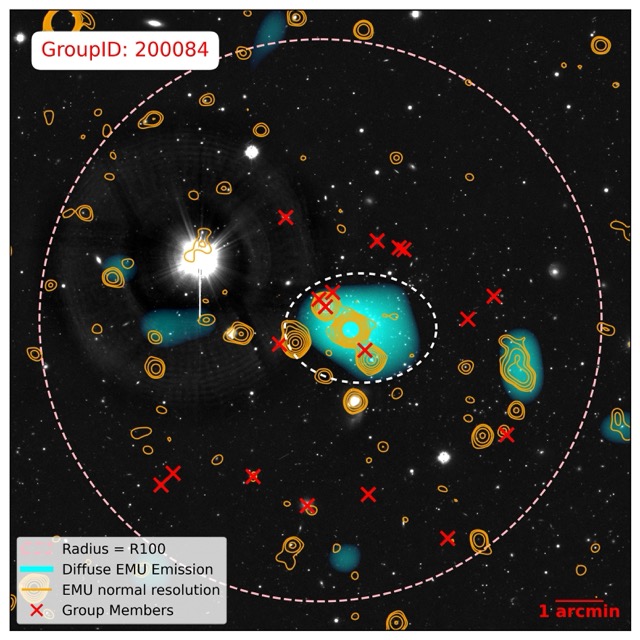}
    \includegraphics[width=0.32\textwidth]{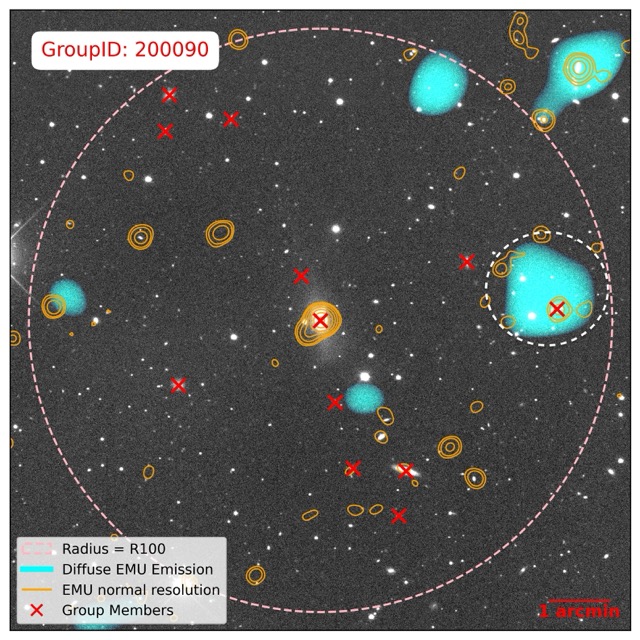}
    \includegraphics[width=0.32\textwidth]{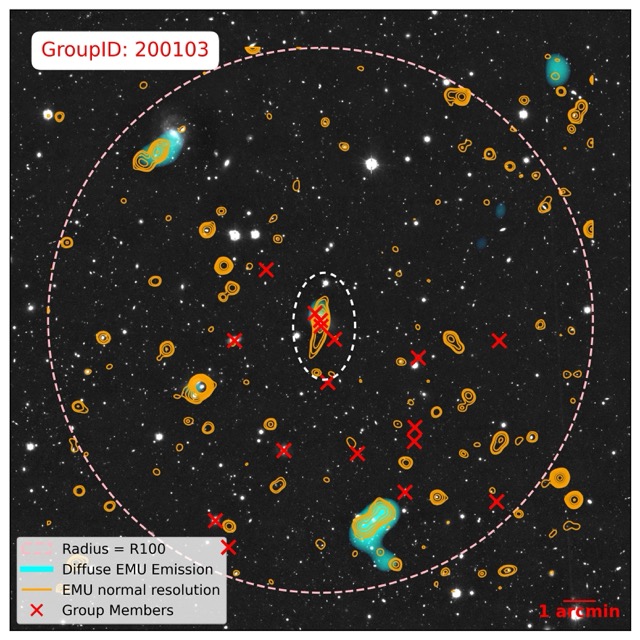}
     \\[1mm] 
     
    \includegraphics[width=0.32\textwidth]{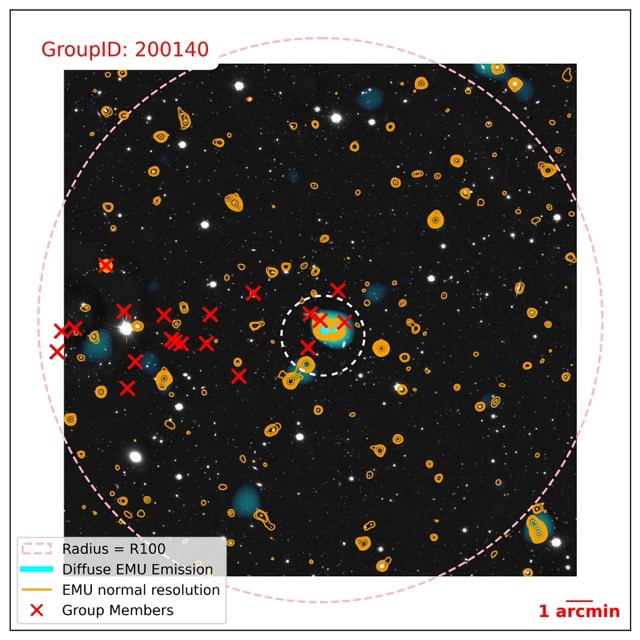}   
    \includegraphics[width=0.32\textwidth]{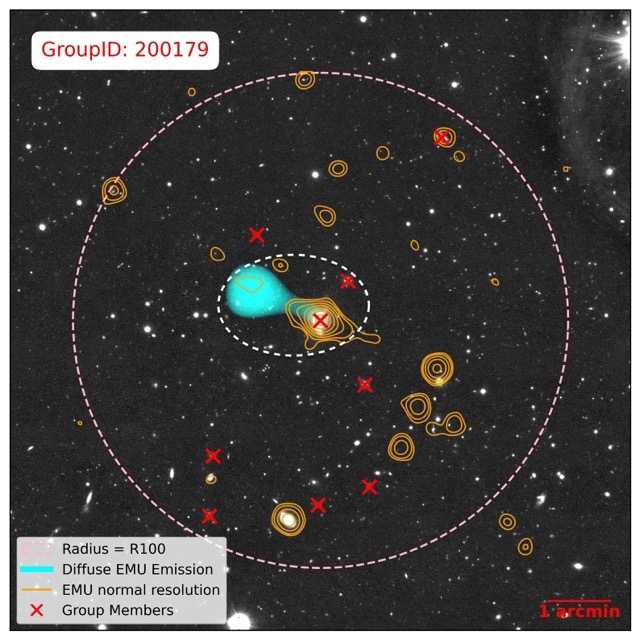}
    \includegraphics[width=0.32\textwidth]{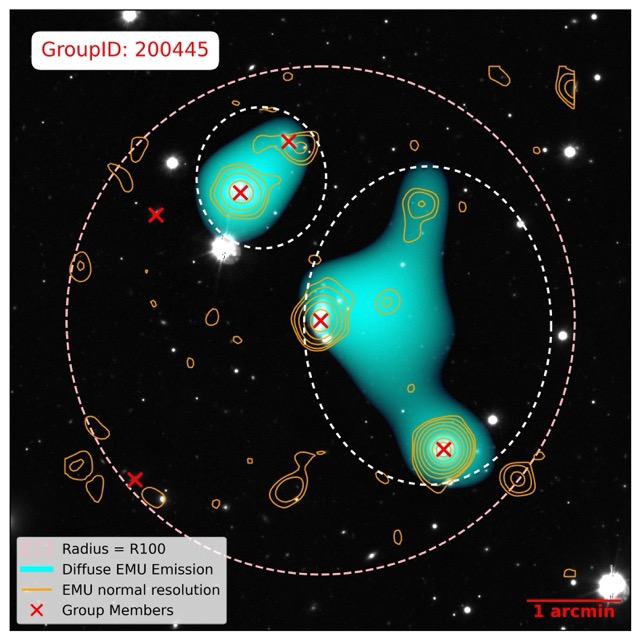}
    \\[1mm] 

    \includegraphics[width=0.32\textwidth]{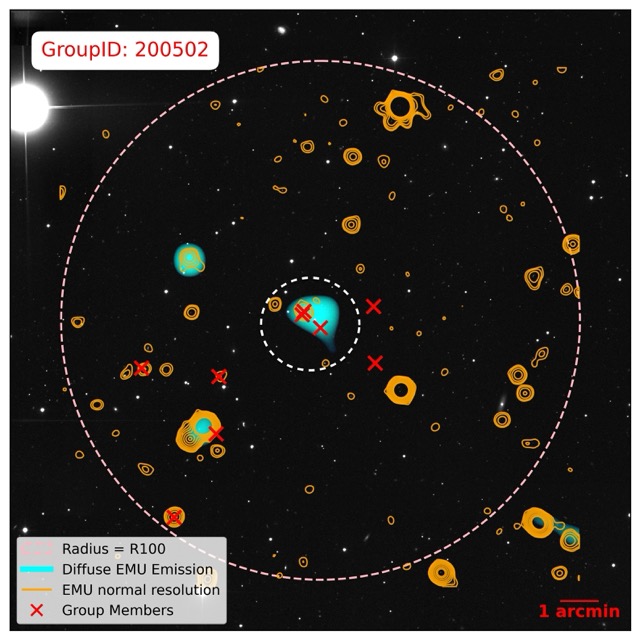}
    \includegraphics[width=0.32\textwidth]{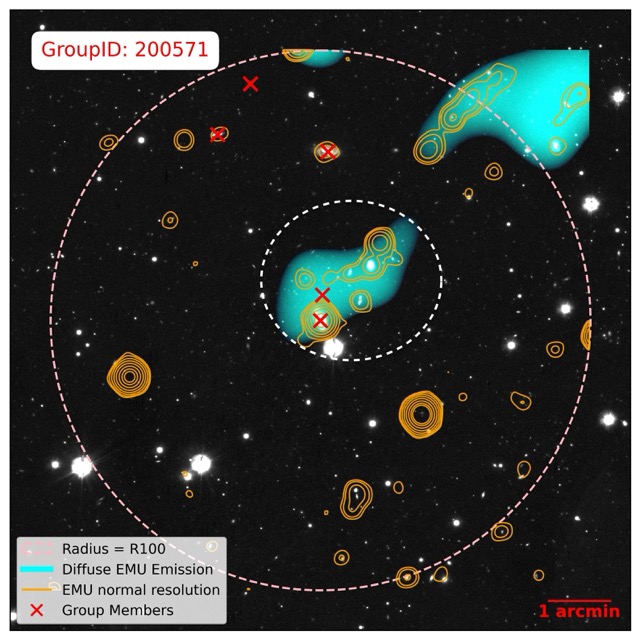}
    \includegraphics[width=0.32\textwidth]{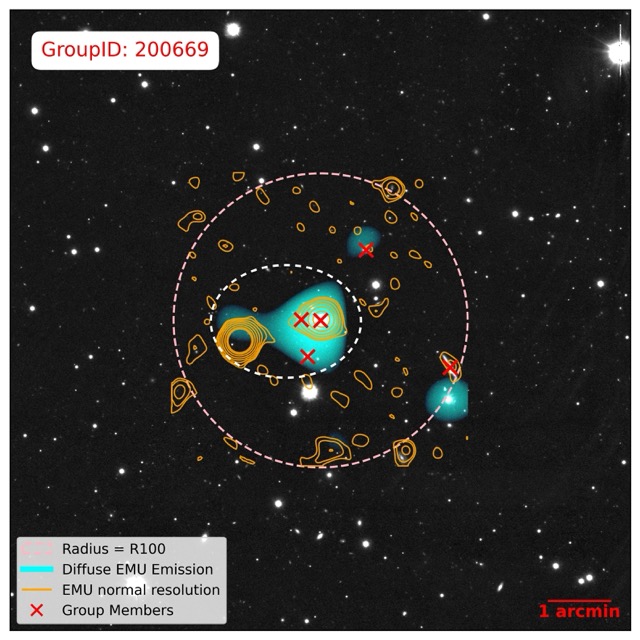}
    \\[1mm] 
    \caption{Multicolor images of candidate diffuse emission groups. Background image is grayscale GAMA map, blue color shows diffuse emission and orange contours show emission from 15$\arcsec$ EMU maps. Red markers show the positions of group galaxies and white circle highlights the diffuse emission region.}
\end{figure*}

\clearpage

\begin{figure*}
    \centering
    \includegraphics[width=0.32\textwidth]{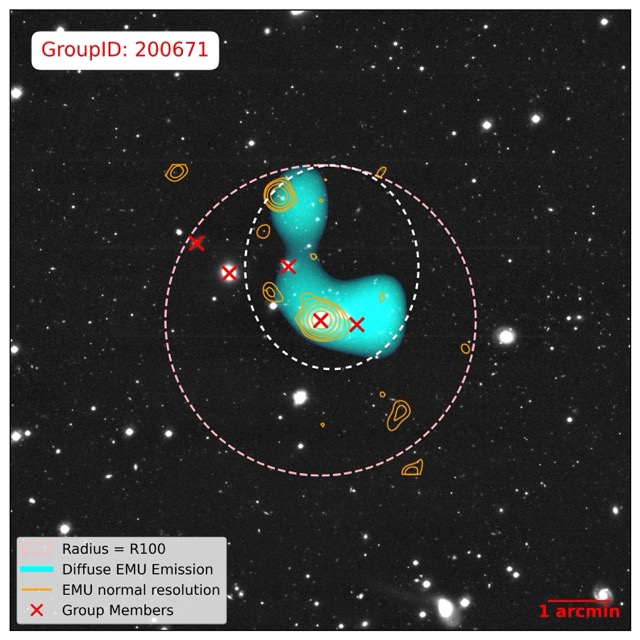}
    \includegraphics[width=0.32\textwidth]{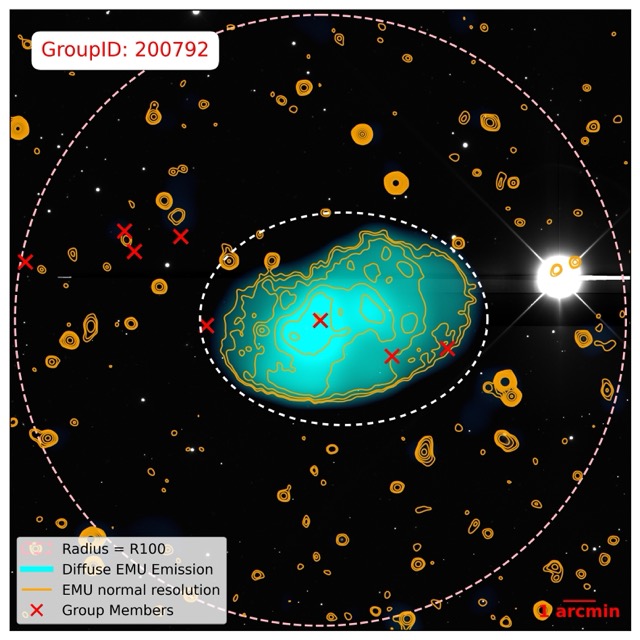}
    \includegraphics[width=0.32\textwidth]{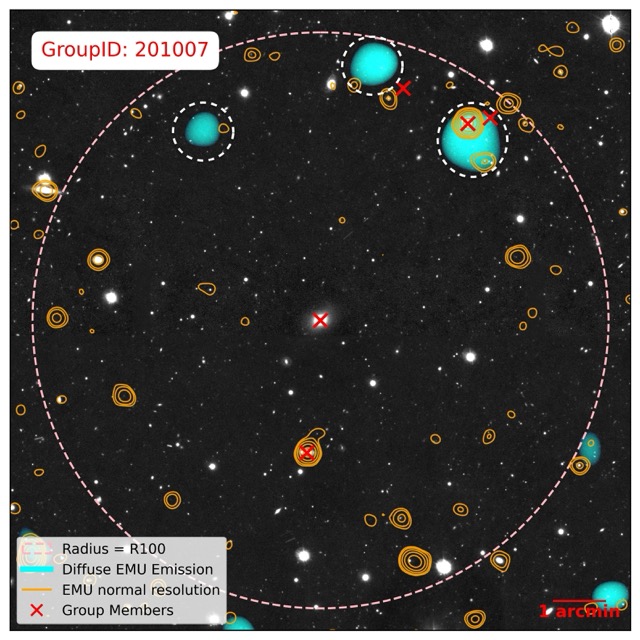}
    \\[1mm] 
    
    \includegraphics[width=0.32\textwidth]{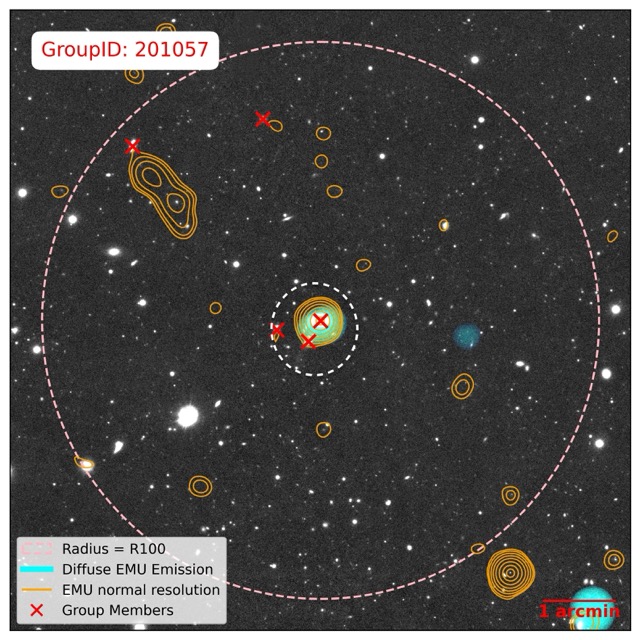}
    \includegraphics[width=0.32\textwidth]{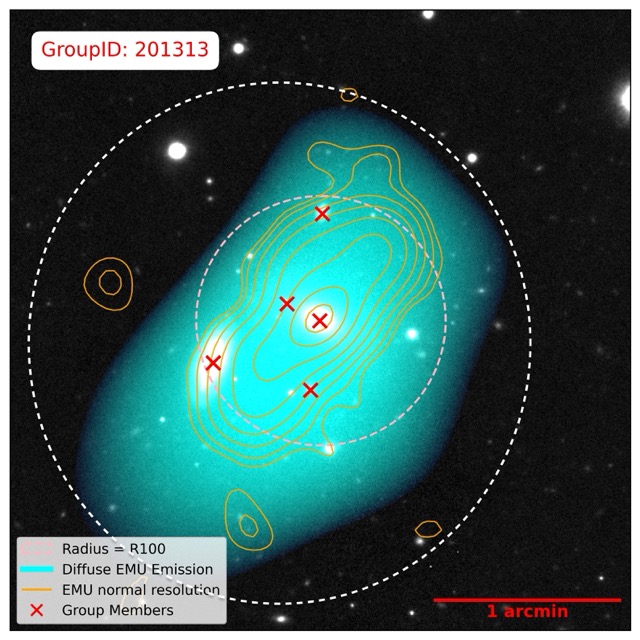}
  \includegraphics[width=0.32\textwidth]{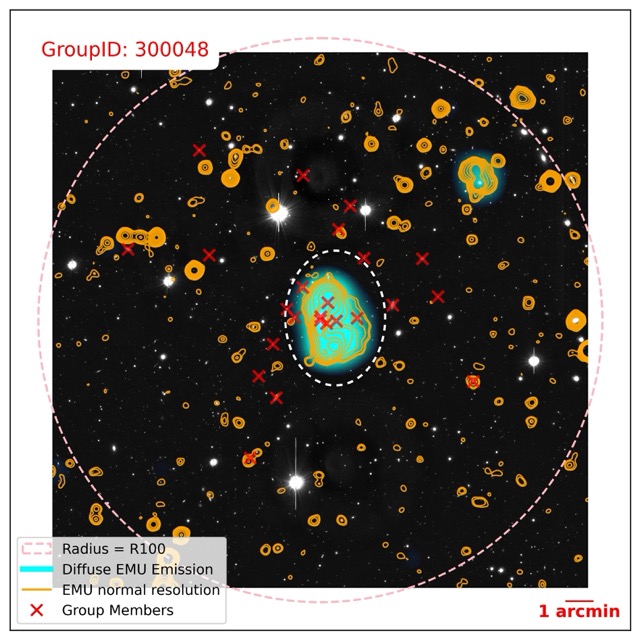}
   \\[1mm] 

    \includegraphics[width=0.32\textwidth]{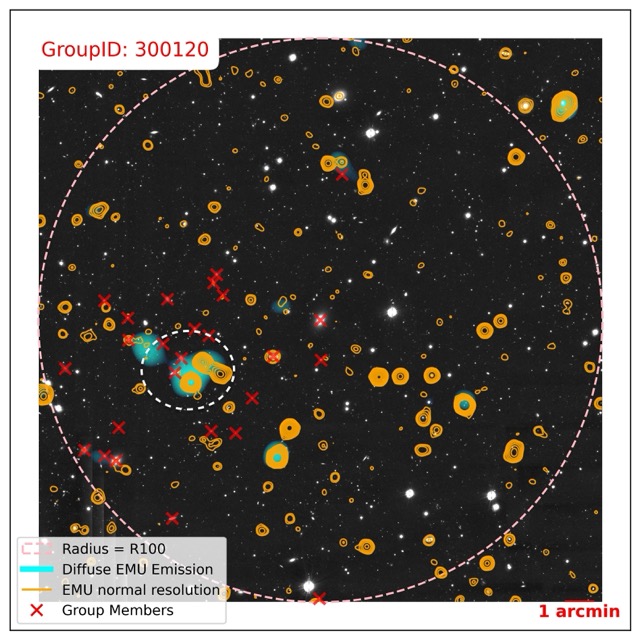}
    \includegraphics[width=0.32\textwidth]{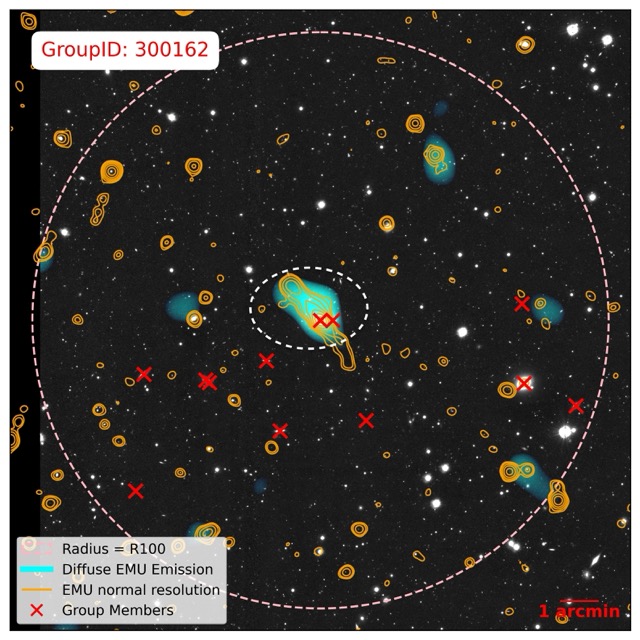}
    \includegraphics[width=0.32\textwidth]{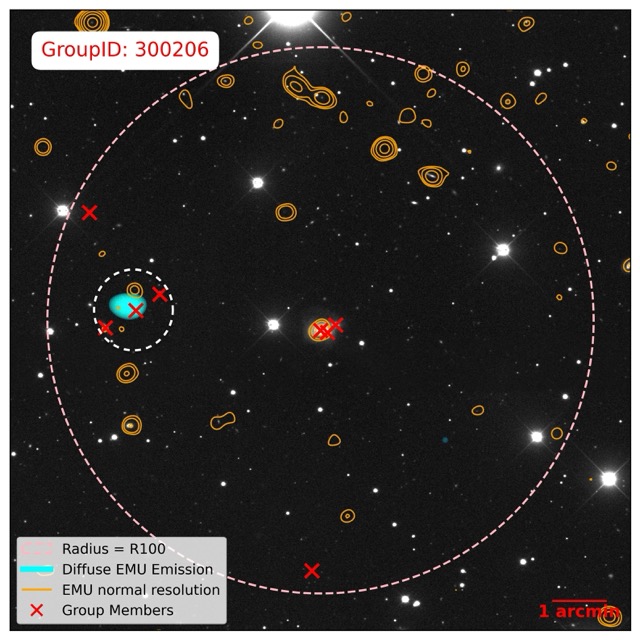}
    \\[1mm] 


    \includegraphics[width=0.32\textwidth]{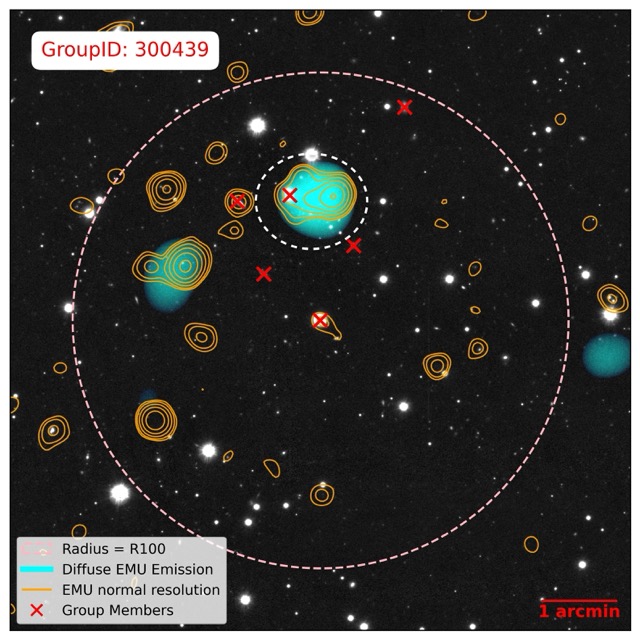}
    \includegraphics[width=0.32\textwidth]{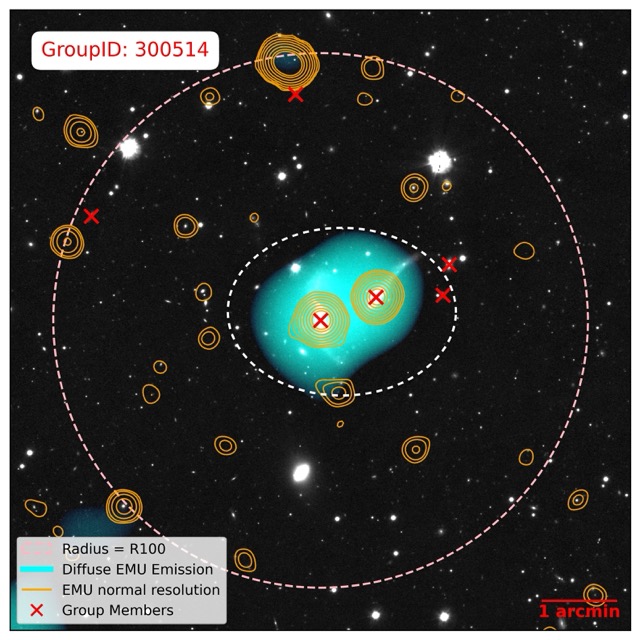}
    \includegraphics[width=0.32\textwidth]{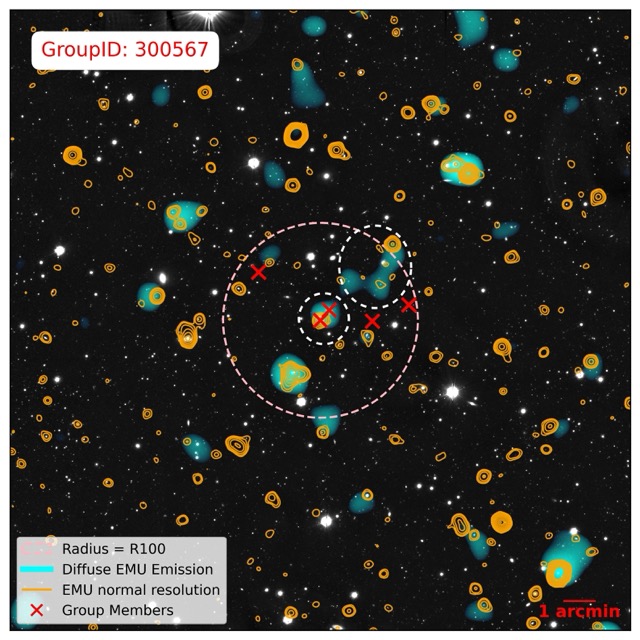}
    \\[1mm] 
    \caption{Multicolor images of candidate diffuse emission groups. Background image is grayscale GAMA map, blue color shows diffuse emission and orange contours show emission from 15$\arcsec$ EMU maps. Red markers show the positions of group galaxies and white circle highlights the diffuse emission region.}
\end{figure*}

\begin{figure*}
    \centering

    \includegraphics[width=0.32\textwidth]{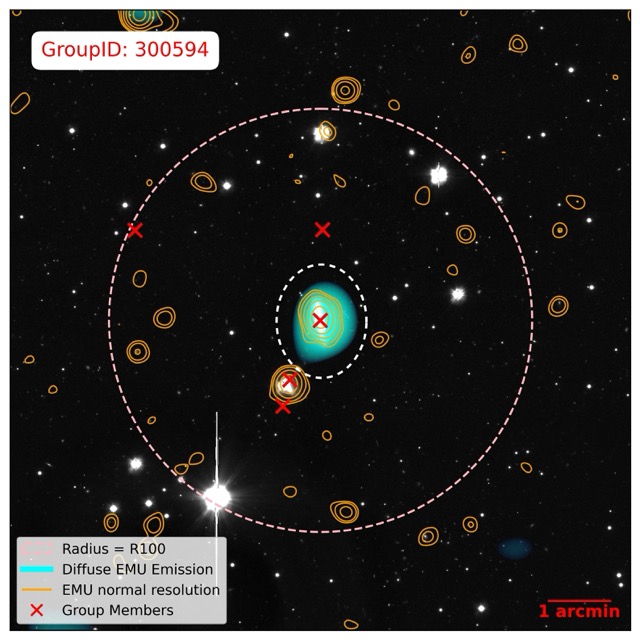}
    \includegraphics[width=0.32\textwidth]{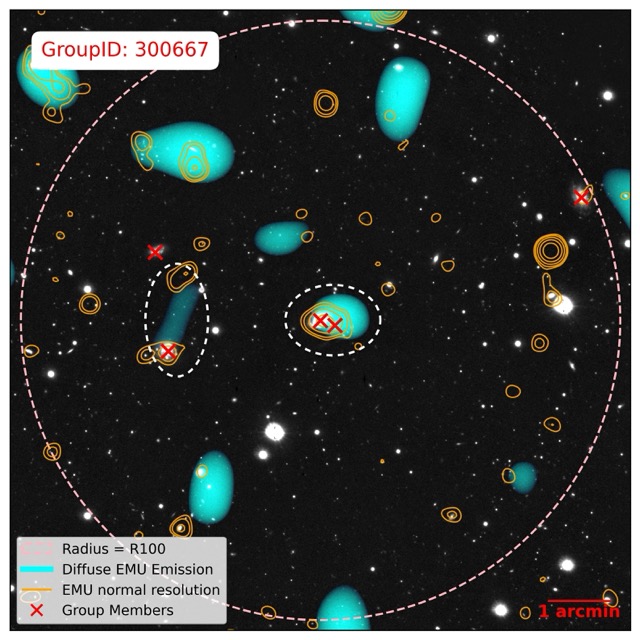}
    \includegraphics[width=0.32\textwidth]{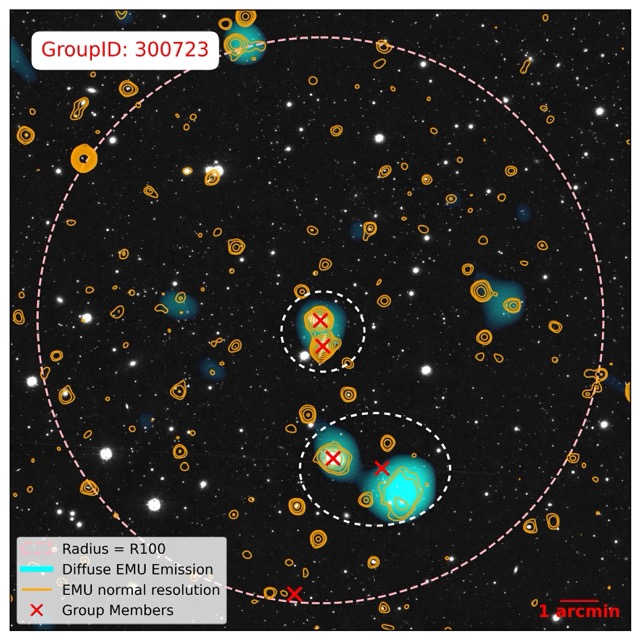}
    \\[1mm] 

    \includegraphics[width=0.32\textwidth]{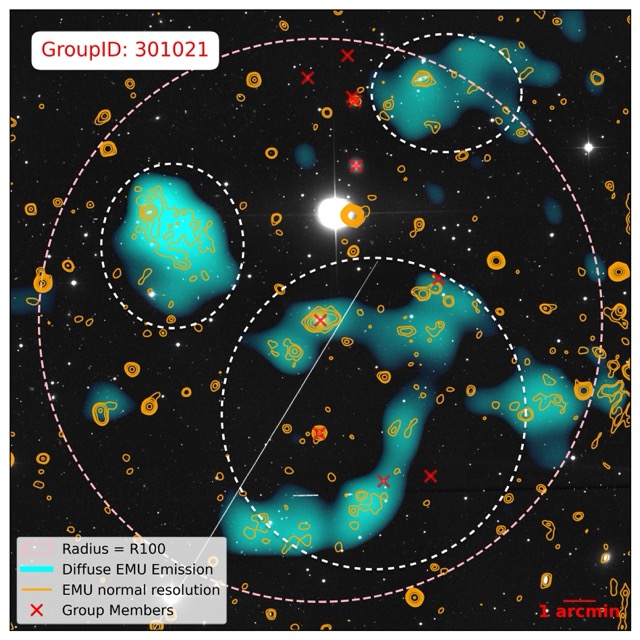}
    \includegraphics[width=0.32\textwidth]{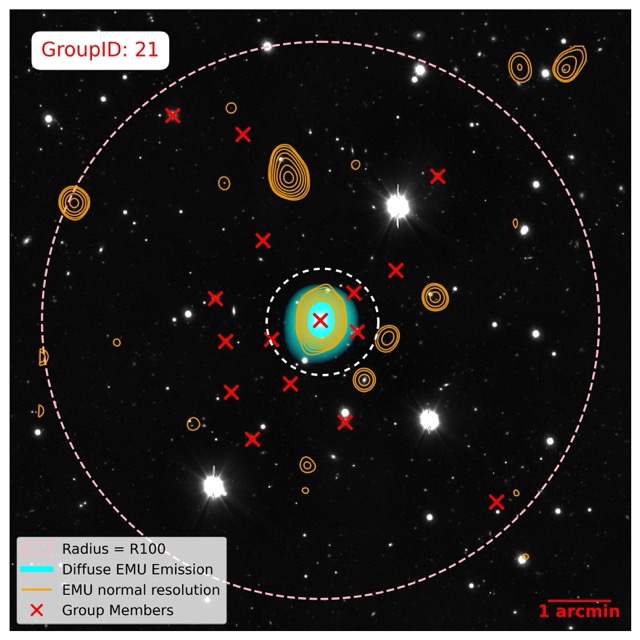}
    \includegraphics[width=0.32\textwidth]{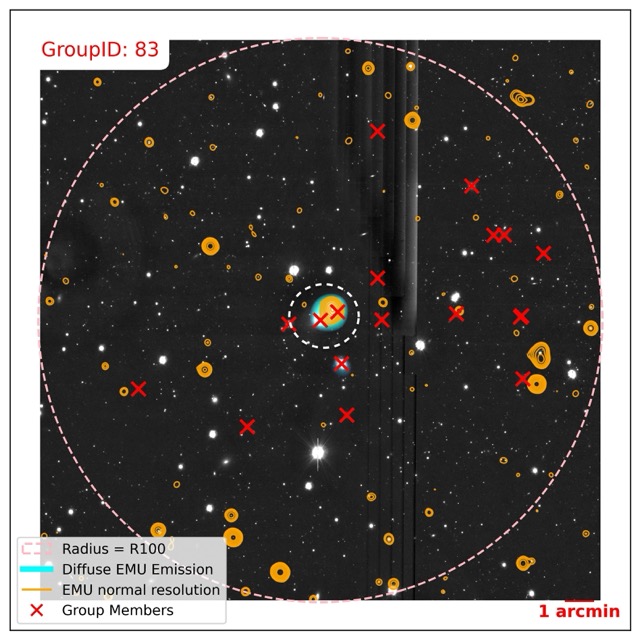}
    \\[1mm] 

    \includegraphics[width=0.32\textwidth]{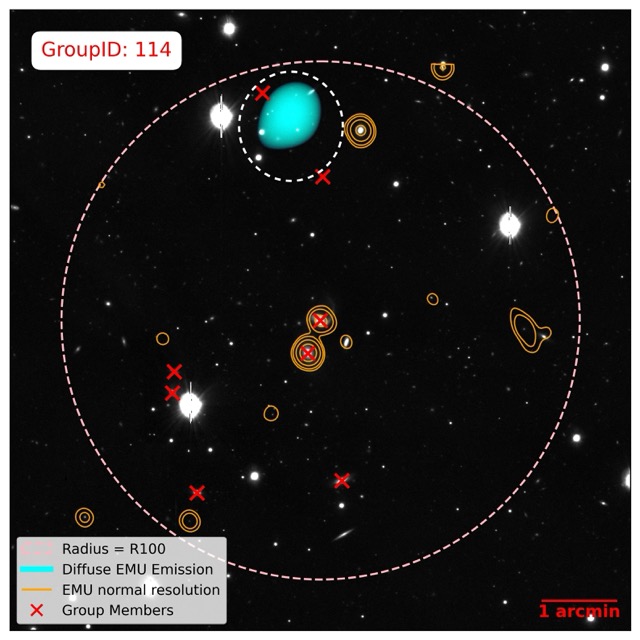}
    \includegraphics[width=0.32\textwidth]{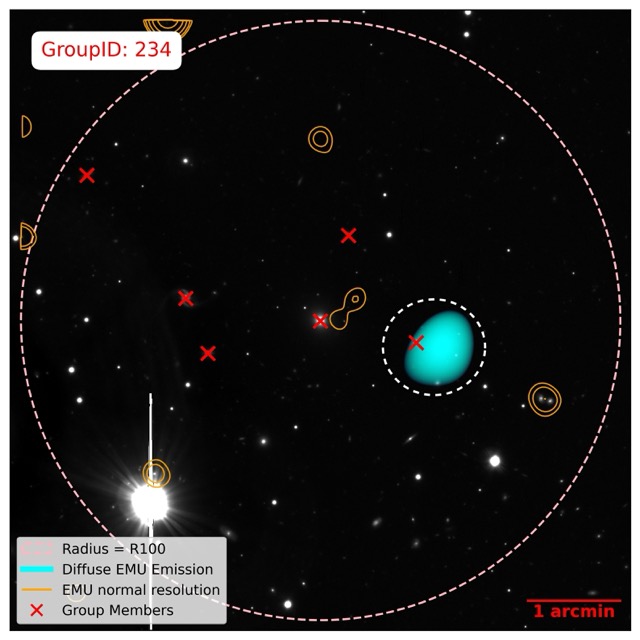}
    \includegraphics[width=0.32\textwidth]{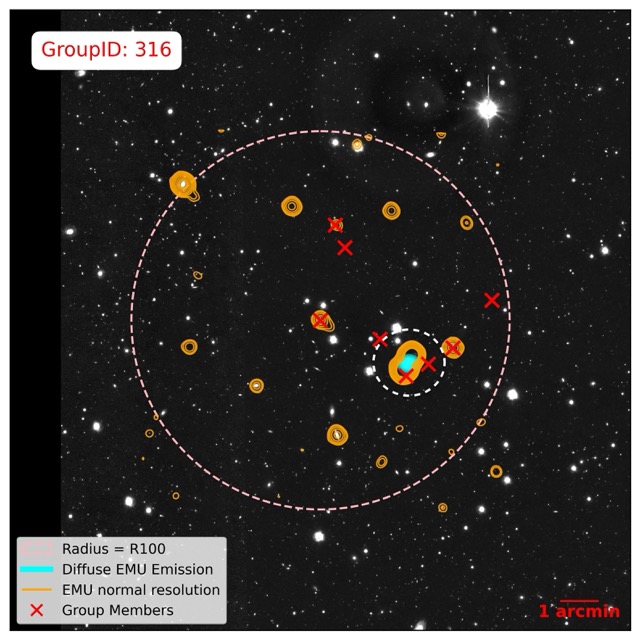}
      \\[1mm] 

    \includegraphics[width=0.32\textwidth]{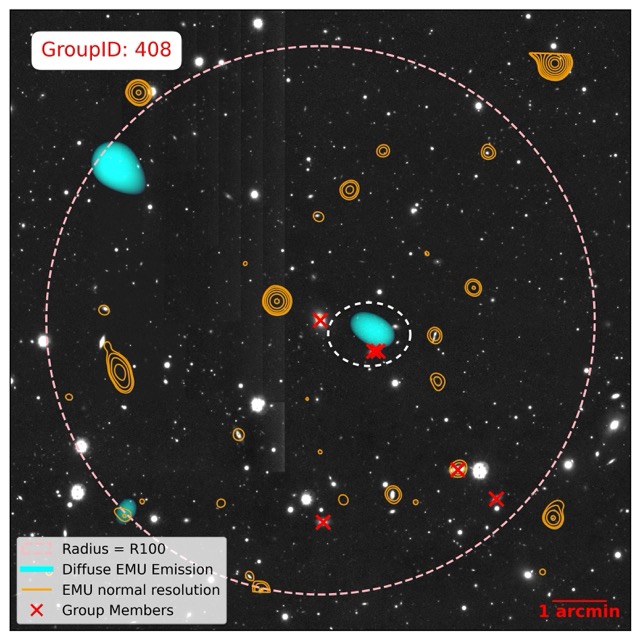}
  
    \caption{Multicolor images of candidate diffuse emission groups. Background image is grayscale GAMA map, blue color shows diffuse emission and orange contours show emission from 15$\arcsec$ EMU maps. Red markers show the positions of group galaxies and white circle highlights the diffuse emission region.}
\end{figure*}
\section{Simulated gaussian in different artefact dominated regions
} \label{Appendix - numerical simulations}
\begin{figure*}
\includegraphics[width= 1\columnwidth]{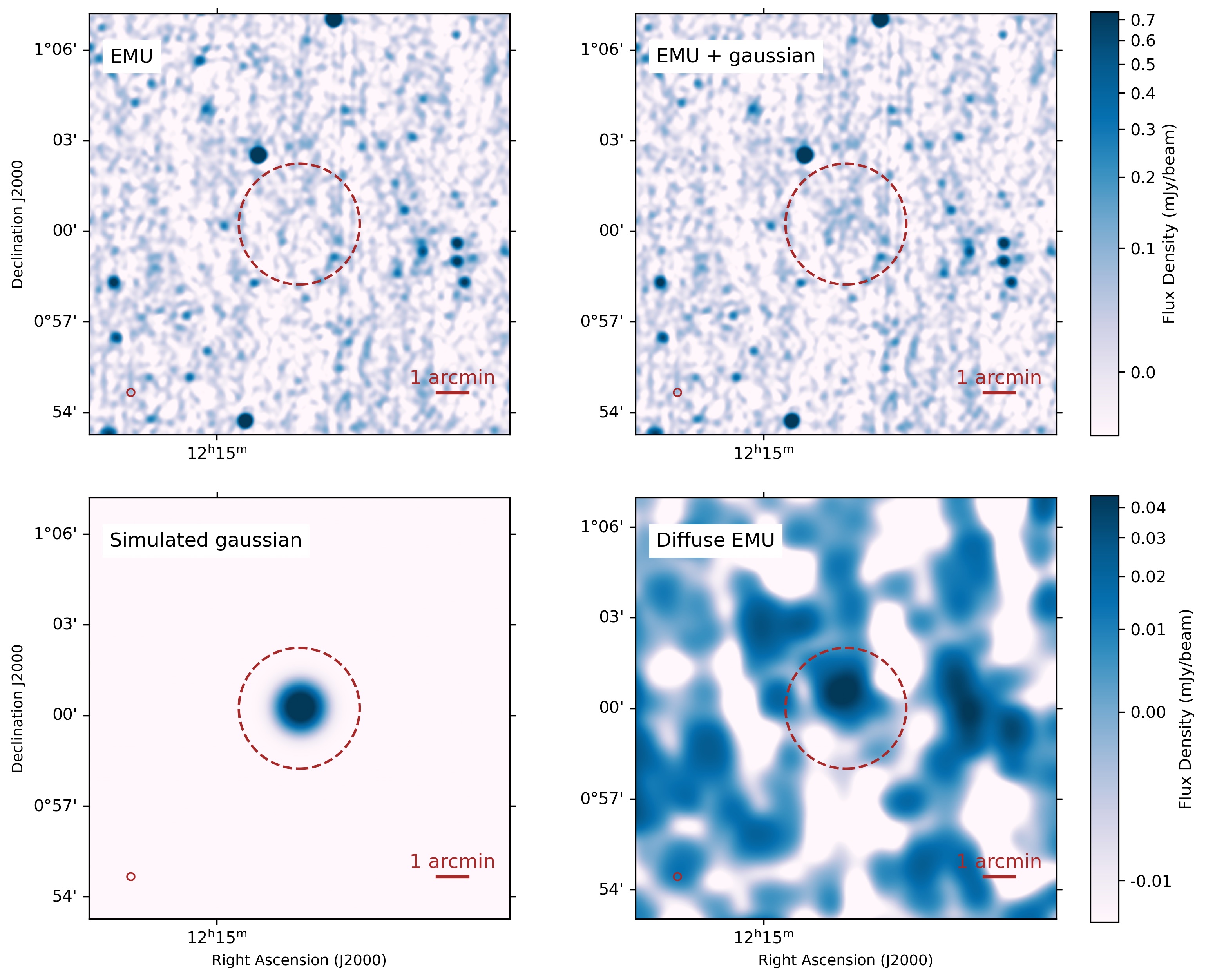}
\includegraphics[width= 1\columnwidth]{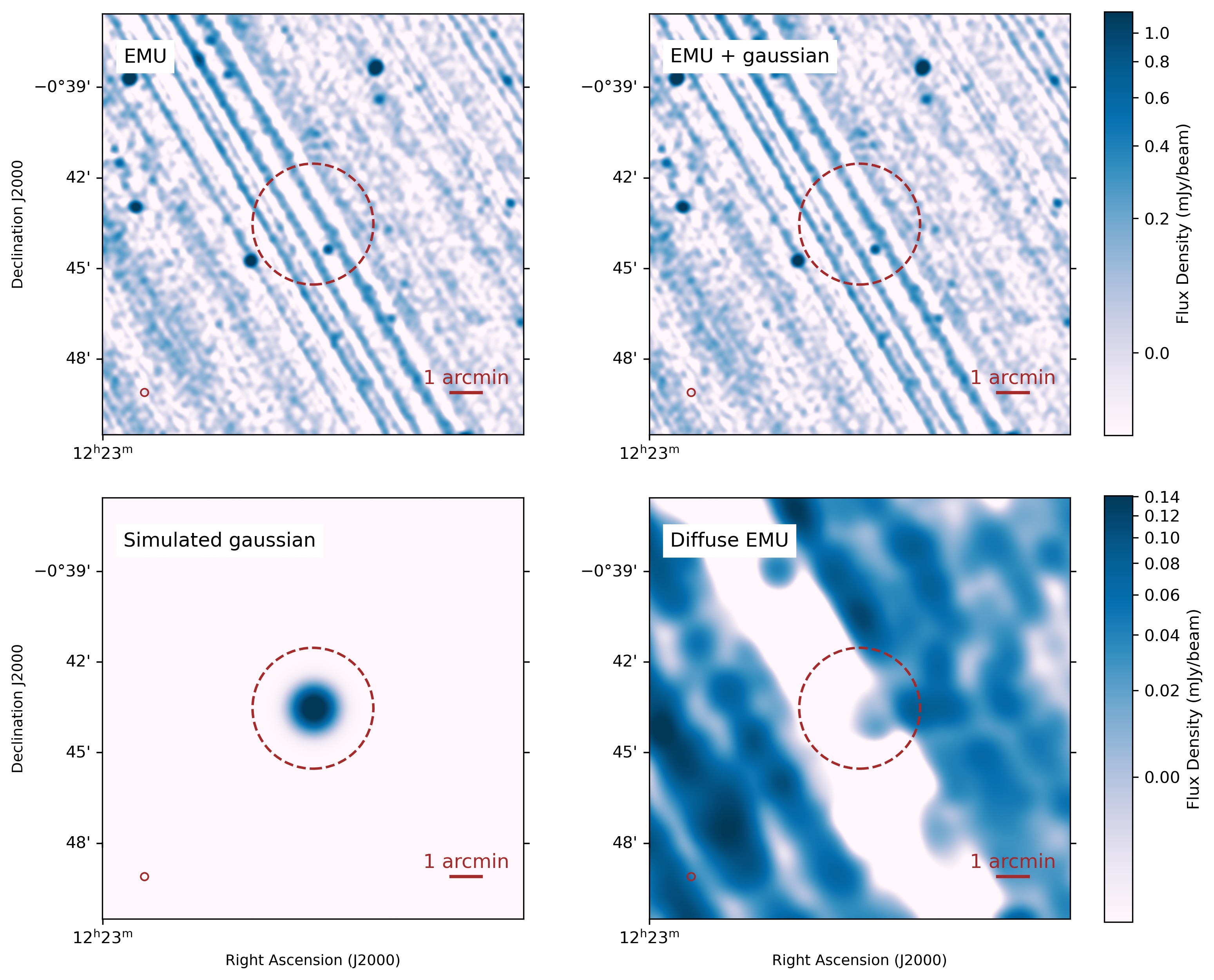}
\\[1cm]
\includegraphics[width= 1\columnwidth]{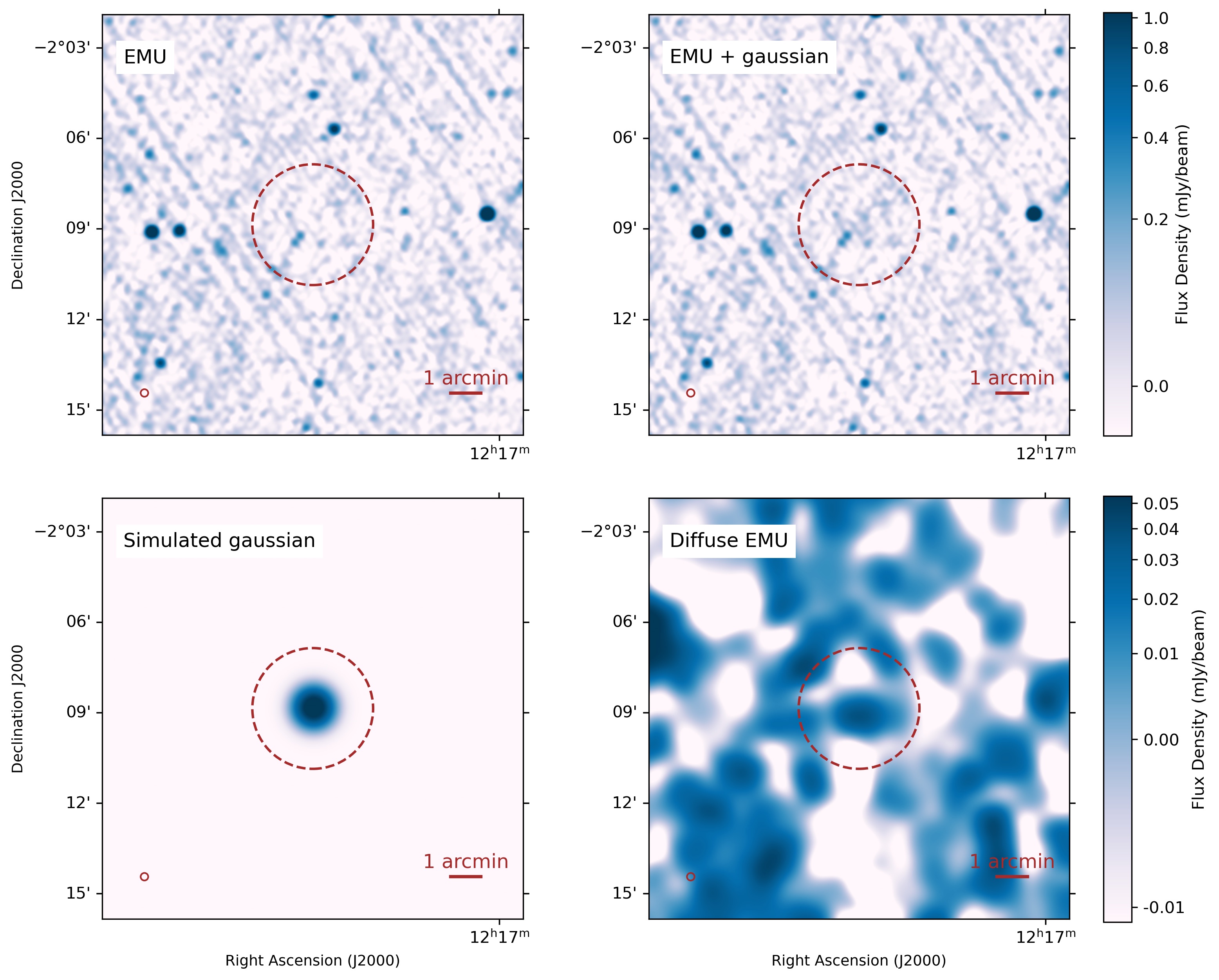}
\includegraphics[width= 1\columnwidth]{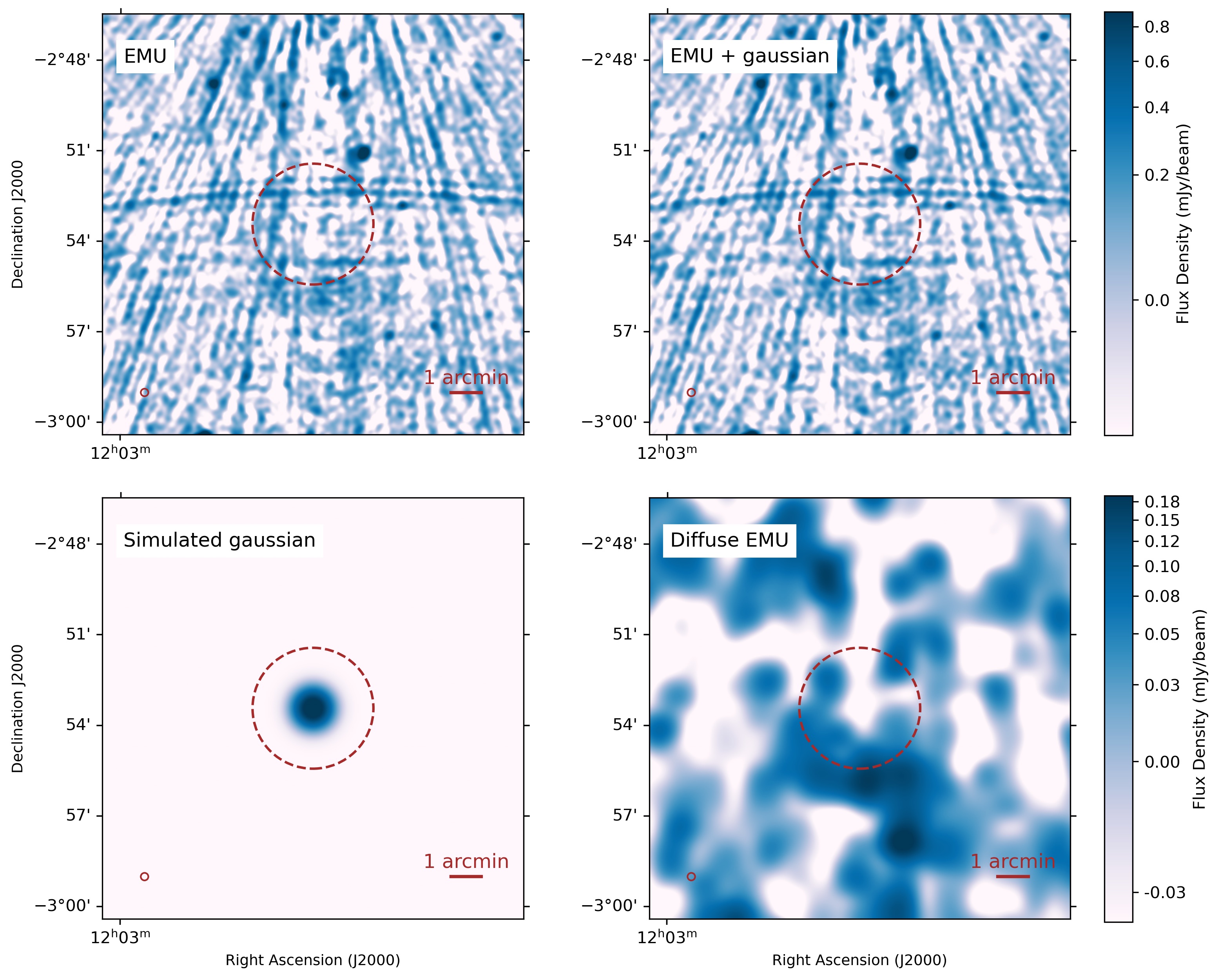}
\caption{Example images of simulated Gaussian profiles in one EMU tile of G12 region. The 2D-Gaussian profile is created with a FWHM of 60$\arcsec$, with an integrated flux density of 1.4 mJy, and a peak flux density of 88 $\mu$Jy beam$^{-1}$. This Gaussian profile is placed in different artefact dominated regions and their behaviour in resulting diffuse maps is shown. }
\label{Simulated_gaussian}
\end{figure*}


\bsp	
\label{lastpage}
\end{document}